\documentclass[final,range]{ar2e}%Put [range] option here for references 

\usepackage{epsfig,latexsym,color}
\usepackage{ulem}
\begin{document}

%\input epsf.tex    %<-If you need EPS figures to be
                   %  called in {figure} environment for PC
%\input epsf.def   %<-If you need EPS figures to be
                   %  called in {figure} environment for Macintosh
\input psfig.sty

\def\beq{\begin{equation}}
\def\eeq{\end{equation}}
\def\beqa{\begin{eqnarray}}
\def\eeqa{\end{eqnarray}}
\def\ddb{$D^0-\bar{D^0}$}
\def\nn{\nonumber}

\def\simlt{\mathrel{\lower2.5pt\vbox{\lineskip=0pt\baselineskip=0pt
           \hbox{$<$}\hbox{$\sim$}}}}
\def\simgt{\mathrel{\lower2.5pt\vbox{\lineskip=0pt\baselineskip=0pt
           \hbox{$>$}\hbox{$\sim$}}}}

%\jname{}
%\jyear{} 
%\jvol{}
%\ARinfo{}

\title{$D^0-\bar{D^0}$ Mixing and Rare Charm Decays\footnote{
To appear in Annual Review of Nuclear and Particle Science, Vol.~53, 2003.}}

\markboth{BURDMAN \& SHIPSEY}{$D^0$--$\bar{D^0}$ MIXING AND RARE CHARM DECAYS}

\author{Gustavo Burdman
\affiliation{Theory Group, Lawrence Berkeley National Laboratory, 
Berkeley, California 94720 \\ email: gaburdman@lbl.gov}
Ian Shipsey 
\affiliation{Department of Physics, Purdue University, 
West Lafayette, Indiana 47907 \\ email: shipsey@physics.purdue.edu}
}


%%
%%
%%   COMP Please format PACS codes same way as Key Words
%%
%%
%% PACS codes     13.20.Fc, 13.25.Ft, 12.60.-i
%
\begin{abstract}
We review the current status of flavor-changing neutral currents in the 
charm sector. We focus on the standard-model predictions
and identify the main sources of theoretical uncertainties in both \ddb~mixing
and rare charm decays. The potential of these observables for constraining 
short-distance physics in the standard model and its extensions is compromised
by the presence of large nonperturbative effects. We examine the possible discovery 
windows in which short-distance physics can be tested and study the
effects of various extensions of the standard model. The current experimental 
situation and future prospects are reviewed. 
\end{abstract}

\maketitle

\section{INTRODUCTION}
\label{intro}
The remarkable success of the standard model  in describing all 
experimental information to date suggests that the quest
for deviations from it should be directed either at higher energy scales
or at small effects in low-energy observables. To the last group belong
measurements (with precision surpassing 1\%) of electroweak observables
at CERN's Large Electron Positron Collider (LEP) and the SLAC Linear Detector 
(SLD) as well as the Tevatron experiments~\cite{lep1}. 
Tests of the standard model through quantum corrections are a powerful
tool for probing the high energy scales possibly related to electroweak
symmetry breaking and the flavor problem. The absence of flavor-changing
neutral currents (FCNC) at tree level in the standard model implies that processes
involving these currents are a primary test of the quantum structure of 
the theory. 
The study of FCNC has been focused on processes involving
$K$ and $B$ mesons, such as $K^0$--$\bar{K}^0$ and $B^0_d$--$\bar{B}^0_d$
mixing, and on rare decays involving transitions such as 
$s\to d\ell^+\ell-$, $s\to d\nu\bar\nu$, $b\to s\gamma$, 
and $b\to s\ell^+\ell^-$.  

The analogous FCNC processes in the charm sector have received 
considerably less scrutiny. This is perhaps because,
on general grounds, the standard-model expectations are very small  for both 
$D^0$--$\bar{D}^0$ mixing
and FCNC decays.     
For instance, no large nondecoupling effects arise from 
a heavy fermion in the leading one-loop contributions.  This is 
in sharp contrast with $K$ and $B$ FCNC processes, which are 
affected by the presence of the top quark in loops.
In the standard model, $D$-meson FCNC transitions involve the rather light down-quark
sector, which implies an efficient Glashow-Iliopoulos-Maiani (GIM) cancellation. 
In many cases, extensions of the standard model may upset this suppression and 
contributions may be orders of magnitude larger than the standard-model 
short-distance contributions.

As a first step, 
and in order to ponder the existence of a window for the observation
of new physics in a given observable in charm processes, we must 
compute the standard-model contributions to such quantities. 
Attention to standard-model backgrounds is particularly important
in this case owing to the presence of potentially large 
long-distance contributions
that are nonperturbative in essence and therefore noncalculable by 
analytical methods. In general, the flavor structure of charm FCNC favors
the propagation of light-quark intermediate states. 
Thus, if it turns out that the charm-quark mass is not heavy enough compared
to a typical scale of hadronic effects,  long-distance effects are likely 
to  dominate. They will obscure the more interesting short-distance 
contributions that
are the true test of the standard model. Large long-distance effects are expected 
in  $D^0$--$\bar{D}^0$ mixing and FCNC charm decays.
In the case of mixing, although the long-distance 
effects dominate over the standard-model short-distance contributions, 
there could still be a significant  
window between 
these and the current experimental limits. 
The predictions of numerous extensions of the standard model lie in this 
window. We examine this possibility in 
Section~\ref{sec_ddbar}, 
where we also show that even the current experimental bounds on 
\ddb~mixing constrain several new-physics scenarios. 

Charm radiative decays are completely dominated
by nonperturbative physics and do not constitute a suitable test of the
short-distance structure of the standard model or its extensions. 
However, semileptonic modes such as $c\to u\ell^+\ell^-$ 
may be used to  constrain 
various standard-model extensions, since their kinematics might allow measurements 
away from the resonance-dominated  region, where the 
bulk of the long-distance contributions lie. We demonstrate this in 
detail in the 
case of  supersymmetric (SUSY) theories with or without  $R$-parity conservation. 
Purely leptonic flavor-violating modes such as $D^0\to\mu^+ e^-$ are free of 
standard-model backgrounds. We review the status and prospects of rare charm decays in 
Section~\ref{sec_rare}.

\section{$D^0$--$\bar{D^0}$ MIXING}
\label{sec_ddbar}
The time evolution of the \ddb\ system is described by the 
Schr\"odinger equation
as 
\beq
i\,\frac{\partial}{\partial t}\,\left(
\begin{array}{c}
D^0(t)\\
\bar{D^0}(t)
\end{array}\right)
=\left({\bf M} - \frac{i}{2}{\bf\Gamma}\right)\;
\left(
\begin{array}{c}
D^0(t)\\
\bar{D^0}(t)
\end{array}\right).
\eeq
Here 
\beqa
\left({\bf M} -\frac{i}{2}{\bf\Gamma}\right)_{ij}
&=&\frac{\langle D_i|H_{\rm eff}|D_j\rangle}{2m_D}\;\;\; 
= \;\;\;m_D^{(0)}\delta_{ij}\nonumber\\
&+&\frac{\langle D_i|H_w|D_j\rangle}{2m_D} 
+ \frac{1}{2m_D}\;\sum_n\,\frac{\langle 
D_i|H_w|n\rangle\langle n|H_w|D_j\rangle}
{m_D^{(0)} - E_n +i\epsilon}.
\label{hamil}
\eeqa
Noticing that
\beq
\frac{1}{m_D^{(0)} - E_n +i\epsilon} = P\left(\frac{1}{m_D^{(0)} - E_n}\right)
+i\pi\delta(E_n-m_D^{(0)}),
\nonumber
\eeq
where $P$ denotes principal value, we see that the absorptive part, ${\bf\Gamma}$, comes from
summing over real intermediate states
\beq
\Gamma_{ij} = \frac{1}{2m_D}\,\sum_n\,\langle D_i|H_w|n\rangle\langle 
n|H_w|D_j\rangle\;\delta(E_n-m_D).
\label{ris}
\eeq 
On the other hand, the contributions to $M_{ij}$ will  include not only 
the second term in Equation~\ref{hamil} but also a dispersive 
piece from off-shell intermediate states. 
Because ${\bf M}$ and ${\bf\Gamma}$ are Hermitian matrices, $M_{12}=M^*_{21}$ and
$\Gamma_{12}=\Gamma^*_{21}$. 
Invariance under  $CPT$ requires
$M_{11}=M_{22}$ and $\Gamma_{11}=\Gamma_{22}$.
The Hamiltonian eigenstates in terms of the weak eigenstates are
\beq
|D_{1,2}\rangle = p\,|D^0\rangle \pm q\,|\bar{D}^0\rangle.
\label{eigen} 
\eeq

Defining the ``right-sign'' amplitudes $A_{\bar f}\equiv A(D^0\to {\bar f})$ 
and 
$\bar{A}_{f}\equiv A(\bar{D^0}\to f)$, we can also define the 
``wrong-sign'' amplitudes
$\bar{A}_{\bar f}\equiv A(\bar{D}^0\to {\bar f})$ and 
$A_{f}\equiv A(D^0\to {f})$, where $f$ and $\bar{f}$ are  $CP$-conjugate  final 
states.
Then the time evolution of states that start as weak eigenstates at $t=0$ 
results
in the time-dependent  wrong-sign rate given by 
\beqa
r(t)&=&\frac{|\langle f|H|D^0\rangle|^2}{|\bar{A}_f|^2} = 
\left|\frac{q}{p}\right|^2\;\left| g_+(t)\,\lambda_f^{-1} + g_-(t)\right|^2,
\label{rt} \\
\bar{r}(t)&=&\frac{|\langle\bar{f}|H|\bar{D}^0\rangle|^2}{|A_{\bar{f}}|^2} = 
\left|\frac{p}{q}\right|^2\;\left| g_+(t)\,\lambda_{\bar{f}} + g_-(t)\right|^2,
\label{rtbar} 
\eeqa
where we have normalized with the right-sign amplitudes and  defined
\beq
g_\pm(t)\equiv \frac{1}{2}\,\left(e^{-i\gamma_1 t} \pm e^{-i\gamma_2 t}\right), \; \lambda_f\equiv\frac{q}{p}\frac{\bar{A}_f}{A_f}.
\label{gpmdef}
\eeq
The eigenvalues of the Hamiltonian are $\gamma_{1,2}=(M_{11}-i\Gamma_{11}/2) \pm(q/p)(M_{12}-i\Gamma_{12}/2)$.
We define
\beq
x\equiv \frac{m_1-m2}{\Gamma}=\frac{\Delta m_D}{\Gamma},
y\equiv \frac{\Gamma_1-\Gamma_2}{2\Gamma}=\frac{\Delta\Gamma_D}{2\Gamma}.
\nn
\eeq
Section~\ref{ddexp} discusses various ways of accessing the mixing parameters $x$ and $y$.

\subsection{Standard-Model Predictions for Mixing Parameter}
\label{smpred}
In this section, we review our understanding of the 
standard-model predictions for \ddb ~mixing. As always in charm FCNC 
processes, 
the main issue is to estimate as accurately as possible the size of the 
long-distance contributions. Only then we can evaluate the potential
of \ddb\ mixing  to test the standard model and constrain its extensions. 

\subsubsection{SHORT-DISTANCE CONTRIBUTION TO $\Delta m_D$}
\label{sdpred}
In the standard model, the short-distance $\Delta C=2$ transition occurs via box diagrams. 
The effective interactions at the $m_c$ scale are described by the
  Hamiltonian~\cite{datta}:
\beq
{\cal H}_{\rm eff.}^{\Delta C=2}=\frac{G_F^2}{4\pi^2}|V_{cs}^{*}V_{cd}|^2
\;\frac{(m_{s}^{2}-m_{d}^{2})^2}{m_{c}^{2}}
\left({\cal O} +2{\cal O'}\right), \label{heffsd}
\eeq
with ${\cal O}\equiv \bar{u}\gamma_\mu(1-\gamma_5)c\;\bar{u}
\gamma^\mu(1-\gamma_5)c$ and 
${\cal O}'\equiv \bar{u}(1+\gamma_5)c\;\bar{u}(1+\gamma_5)c$. The additional operator $\cal O'$ arises from the nonnegligible external 
momentum. 
The matrix elements of the operators can be parameterized by
\beq
\langle D^0 |{\cal O}|\bar{D}^0\rangle = \frac{8}{3}m_D^2 f_{D}^{2} B_D 
\quad ;\quad
\langle D^0 |{\cal O}'|\bar{D}^0\rangle = -\frac{5}{3}\left(\frac{m_D}{m_c}
\right)^2 
m_D^2 f_{D}^{2} B'_D. \label{mat_par}
\eeq
In the vacuum-insertion approximation,  one has  $B_D=B'_D=1$ and the 
box  diagrams' 
contribution to the mass difference is 
\beq
\Delta m_{D}^{\rm box}\simeq 1.4\times 10^{-18}\;{\rm GeV}\left(\frac{m_s}{0.1 
{\rm GeV}}\right)^4\; \left(\frac{f_D}{0.2 {\rm GeV}}\right)^2. \label{dmd_sd}
\eeq
The $b$-quark contribution is much smaller owing to additional Cabibbo-Kobayashi-Maskawa (CKM) suppression. Thus, for typical values of $f_D$ and $m_s$, the box diagrams contribute
with approximately $x_{\rm box}\simeq ({\rm few}\times10^{-6})-(10^{-5})$.
In contrast to $K$ and $B$ mixing, the internal quarks in the 
box diagrams here are down-type quarks. The $b$-quark contribution, 
which would give in principle
the largest GIM violation, is suppressed by small CKM mixing factors. 
The leading contribution, 
as shown in Equation~\ref{heffsd}, is given by the strange quark and 
therefore results in a very effective GIM suppression. 

The box-diagram contributions to $\Delta\Gamma$ are further suppressed by 
$m_s^2$. This can be seen as the two powers of 
the helicity suppression factor from the 
matrix element of a $0^-$ meson to a pair of quarks.

\subsubsection{LONG-DISTANCE CONTRIBUTIONS}
\label{predld}

The long-distance contributions to $D^0$--$\bar{D^0}$ mixing are inherently nonperturbative and cannot be calculated from first principles. It is however extremely important to estimate their size in order to understand the origin of a possible experimental observation. 

A first step is to recall that \ddb ~mixing is an SU(3)-breaking 
effect~\cite{wolf,dght}: 
\beq
x,~y\sim \sin\theta_c^2\times [{\rm SU}(3) {\rm ~breaking}].
\label{naive}
\eeq
Thus, the task is to estimate the amount of SU(3) breaking. 
An important observation \cite{fglp,georgi}  is that
the  SU(3) breaking is a second-order effect in the quark masses. 
This circumstance would lead to the naive estimate
\beq
x,~y\sim \sin\theta_c^2\times\,\left(\frac{m_s}{\Lambda_{\rm hadr.}}\right)^2
\,\simlt O(10^{-3}),
\label{naive2}
\eeq
with $\Lambda_{\rm hadr.}\sim O(1)$~GeV a typical hadronic scale. 
There are two basic methods to estimate the contributions to mixing
beyond box  diagrams: the  heavy-quark effective theory (HQET) approach pioneered by Georgi~\cite{georgi}, 
which is essentially rooted in the 
operator product expansion (OPE); and an exclusive approach computing
the contributions of hadrons in complete SU(3) multiplets to the dispersive
and absorptive parts. 

%
%
%%% COMP: \paragraph translates into a fourth-level heading.
%
%
\paragraph{\ddb~Mixing in Heavy-Quark Effective Theory:}
The applicability of the HQET ideas to \ddb~mixing rests on the
assumption that the charm-quark mass is much larger than the typical scale
of the strong interactions. Georgi pointed out~\cite{georgi}
that in this case there are no nonleptonic transitions to leading order in
the effective theory, since they would require a large momentum transferred
from the heavy quark to the light degrees of freedom. This means that, in the
effective low-energy theory,  mixing is a consequence of matching the full
$\Delta C=2$ theory at the scale $m_c$ with the HQET and then running down to
hadronic scales ($\ll m_c$). No new operators enter at
low energy. The ``long-distance" effects are provided by the tree-level matching
of operators that are nonleading in the $1/m_c$ expansion but that are less 
GIM-suppressed than the box diagrams.

First let us consider the four-quark operators generated from the box diagrams
by integrating out the $W$s. 
The contribution of these operators
to the mass difference behaves like \cite{georgi}
\beq
\Delta m_{D}^{(4)}\simeq\frac{1}{16\pi^2}\frac{m_{s}^{4}}{m_{c}^{2}}, 
\label{fq_op}
\eeq
where the first factor comes from the loop and $m_d$ is neglected. This presents
the same amount of GIM suppression as the box diagrams and therefore causes no enhancement. 

There will also be higher-dimension operators that, although
suppressed by additional powers of $1/m_c$, can give important contributions. 
can give important contributions. For instance, six-quark operators arise by ``cutting" one of the
light-quark lines in the loop and then shrinking the connecting
line leftover. 
As a consequence, 
two powers of $m_s$ are lost and the contribution from six-quark operators 
goes like
\beq
\Delta m_{D}^{(6)}\simeq\frac{1}{m_c}\frac{m_{s}^{2}}{m_{c}^{2}}
\left(m_sf^2\right),
\label{sq_op}
\eeq
where the last factor comes from taking the hadronic matrix elements and
$f$ is the  pseudo--Goldstone-boson decay constant. This constitutes an enhancement
of $(4\pi f)^2/m_c\,m_s$ with respect to the four-quark operator contribution. 

Finally, eight-quark
operators are obtained by cutting the remaining light-quark line and
bridging the two four-quark pieces with a gluon. The resulting contribution
goes like
\beq
\Delta m_{D}^{(8)}\simeq\frac{\alpha_s}{4\pi}\frac{1}{m_{c}^{2}}
\frac{\left(m_sf^2\right)^2}
{m_{c}^{2}}. \label{eq_op}
\eeq
This is the least GIM-suppressed
contribution, although it is suppressed by $1/m_{c}^{2}$ and, most important,
by the  factor $\alpha_s/4\pi$. Therefore, 
no enhancement is expected from these operators.
 
Ohl et al.\ performed a very detailed calculation in this approach,
including QCD corrections to one loop~\cite{ohl}. 
Their results can be summarized as
\beqa
\Delta m_{D}^{(4)}&\simeq &(0.8-1.5)\times10^{-16}~{\rm GeV}
\left(\frac{m_s}{0.1~{\rm
GeV}}
\right)^4 \nn\\
\Delta m_{D}^{(6)}&\simeq &(0.6-1.6)\times10^{-16}~{\rm GeV}
\left(\frac{m_s}{0.1{\rm
GeV}}
\right)^3 \label{dmops}\\
\Delta m_{D}^{(8)}&\simeq &(0.05-0.25)\times10^{-16}{\rm GeV}
\left(\frac{m_s}{0.1{\rm
GeV}}
\right)^2. \nn
\eeqa
The lower numbers correspond to adding the various coefficients in 
quadrature, since naive 
dimensional analysis does not predict the relative signs. The upper 
numbers come from adding
all contributions coherently. 
This leads to 
\beq
x\simeq (0.6-2)\times10^{-4}
\label{hqet_dmg}
\eeq
for $m_s=0.1$~GeV.
Thus, the ``long-distance'' enhancement
observed in this approach is of about one order of magnitude.

More recently, Bigi \& Uraltsev made similar arguments also in the context of 
the OPE~\cite{bigi_ural}. However, they found a larger eight-quark operator 
contribution than Equation \ref{dmops} indicates.
This results in $x\simeq 10^{-3}$. 
Bigi \& Uraltsev also undertook~\cite{bigi_ural} a study of $\Delta\Gamma$ in the 
OPE  approach.
They found that the most important contribution to $y$ comes from the 
eight-quark 
operators, where the gluon connecting the two sets of four-quark operators 
is split to generate an imaginary part in the diagram. The ``suppression'' 
to be paid (relative to $\Delta m_D$)
is given by $\beta\,\alpha_s/4\pi\simeq O(1)$, where  $\beta=(11-2n_f/3)$ is
the QCD beta function.
This results in $y\simeq x\simeq 10^{-3}$.

\begin{figure}
\centerline{
\psfig{figure=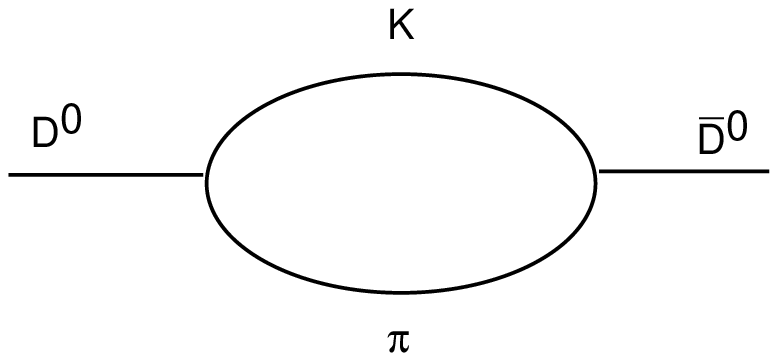,width=3.2in,angle=0}}
\caption[]{ Long-distance contribution from two charged pseudoscalar 
intermediate states.}
\label{pp_cont}
\end{figure}

%
%
%%% COMP: \paragraph means a fourth-level heading.
%
%
\paragraph{Exclusive Approach:} 
An estimate of the long-distance contributions can be obtained
 by assuming they come from the propagation of hadronic states to which 
both $D^0$ 
and $\bar{D}^0$ can decay. There will be one-, two-, three- $\ldots$ particle 
intermediate 
states. Each of these groups can be further separated into sets whose 
contributions 
vanish separately in the  SU(3) limit. For instance, one of these sets is
 formed by  the two-charged-pseudoscalar intermediate states\footnote{Strictly speaking, 
this is a U-spin set.} $\pi^+\pi^-$, 
$K^+K^-$, 
$K^-\pi^+$, and $K^+\pi^-$. 
Thus, computing 
their contribution to the mass difference, as shown schematically 
in Figure~\ref{pp_cont}, gives a  concrete realization of the estimate in 
Equation~\ref{naive} for an  SU(3) set for which data are available. 
This was first done~\cite{dght} in the massless limit. 
Although these ``self-energy" diagrams will depend on the interaction 
chosen for the vertices,  they have a universal imaginary part that
typically comes from a logarithm. 

The contribution 
of Figure~\ref{pp_cont} to the mass difference obeys a dispersion relation of 
the form~\cite{dght}
\beq
\Sigma(p^2)=\frac{1}{\pi}\;\int_{s_0}^{\infty} \frac{Im[\Sigma(s)]\; ds}{(s-
p^2+i\epsilon)}, \label{dis_rel}
\eeq
where $s_0\equiv (m_1+m_2)^2$, and 
$m_1$ and $m_2$ are the masses in the loop. 
The mass difference is then given by $\Delta m_D = -Re[\Sigma(m_D^2)]/2m_D$.
Taking into account a subtraction 
forcing the condition $\Sigma(0)=0$, and keeping the masses, the implementation 
of a cutoff $\Lambda$ in the dispersive integral (Equation~\ref{dis_rel})
gives~\cite{gb_tcf}
\beqa
x&\simeq&\frac{m_D}{4\pi}\left\{ \frac{B(\pi^+\pi^-)}{{\bf 
p}_{\pi\pi}}\;I(m_\pi,m_\pi,\Lambda)+
\frac{B(K^+K^-)}{{\bf p}_{KK}}\;I(m_K,m_K,\Lambda) \right. \nn \\
& &\left. \quad\quad -2\cos\delta\frac{\sqrt{B(K^-\pi^+)\;B(K^+\pi^-)}}
{{\bf p}_{K\pi}}\;I(m_\pi,m_K,
\Lambda)
\right\}, \label{dm2_full}
\eeqa
where ${\bf p}_{ij}$ is the magnitude of the three-momentum in the actual 
decay, $\delta$ is the strong phase 
%defined in Equation~\ref{rddef}
between the Cabibbo-allowed and the doubly Cabibbo-suppressed decays,
$B$ is the branching fraction, and the 
integrals are 
\beq
I(m_1,m_2,\Lambda)=-\int_{s_0}^{\Lambda^2} \frac{\sqrt{1-\frac{s_0}{s}}\;ds}{
s-m_{D}^{2}}. \label{int_def}
\eeq
By taking the massless limit in Equation~\ref{dm2_full}, one recovers the result in Reference~\cite{dght} with 
the identification $\mu^2=2\,m_D\,(\Lambda-m_D)$. Although the result depends 
strongly on 
the cutoff $\Lambda$, this can be interpreted as the value of $s$ for which 
the internal momentum reaches its maximum. Not surprisingly, 
the value of $\Lambda$ that gives an internal momentum of $\sim 1$~GeV 
also yields $\mu\sim 1$~GeV. This is $\Lambda\simeq (2$--2.2)~GeV, 
not far 
above $m_D$. 
Using this cutoff results in a contribution to the mass difference of
\beq
x\simeq 6.5\times 10^{-4}\times [1-\cos\delta~b],
\label{cha_est}
\eeq
where we  define $b$ by
\beq
 \frac{B(K^+\pi^-)}{B(K^-\pi^+)}=b^2\;\left|\frac{V_{cd}^*V_{us}}{V_{cs}^*V_{ud}}\right|^2,
\label{def_b}
\eeq
as a measure of the amount of  SU(3) breaking in the absolute value of the 
amplitudes, in addition to the  SU(3)-breaking relative phase $\delta$. 
Measurements in  doubly Cabibbo-suppressed (DCS) decays 
result in  $b\sim 1.22$ \cite{pdg02}.  

It is clear that the 
cancellation in this  SU(3) set is quite effective.
Although the dependence on the cutoff $\Lambda$ appears unsatisfactory, 
it has a clear
physical interpretation. In order to derive Equation~\ref{dm2_full}, we 
assumed that the
couplings at the vertices in Figure~\ref{pp_cont} are pointlike and 
therefore constant 
in $s$. However, the  hadrons participating in these  interactions are 
composite, with the 
compositeness scale $\Lambda_{\rm hadr.}\sim O(1)$~GeV, a typical 
hadronic scale. 
Thus, at or near $\Lambda_{\rm hadr.}$, the couplings should develop a 
form-factor suppression
determined by this energy scale that in turn would render the integrals 
in Equation~\ref{int_def} finite without ad~hoc cutoff. 
Then we should replace $I(m_1,m_2,\Lambda)$ with  
\beq
I(m_1,m_2)= -\int_{s_0}^{\infty} g^2(s)\frac{\sqrt{1-\frac{s_0}{s}}\;ds}{
s-m_{D}^{2}}, \label{int_nolam}
\eeq
with the coupling of the interaction vertex normalized so that 
$g(s)\sim 1$ for $s<m_D^2$ and it falls off with $\Lambda_{\rm hadr.}$ for 
$s\geq m_D^2$. Of course, the dependence on the cutoff is now hidden in $g(s)$, but its physical meaning is more transparent.  

In principle, pseudoscalar-vector (PV), 
vector-vector (VV), and all other
possible intermediate states can be treated similarly. However, in many cases, the data are incomplete. In some cases, phase space suppresses some of the  modes, but unlike 
in $\Delta\Gamma$~\cite{fglp} (see discussion below), 
there is no  SU(3)-breaking effect induced by running out of phase space. 
Modes that are not accessible to the actual $D$ decay (and thus to the 
absorptive part)  still contribute to the 
dispersive  part. We cannot express these in terms of 
branching ratios
because the decays are not physical. But in principle the amplitudes for these 
``off-shell'' contributions are well-defined. 
We then conclude that estimates provided by lighter, far-from-threshold 
sets, such as the one in  Equation~{\ref{dm2_full}, are likely to be the most reliable ones in this approach. The underlying assumption is that the contributions of 
other sets, their signs  being  not fixed, should not conspire to add coherently 
and change significantly the order of magnitude of the effect. 
This reasoning results in 
\beq
x\simlt 10^{-3},
\label{ldxd}
\eeq
with large uncertainties. In sum, the exclusive approach 
seems to be consistent with 
our naive estimate of Equation~\ref{naive2}.
Additional sources of enhancement in \ddb~mixing are discussed in Reference~\cite{gp}.

The exclusive approach can also be applied to $\Delta\Gamma_D$. 
For instance, for the PP (two pseudoscalars) set considered above, the contribution to 
$y$ has an expression similar to Equation~\ref{dm2_full}, so that 
the cancellation is still quite efficient~\cite{fglp}.
However, $y$ receives
contributions only from real intermediate states, so phase space could be 
a considerable source of  SU(3) breaking, in a way that did not arise for the $x$ case \cite{fglp}. 
In the most extreme cases, some states belonging to a given set would 
simply be absent 
because $D$s are not allowed to decay into them. This could upset the 
SU(3) cancellations, perhaps to a large extent.
This source of  SU(3) violation is estimated by considering the effect of 
phase space in the contribution of a given intermediate state $F$ and 
within a given  SU(3) representation $R$. Thus, one can write
\beq
y_{F,R}=\frac{\sum_{n}\langle\bar{D}^0|H_w|n\rangle \rho_n
\langle n|H_w|D^0\rangle}
{\sum_{n}\langle D^0|H_w|n\rangle \rho_n
\langle n|H_w|D^0\rangle} 
= \frac{\sum_{n}\langle\bar{D}^0|H_w|n\rangle \rho_n
\langle n|H_w|D^0\rangle}{\sum_{n}\Gamma(D^0\to n)},
\label{yfr1}
\eeq   
where the sums are performed over intermediate states $n$ belonging to the 
class $F$ (PP, PV, etc.) and the representation $R$.
Here, $\rho_n$ is the phase space of state $n$ and depends only on the 
masses of the particles in this intermediate state. 
The expression in Equation~\ref{yfr1} would be $y$ if the states $n$ of $(F,R)$ 
were the only states available for the $D^0$ to decay into. Then, in order to 
obtain 
the corresponding contribution to $y$, $y_{F,R}$ must be rescaled to the 
total branching ratio
of the states in $(F,R)$:
\beq
y=\frac{1}{\Gamma}\,\sum_{(F,R)}y_{F,R}\,\left[\sum_{n}\Gamma(D^0\to n)\right].
\label{ytot}
\eeq 

The scarcity of precise data makes it difficult to disentangle the various 
SU(3) contributions.
A very rough estimate of the relevant branching ratios has been made~\cite{fglp} in order
to estimate $y$ in this approach. The states considered  are PP, PV, VV, 3P (three pseudoscalars), and 4P. 
In most cases, the contributions to $y$ are  $O(10^{-3})$ at most.
The only exception to this is the 4P  intermediate 
states, where
very large values of $y_{4P,R}$ are found.  
This dramatic effect comes mostly from the fact that some members of the 
SU(3) multiplets that should cancel have run out of phase space, since the 
average mass of the multiplet is not far from threshold to begin with.
In this case, it is found that $y\simeq 10^{-2}$ can be obtained, 
although with large 
uncertainties. 
%\noindent
%KT added paragraph break here

In obtaining this estimate of the long-distance contributions to 
$y$, it was assumed that
phase space is the only source of  SU(3) breaking. This constitutes a good estimate if additional sources of breaking do not 
cancel the effect
of phase space significantly. However, this cancellation would naturally occur
if the charm quark is heavy enough so that quark-hadron duality, in the spirit
of the OPE approach discussed above, is valid. 
Furthermore, in some cases SU(3) breaking occurs in the 
opposite direction  to  the one dictated by phase space: 
$B(D^0\to K^+ K^-)/B(D^0\to\pi^+\pi^-)\simeq 2.9$. 
In any case, the effect found in Reference~\cite{fglp} is a   reminder of the difficulty of estimating reliably the size of the standard-model effect.

In this examination of the long-distance contributions to mixing, we have seen that 
\beq	
x\simlt 10^{-3},~~~~y\simlt 10^{-2}.
\label{ldfinal}
\eeq	
The possibility of having $y$ at the $1\%$ level raises a serious question 
as to whether it will be possible to extract a smaller value of $x$ from 
data with enough precision to  constrain the short-distance physics 
contributions from the standard model and/or its extensions. 

\subsection{$CP$ Violation}
\label{cpv}
In the standard model, the $D$ system is not as sensitive to $CP$ violation 
as the $K$ and $B$ mesons are. Once again, the small effects predicted in the standard model could leave open a window to the observation of new-physics effects. Here---rather than going into specific calculations---we discuss some general features of $CP$ violation in $D$ mesons, both in the standard model and beyond.

\subsubsection{DIRECT $CP$ VIOLATION}
\label{directcp}

Direct $CP$ violation requires the presence of 
both weak and 
strong relative phases between two or more amplitudes contributing 
to a given final state. In the standard model, relative weak phases can only be obtained in 
Cabibbo-suppressed decays, for instance,  
via the interference between spectator and penguin amplitudes. 
In order to estimate the size of the $CP$ asymmetries this would generate, we write
\beqa
A_{CP}&\simeq& \frac{Im[V_{cd}\;V_{ud}^{*}\;V_{cs}\;V_{us}^{*}]}
{\lambda^2}\;\;
\sin\delta_{\rm st}\;\;\frac{P}{S} \label{asy_dir} \\
&\simeq &A^2\;\eta\;\lambda^4\;\sin\delta_{\rm st}\;\frac{P}{S} 
\quad\quad\simlt 10^{-3}, \nonumber
\eeqa
where $\delta_{\rm st}$ is the strong relative phase between the penguin and the spectator amplitudes, $P$ and $S$, and $A\sim 1$ and $\eta$ are CKM parameters in the 
Wolfenstein parameterization. 
Specific model calculations \cite{yb58pg56} for $D\to KK, \pi\pi, K^*K$, three-body modes, etc.\ 
yield this order of magnitude for the effect. New physics could enter, 
for instance, through large phases in the penguin diagram. This could give 
asymmetries of the order of 1\% or larger. 
On the other hand, Cabibbo-allowed decays
do not have two amplitudes with different weak phases, and therefore the $CP$ 
asymmetry is zero in the standard model. Some new-physics scenarios, e.g., some left-right--symmetric models, provide 
extra phases and could give asymmetries as  large as 1\%.

\subsubsection{INDIRECT $CP$ VIOLATION}
\label{indirectcp}
When $CP$ violation is negligible, the $CP$ asymmetry is proportional to 
\beq
A_{CP}\sim -2\,(x\cos\delta + y\sin\delta)\,\sin\phi\,(\Gamma t).
\label{icpv}
\eeq
In the standard model, $\phi\sim 2\,A^2\lambda^4\eta\simlt 10^{-3}$. 
New physics in $x$ could induce a large $\sin\phi$.  
Thus, even if $x\ll y$, provided there is a significant
relative strong phase between the $K\pi$ final states, there could be 
sensitivity
to $CP$-violating phases from new-physics contributions to $x$.

\subsection{Constraints on Physics Beyond the Standard Model}
\label{secddb_new}
Because the predictions for \ddb~mixing 
in the standard model are quite uncertain, the existence of a discovery window between the 
current experimental limits and the standard-model prediction cannot be established  accurately.
On the other hand, several extensions of the standard model predict large enhancements 
in $\Delta m_D$. In many cases, even the current experimental limit 
is enough to severely constrain models.
The situation is illustrated in Figure~\ref{hnelson}, 
where standard-model predictions for $x$ and $y$  are collected along with predictions from extensions of the standard model for $x$. Current experimental bounds already 
exclude or severely 
limit the parameter space of some standard-model extensions.  
In some of these models, theorists have been forced 
to choose either theoretical mechanisms or parameters that avoid FCNC 
signals in  $K$ and $B$ physics. However, this often leads to large \ddb~mixing effects.
Thus, although a positive observation of 
$x\simlt 10^{-2}$ may not be a clear indication of new physics, pushing 
the experimental 
bound as low as possible has a great impact on theoretical model building. 
We illustrate this point in two paradigmatic cases: 
weak-scale  supersymmetry (SUSY) and strong  dynamics at the TeV scale. 

\subsubsection{THE MINIMAL SUPERSYMMETRIC STANDARD MODEL and $\Delta m_D$}
\label{secddb_mssm}
Weak-scale  SUSY is a possible solution to the hierarchy problem. 
The  minimal supersymmetric standard model (MSSM), the simplest 
SUSY extension of the standard  model, involves a doubling of the 
particle spectrum by putting all standard-model fermions in chiral supermultiplets 
and placing the standard-model gauge bosons in vector supermultiplets. 
Many new parameters are introduced. The soft  SUSY-breaking sector generally includes three gaugino masses, as well as trilinear 
scalar  interactions and Higgs and sfermion masses. 

In general, 
sfermion masses are not related to fermion masses. In particular, if
we choose to rotate the squark fields by the same matrices that 
diagonalize the quark mass matrices, squark mass matrices 
are not diagonal~\cite{susyfcnc}.  
In this ``super-CKM'' basis, squark propagators can be expanded 
so that nondiagonal mass terms result in mass insertions that change 
the squark flavor. 
These mass insertions can be 
parameterized in a model-independent fashion via
\begin{equation}
(\delta^u_{ij})_{\lambda\lambda'}={(M^u_{ij})^2_{\lambda\lambda'}
\over M^2_{\tilde q}}\,,
\label{softsusy}
\end{equation}
where $i\neq j$ are generation indices, $\lambda,\lambda'$ denote the
chirality, $(M^u_{ij})^2$ are the off-diagonal elements of the up-type squark 
mass matrix, and $M_{\tilde q}$ represents the average squark mass.
This effect 
can be avoided in specific  SUSY-breaking  scenarios such as gauge mediation
or anomaly mediation, but is present in most situations, e.g., if SUSY breaking is 
mediated by gravity.
In most cases, the main contribution from these soft SUSY-breaking terms  
to FCNC transitions comes from the gluino-squark loops. 
Such contributions to rare $K$ and $B$ transitions have led to
strong universality constraints on the charged $Q=-\frac{1}{3}$ squark
sector (see, e.g., \cite{ellis}). 
The most stringent bounds that apply to the nonuniversal soft breaking
terms $(\delta^u_{12})_{\lambda\lambda'}$ come from the experimental 
searches for $D^0$--$\bar{D^0}$ mixing.\footnote{Limits obtained from 
charge and color breaking and bounding the potential from 
below~\cite{savas} apply to the trilinear terms but not to the 
squark mass terms. Thus, unless the squark mass  matrices are kept diagonal, 
such arguments cannot be used to constrain the nonuniversal mass
insertions.} For instance, the CLEO limit~\cite{cleodd} 
implies~\cite{wyler} 
\begin{equation}
\frac{1}{2}\left\{\left(\frac{\Delta m_D}{\Gamma_{D^0}}\right)\cos\delta
+\left(\frac{\Delta\Gamma_D}{2\Gamma_{D^0}}\right)\sin\delta\right\}^2
<4 \times 10^{-4}\%,
\label{ddbound}
\end{equation}  
where $\delta$ is a strong relative phase between the Cabibbo-allowed and 
the  DCS  $D^0\to K\pi$ decays. Neglecting this phase 
results in the constraints obtained in Reference~\cite{wyler}, 
which we collect in Table~\ref{numssm}. 
These bounds were obtained assuming 
that $(\delta^u_{12})_{RR}=0$ and $(\delta^u_{12})_{LR}=(\delta^u_{12})_{RL}$.

Similar bounds on $(\delta_{ij}^d)_{\lambda\lambda'}$ are obtained in $K$ and $B$ processes. 
In general, SUSY models naturally give $(\delta_{ij}^q)_{LL}$ and 
$(\delta_{ij}^q)_{LL}$
of order 1, whereas $(\delta_{ij}^q)_{LR}\simeq (m_Z/M_{\tilde{q}})$ is expected.
This can be avoided if squark masses are assumed to be universal at some high energy
scale. However, the renormalization group running will generate
nondiagonal values of the $(\delta_{ij}^q)$s in general.
The mechanism of quark-squark  alignment~\cite{ns}, in which 
squark mass matrices are diagonalized along with the quarks, are an alternative to 
squark degeneracy. This alignment occurs naturally in theories with an 
Abelian horizontal
symmetry~\cite{ns2}. In order to avoid $K^0$--$\bar{K^0}$~mixing bounds, the down-squark 
mass matrices are chosen to be diagonal. In order to obtain 
the correct 
Cabibbo angle, an off-diagonal term in the up-squark mass matrices must 
be allowed. 
This leads to the prediction of large \ddb~mixing, saturating the 
experimental bound.  

\subsubsection{CONSTRAINTS ON STRONG DYNAMICS FROM \ddb~MIXING}
\label{ddb_strong} 
We briefly address the constraints arising from $\Delta m_D$ on 
theories where the electroweak symmetry is broken by new strong interactions
at the TeV scale (for a review, see \cite{hillsimmons}). 
These include technicolor~\cite{tc} and  extended technicolor (ETC)~\cite{etc}, 
top condensation models~\cite{topcond,topc}, and 
composite Higgs models~\cite{comph}, 
among others. In fact, 
this may also cover the standard model if the Higgs is not light enough~\cite{cds}.

If we consider the standard model as an effective theory, then its scalar sector depends
on two quantities~\cite{georgi93}: $\Lambda$, the scale of the underlying
physics (or scalar ``compositeness'' scale), and $f$, the amplitude for producing
the scalars from the vacuum. The ratio of these two scales $\kappa=\Lambda/f$ is
estimated to be $\kappa\simeq 4\pi$ in naive dimensional analysis 
\cite{nda}.  
On the other hand, in order for fermions to acquire masses from their Yukawa
couplings to the Higgs, they may interact with its constituents through 
additional
interactions characterized by the mass scale $M$. Assuming that this 
flavor scale is 
associated with  new gauge interactions, it is possible to make explicit the 
dependence of the Yukawa couplings on $\Lambda$ and $M$~\cite{cds}: 
\beq
-\frac{g^2}{M^2}\;\frac{\Lambda^2}{\kappa}\,\bar{f}_R\phi\psi_L\sim \sqrt{2}\,
\frac{m_f}{v}\bar{f}_R\phi \psi_L,
\label{lamvsmf}
\eeq
where $\psi_L$ is an SU(2)$_L$ doublet containing $f_L$ and 
$v\simeq 246$~GeV is the 
vacuum expectation value of the Higgs field. 
Thus, in order to have the correct fermion mass, one has
\beq
\Lambda\simgt\frac{M}{g}\sqrt{\sqrt{2}\kappa\frac{m_f}{v}}.
\label{lbound}
\eeq
The relation in Equation~\ref{lbound} between the Higgs
compositeness scale and the scale of flavor physics will result in bounds on 
$\Lambda$. In particular, in addition to the interactions
coupling the constituents of the Higgs to the standard-model fermions, there  are diagonal interactions involving only 
the standard-model fermions leading to dimension-six operators of the form
\beq
\frac{g^2}{M^2
}\,(\bar{f}_{L,R}\gamma_\mu f_{L,R})\, 
(\bar{f}_{L,R}\gamma_\mu f_{L,R}).
\label{dim6}
\eeq
For instance, for quarks, the interactions in Equation~\ref{dim6} result in 
FCNC once the quark mass matrix is diagonalized. Thus, operators  such as
\beq
\frac{g^2}{M^2}\,U_{L,R}^{cu*}\,U_{LR}^{cu*}\,
(\bar{u}_{L,R}\gamma_\mu c_{L,R})\, 
(\bar{u}_{L,R}\gamma_\mu c_{L,R}),
\label{dim6fcnc}
\eeq
where the matrices $U_{L,R}$ rotate the left-handed and right-handed up-quarks
to their mass eigenbasis, will induce \ddb~mixing. 
Similarly, there will be FCNC operators acting in the $K$ and $B$ 
sectors. Bounds on $M/g$ then result in bounds on $\Lambda$. For instance, 
assuming $U_{L,R}^{cu}\simeq \sin\theta_C$, one obtains
\beq
\Lambda\simgt22~{\rm TeV}\,\sqrt{\kappa\,\left(\frac{1.5~{\rm TeV}}{m_c}\right)}
\label{ddblbound}
\eeq
from the LR product of color-octet currents. For the  naive-dimensional-analysis value $\kappa\sim 4\pi$, 
we have $\Lambda\simgt 78$~TeV. 

In principle, this bound could be 
weakened considerably by assuming $U_{L,R}\ll~1$. But to do so would invite bounds
from the down-quark sector. For instance, if $D_L^{ds}\sim\sin\theta_C$, then
$\Lambda\simgt 15$~TeV. 
This is a general result that applies not only to ETC theories but also to a host of 
other scenarios. 

The constraints of FCNCs  apply not only to the flavor-physics
scale but also to the Higgs-compositeness scale. If we interpret the FCNC constraint as a bound on the cutoff scale, these theories are fine-tuned.
Whatever the way out of the bounds, this analysis suggests that these scenarios
tend to have plenty of flavor physics. 
Furthermore, we can interpret the cutoff scale $\Lambda$  as an implied upper bound on the Higgs mass
by making use of the triviality constraint~\cite{cds}:
\beq
m_H^2\ln\left(\frac{\Lambda}{m_H}\right)\leq \frac{4\pi^2 v^2}{3}.
\label{mh2}
\eeq 
The typical flavor-physics bounds on $\Lambda\simlt 80$~TeV yield 
$m_H\leq 400$~GeV.
Then, if we consider the standard model as an effective theory valid up to a scale $\Lambda$, 
a light Higgs results in a more fine-tuned scenario, i.e., $m_H/\Lambda\ll 1$. 
In heavier-Higgs scenarios, the cutoff scale is lower and  
flavor physics is likely to saturate the experimental
bounds on a variety of flavor-physics observables. This circumstance is typical in 
theories where the 
Higgs is a composite field and the standard-model fermions feel the interaction binding it.
On the other hand, these bounds can be avoided 
in models with a GIM mechanism built in~\cite{tcgim}. 

%
%
%%% COMP: \paragraph means a fourth-level heading.
%
%
\paragraph{Topcolor:}
In top-condensation models~\cite{topcond}, the constituents of the Higgs are
the third-generation left-handed quarks as well as $t_R$. Hill~\cite{topc}  proposed that a new gauge interaction strongly coupled to the third-generation quarks
is responsible for top condensation. The topcolor interactions break 
at the TeV scale 
as SU(3)$_1\times$ SU(3)$_2\to$ SU(3)$_{\rm color}$, leaving, besides the massless gluons, 
a set of color-octet gauge bosons (the top-gluons) 
leading to the Nambu--Jona-Lasinio effective interactions
that result in top condensation. This leads to electroweak symmetry breaking as well as to a large ``constituent'' 
top mass. 

Tilting the vacuum in the top directions to avoid a 
large  $b$-quark mass is typically accomplished through additional Abelian interactions
that leave a $Z'$ strongly  coupled to third-generation fermions. In some 
models, the tilting is
done by simply arranging that $b_R$ not couple to the  topcolor interactions. 
The top-gluon interactions (as well as the $Z'$s if present) are 
nonuniversal, 
leading to FCNC at tree level. 
These arise after quarks are rotated to their mass eigenbasis by the rotations
\beq
U_{L,R}^i\to{\cal U}_{L,R}^{ij}\,U_{L,R}^j, ~~~~~~~~
D_{L,R}^i\to{\cal D}_{L,R}^{ij}\,D_{L,R}^j,
\label{rot2mass}
\eeq
where the rotation matrices ${\cal U}_{L,R}$ and ${\cal D}_{L,R}$ 
are unitary. The CKM matrix is then $V_{\rm CKM}={\cal U}_L^\dagger\,{\cal D}_L$.

Constraints on  topcolor models are reviewed in Reference~\cite{bbhk}. The bounds 
from the 
down-quark sector impose severe constraints on the entries of ${\cal D}_{L,R}$ 
mainly coming from the exchange of bound states that couple strongly to the $b$ quark. 
There are several contributions to $\Delta m_D$. 
First, if we consider theories where  topcolor only partly breaks the 
electroweak symmetry, there will be a set of pseudo--Nambu-Goldstone bosons in the spectrum 
with masses of a few hundred GeV. They contribute ~\cite{bbhk}
\beq
\Delta m_D\simeq \frac{f_D^2\,m_D^2}{2}\,\frac{m_t^2}{f_{\tilde{\pi}}^2
m_{\tilde{\pi}}^2}
({\cal U}_L^{tu*}{\cal U}_R^{tc})^2,
\label{dmdtpion}
\eeq
where $f_{\tilde{\pi}}\approx$(50--70)~GeV is the top-pion decay constant.
If it is assumed that the entries 
of ${\cal U}_{L,R}$ are similar to those of $V_{\rm CKM}$, we then obtain
\beq
\Delta m_D \simeq 2\times10^{-14}\,\left(\frac{200~{\rm GeV}}{m_{\tilde{\pi}}}
\right)^2 {\rm ~GeV},
\eeq 
which is just below the current experimental bound even if we consider $\delta=0$.

However, the scalar sector of  topcolor theories is somewhat
model-dependent. For instance, in some scenarios there are no top-pions in the spectrum 
and  topcolor interactions  fully break the electroweak symmetry. 
The most model-independent aspect of  topcolor theories is 
the top-gluon sector. The constraint that the third-generation
coupling be supercritical leaves the top-gluon mass as the only free parameter, 
beyond the quark-rotation matrices. Top-gluon exchange also  contributes 
to $B^0$--$\bar{B^0}$~\cite{blr}~mixing, from which important bounds on  topcolor 
models can be derived. 

Here we examine the top-gluon contributions to \ddb~mixing.
In principle, many four-quark  operators, upon quark rotation to the mass basis,  
contribute to $\Delta m_D$ as a result of top-gluon exchange. 
Before rotation, we have the product of third-generation currents, which is 
enhanced by a factor of~\cite{bbhk} $\cot^2\theta\simeq 22$; the product
of one third-generation current and one light current, proportional to 
$\cot\theta\times\tan\theta=1$; and the product of two light currents, which 
is suppressed by $\tan^2\theta$. It appears at first that---although suppressed by 
$\tan^2\theta$---the last group would result in the largest contribution, since
the quark rotations in the first two generations are expected to be of 
the order of
the Cabibbo angle. 
A typical contribution to $\Delta m_D$ from the last group gives~\cite{bghp2}  
\beq
\Delta m_D \simeq 4\pi\alpha_s(M_G)\,\tan^2\theta\frac{({\cal U}_{L}^{cu})^2}
{2M_G^2}\,
(\bar{u}_L\gamma_\mu u_L)(\bar{u}_L\gamma^\mu u_L),
\label{dmdtg}
\eeq
which by itself gives a very stringent constraint,
$M_G/U_L^{cu}\geq 100$~TeV. 

This inference is not necessarily correct. 
Not only are the matrices ${\cal U}_{L,R}$ unitary, but it is also very  plausible that
the $2\times 2$ block of ${\cal U}_{L,R}$ rotating the  first two generations
is nearly unitary. This would be the most 
likely situation if both ${\cal U}_{L,R}$ have a form similar to $V_{\rm CKM}$. 
If this is the case, then 
\beq
{\cal U}_{L,R}^{uu*}{\cal U}_{L,R}^{uc}\simeq -
{\cal U}_{L,R}^{cu*}{\cal U}_{L,R}^{cc}.
\label{unit}
\eeq
This implies that all contributions involving one or two
light currents before rotation will almost cancel, thus leading to 
no real bound. We conclude that the only significant contributions 
to \ddb~mixing come from 
\beq
\Delta m_D\simeq 4\pi\alpha_s(M_G)\,\frac{\cot^2\theta}{2M_G^2}\,
({\cal U}_{L,R}^{tu*}{\cal U}_{L,R}^{tc})^2\,
(\bar{u}_L\gamma_\mu u_L)(\bar{u}_L\gamma^\mu u_L).
\label{dmdtg2}
\eeq
Using $f_D=0.2$~GeV, $\alpha_s(1{\rm ~TeV})=0.09$, and $\cot^2\theta=22$, this leads
to 
\beq
\Delta m_D \simeq 1.8\times10^{-13}\;\left(\frac{{\cal U}_{L,R}^{tu*}
{\cal U}_{L,R}^{tc}}
{\sin^5\theta_C}\right)^2\;\left(\frac{1 {\rm ~TeV}}{M_G}\right)^2\; {\rm ~GeV},
\label{topgres}
\eeq
where we  assume that the elements of ${\cal U}_{L,R}$ scale with the 
Cabibbo angle
$\theta_C\sim 0.22$ just as the elements of $V_{\rm CKM}$ do. 
This is compatible with the current experimental limit for $\Delta m_D$,
e.g., for $M_G\geq 2$~TeV or with some of the elements of 
${\cal U}_{L,R}$ 
being slightly smaller that the naive scaling.
Thus we see that generically the contributions of top-gluons, 
although not in conflict with bounds from \ddb~mixing, are not far from 
the current experimental sensitivity.

\subsubsection{OTHER NEW-PHYSICS SCENARIOS}
\label{ddb_othernp}
As is obvious from Figure~\ref{hnelson}, there is no shortage of 
theories beyond the standard model whose parameter space is either already constrained by
the current experimental limit on \ddb~mixing or sits just below it. 
Here we mention just a few more.

%
%
%% COMP: \paragraph means a fourth-level heading
%
%
\paragraph{Extended Higgs sectors without natural flavor conservation:} In these scenarios,
scalars have tree-level FCNCs that may saturate the current limits on mixing
just as in the case of topcolor scalars. 
%
%
%% COMP: \paragraph means a fourth-level heading
%
%

\paragraph{Fourth-generation down quarks:} Whether a fourth-generation down quark  is a member of a 
doublet or a singlet, its contributions to the box diagram dramatically break the 
GIM cancellation in Equation~\ref{heffsd} if it has a large Yukawa coupling
to the Higgs field. 
It could saturate the bound if its 
mixing with the standard-model quarks is judiciously chosen.
%
%
%% COMP: \paragraph means a fourth-level heading
%
%

\paragraph{Extra dimensions:} If gauge or matter fields propagate in 
compact extra dimensions, the low-energy effective theory below the 
compactification scale $1/R$ will contain the massless states, the 
zero-modes. Starting at the compactification scale there are massive 
Kaluza-Klein (KK) excitations of these fields. These towers may contribute 
to flavor observables and in particular to \ddb~mixing through loops.
In addition, if fermions propagating in the bulk have different bulk mass terms,
their wave function in the extra dimensions will differ (see, e.g., \cite{extradd}). 
This leads to 
nonuniversal couplings to the KK excitations of the gauge
fields and therefore to tree-level FCNC interactions. For instance, the 
interactions with the first KK excitations of gluons may mimic those of 
top-gluons if the third-generation quarks are localized toward the TeV 
brane in a compact five-dimensional model with Randall-Sundrum metric. 
Of course, as discussed above, the observation of any nonvanishing 
value of $x$ cannot be interpreted as 
evidence of new physics because of the proximity of the highest estimates
from  long-distance, hadronic physics. 
However, this observation may be complementary to any deviation from 
the standard model observed in other flavor-physics observables as well as  at  the energy frontier.

\subsection{Experimental Status and Prospects}
\label{ddexp}

The charm quark was discovered in 1974 in $e^+e^-$
annihilation at SPEAR at the Stanford Linear Accelerator Center and simultaneously in hadronic collisions at
Brookhaven National Laboratory~\cite{CHARM}. Since then, charm hadrons have been produced
by a wide range of experimental facilities. The principal
production processes are $e^+e^- \rightarrow c \bar c$ at $\sqrt
s= m[\Upsilon(4S)]$; $Z^0 \rightarrow c \bar c$;  hadroproduction, both at fixed-target experiments and at the Fermilab
Tevatron; photoproduction; and threshold production, $e^+e^- \rightarrow \psi(3770)
\rightarrow D \bar D$. The cross sections vary from 1.3~nb in
$e^+e^- \rightarrow c \bar c$ at $\sqrt s= m[ \Upsilon(4S)]$ to
microbarns for photoproduction, and to of order a millibarn at
the Tevatron. However the ratio of signal cross section to
background cross section varies from 1/2.5 in $e^+ e^-$
annihilation at the  $B$ factories to 1/100 at the Tevatron.

Trigger efficiencies vary by orders of magnitude between
experiments, and detection efficiencies vary as well. In a typical (30~fb$^{-1})$
year at a  $B$ factory, $ 8\times 10^7$ charm quarks
are directly produced, with a similar number produced through the
decay of $B$  mesons. The trigger efficiency is close to unity, and
the detection efficiency is of order $\sim 50\%$ for a two-body
hadronic decay. Because of the high relative production rate of charm
compared to background, excellent mass resolution, and excellent
particle identification, there is no need to exploit the long
lifetime of the charm hadrons to isolate a clean charm data
sample. At the $Z^0$, the environment is similar but the charm
hadrons have higher  momenta, $\sim 30 $~GeV/$c$ compared to 3  GeV/$c$ 
near  $B$ threshold~\cite{ALEPH}. In contrast, fixed-target
experiments measure the  $c$-hadron decay time very  precisely, which is crucial to isolate clean event samples from the large
noncharm backgrounds. The early part of Run IIa at the
Tevatron produced $\sim 3 \times 10^{11}$ charm quarks. Until recently, most of these particles would not
have been observed. However, there was a tremendous breakthrough
in 2002, when CDF~\cite{CDF_trigger} demonstrated the ability to trigger on
the detached vertices produced by hadronic $B$- and $D$-meson
decays~\cite{CDF_trigger}. Some 450,000 $D^0 \rightarrow K^- \pi^+$
decays have now been reconstructed. 

To give another idea of how significant an achievement this is, it is 
useful to define an effective reconstruction cross section for a given 
mode at a given experiment
as the quantity which, when multiplied by the integrated luminosity, 
yields the number of reconstructed events in that mode for that 
experiment. For CDF the effective reconstruction cross section for $D^0 
\rightarrow K^- \pi^+$ is about 7 nb.  This should be compared to the 
effective reconstruction cross section of about 13 pb for $D^0 
\rightarrow K^- \pi^+$  in  $e^+e^- \rightarrow D^0 X$ at 10 GeV.

The Cabibbo-allowed two-body decay $D^0 \rightarrow K^- \pi^+$ is
easy to reconstruct experimentally and has a large branching
ratio, 3.85\%~\cite{pdg02}. The number of reconstructed $D^0
\rightarrow K^- \pi^+$  decays therefore serves as a benchmark to quantify
the size of charm data samples and facilitates comparison between
experiments. Using this benchmark, Table ~\ref{tab:summary}
compares experiments that have recently published results on charm
mixing, rare decays, and searches for $CP$ violation. 
Table~\ref{tab:summary} also lists the time resolution for these
experiments, which is important in the measurement of
lifetimes and mixing. Fixed-target experiments have excellent
vertex and proper time  resolution, five to eight times better than those of the $e^+ e^-$ 
experiments operating near $B$--$\bar{B}$ threshold (for a discussion of the measurement 
of proper time
in heavy-flavor physics, see \cite{PROPER}). At the
Tevatron, the proper time resolution is similar to that achieved in fixed target experiments.

These data samples have led to many beautiful
measurements of the properties of charm hadrons~\cite{PDG}. The
lifetime of charm mesons is now known with exquisite  precision; for example, 
the lifetime of the $D^0$ and $D^+$ are known to 3 and
6 per mille, respectively. However, the data samples collected so
far have been too small to allow detection of charm mixing, $CP$
violation, and (with one recent exception) 
rare decays. Although the fixed-target program at
Fermilab is now complete, the outlook for much larger samples of
charm hadrons in the near future is very promising. The experiments  BaBar, Belle, 
and CDF expect to increase their data
samples, compared to those used in current publications, by
factors of 5, 10, and 30, respectively, over the next few years. Each
of these experiments also has an upgrade path. There are plans to
increase the luminosity of the KEK-B accelerator to $\sim 10^{35}$
${\rm cm}^{-2} {\rm s}^{-1}$ by about 2007, an order of magnitude
greater than  at present. The configuration is called
Super-KEK-B.
%\cite{Super-KEK-B}. 
(The plans and current status of this project
are described at http://www-kekb.kek.jp/SuperKEKB/Workshop.html.)
This impressive luminosity
corresponds to a production rate of $ 2 \times 10^9$ charm hadrons
in a Snowmass year of  $10^7$~s. A five-year program is
envisaged at KEK-B. The SLAC group is developing a proposal for a ``super $B$ factory'' with a luminosity goal of $\sim 10^{36}$ ${\rm cm}^{-2}
{\rm s}^{-1}$, corresponding to $2 \times 10^{10}$ charm hadrons
in  $10^7$~s. This requires a new machine and a very
significant upgrade of the BaBar detector, known as
SuperBaBar~(see, e.g., \cite{SuperBABAR}). 

At CDF there is great potential, if charm stays within the trigger
bandwidth, to reconstruct 30 million $D \rightarrow K^- \pi^+$ in the $4.4~fb^{-1}$ 
expected in
Run II (our estimate is based on the first 65 pb$^-1$ of integrated luminosity), 
which would be the largest charm sample so far recorded.
The sample could grow by another factor of five during Run IIb.
The success of the CDF detached vertex trigger augurs well for
future hadron collider experiments such as the proposed
BTeV experiment (see, e.g., \cite{BTeV,Kaplan}), which expects to collect $\sim 8
\times 10^8$ charm hadrons in  $10^7$~s. At the Large Hadron Collider (LHC), 
the charm
cross section is twice as  large as at the Tevatron, but most
of the increase occurs at inaccessibly small angles to the beam
direction. Detector background rates also increase. The design of
the trigger for the dedicated $B$-physics experiment
LHCb (see, e.g., \cite{LHCb}) is not expected to have significant acceptance
for charm. At the LHC experiments, ATLAS and CMS, the emphasis is placed on triggers for high
$p_T$ physics. The experiments do not have detached vertex triggers, although they may
be added as upgrades a few years after turn on. No projections for charm yields
exist at this time~\cite{lhc_c}. 

The prospect of a multi-hundred-GeV $e^+ e^-$ linear collider has
led to an investigation of the possibility of running at the $Z$  pole, where
the charm-quark cross section is $\sim 6$~nb. Although the linear
collider would run well above the $Z$ pole, a scheme exists to
allow continuous $Z$ running simultaneously, a so called Giga-Z
machine (see, e.g., \cite{GIGA-Z}). The production statistics of Giga-Z do not,
in our view, make it competitive with other future facilities, so we do not discuss it here.

The reaction $\psi(3770) \rightarrow D \bar D$ occupies a special
place in the charm experimentalist's and theorist's arsenal. It is
the only charm production process in which the  charm-quark/anticharm-quark pair produces two charm hadrons that are quantum-mechanically correlated. It is also the only experimental environment in which the probability to reconstruct two charm
hadrons in a single event is large. The $\psi(3770)$ therefore
offers crucial experimental advantages for the determination of
absolute charm branching ratios, charm-decay constants and charm semileptonic form  factors, and the CKM matrix elements $V_{cs}$ and
$V_{cd}$. This suite of measurements is important to the
international program in precision flavor
physics (see, e.g., \cite{GIGA-Z}) and is widely held to be the main motivation
for a charm factory. Particularly pertinent to this review, the $\psi(3770)$ offers unique opportunities to search for charm
mixing and $CP$  violation by exploiting quantum coherence and to search for rare
decays by exploiting a background-free environment. 

Very limited data have been taken at the $\psi(3770)$ since 1984, when Mark
III~\cite{MARKIII} ran there and accumulated 9.6 pb$^-1$. More recently, 
the BESII experiment obtained $20~{\rm pb}^-1$ at and nearby the resonance~\cite{besII_c}. 
At the 
$\psi(3770)$, CLEO-c \cite{YB} expects to accumulate a data sample 300 
times larger than MARK III, whereas BES III\footnote{The plans and current 
status of this project are described at 
http://bes.ihep.ac.cn/besIII/index.html.} will have a sample 10 times larger 
than CLEO-c. To demonstrate the power of $\psi(3770)$
running, we also consider the reach of a purely hypothetical ``Super Charm factory" 
with a luminosity 17 times larger than that of 
BEPC II, although  it is considered unlikely that such a machine could be
built. Table~\ref{tab:summaryfuture} summarizes the future charm
data samples we use for projections in this review.

In the kaon and $B_d^0$ meson systems, the mixing rates
are large, $(\Delta m/\Gamma)~\simeq O(1)$.  
The mixing
parameters can be measured directly from observation of flavor
oscillations as a function of time. In the $B_s^0$ system, the
mixing rate is expected to be so large in the standard model  
that excellent time resolution is required
to detect the oscillations. The situation is different for $D$
mesons, where the mixing rate is expected to be very small, precluding the
direct observation of flavor  oscillations. Alternative methods include the following: 

\begin{enumerate}
    \item Measurements of wrong-sign $D^0$ decays, using either
semileptonic final states or  hadronic final states.
    \item Comparison of the lifetime of the $D^0$ measured in hadronic
decays and to final states that are $CP$ eigenstates.
    \item Time-integrated measurements exploiting quantum coherence at the
$\psi(3770)$.
\end{enumerate}
We now consider each of these methods in turn. Only the first
two have been attempted so far. A detailed discussion of recent
results can be found elsewhere~\cite{Grothe_review}. 

In order to detect mixing, which is a small effect, 
it is necessary to know the initial
flavor of the $D$ meson and the flavor at the time it decayed. In
most measurements  that have been performed so far, both at fixed-target 
and at $e^+e^-$ facilities, the initial flavor of the $D^0$ is
determined from the observation of the sign of the ``slow pion''
(also called the ``soft pion") in the decays $D^{*+} \rightarrow
D^0 \pi_{s}^{+}$ and $D^{*-} \rightarrow \bar{D^0} \pi_{s}^{-}$,
where $s$ denotes soft. This strong decay is  extremely useful in  charm and beauty experiments because  the branching ratio is
large $D^{*+} \rightarrow D^0 \pi_{s}= 68.1\%$, and $Q$, the mass
difference between the initial and final states,   
is small---$Q=m(D^{*+})-[m(D^0)+m(\pi_s)]=6~{\rm MeV}$. Consequently,
$D^{*+}\to D^0\pi^+$ produces a narrow peak in the $Q$ distribution.
In addition, the background is extremely low, so the
miss-tag rate is small, about one per thousand in current experiments.

\subsubsection{SEMILEPTONIC FINAL STATES}
\label{sec_slfs}
In semileptonic decays, $A_f=\bar{A}_{\bar{f}}=0$ and we can write (in
the small mixing limit)
\beq
r(t)=\left|\frac{q}{p}\right|^2\,\left|g_-(t)\right|^2\simeq
\frac{e^{-\Gamma t}}{4}\,\left|\frac{q}{p}\right|^2\,(x^2+y^2)\;(\Gamma t)^2.~
\label{rmixt}
\eeq 
The value of $\bar{r}(t)$ is obtained by exchanging $q$ and $p$ in Equation~\ref{rmixt}.
In the absence of $CP$ violation, the integrated mixing rate is 
$R_M\simeq(x^2+y^2)/2$. 
The charge of the soft pion tags the flavor of the initial state as a $D^0$
or $\bar{D^0}$. The Cabibbo-favored decay [with $B(D^0\rightarrow
K^{-}l^{+}\nu)\simeq 7\%$] gives a $(pi_s^+\ell^-)$ ``right-sign" (RS) charge
correlation. If the $D^0$ mixes, before it decays it is a source
of ``wrong-sign" (WS) decays: $D^0 \rightarrow {\bar D^0} \rightarrow
K^{+} l^{-}\nu$. In the standard model, only one diagram exists for semileptonic decays and so there is no direct decay route for
$D^0\rightarrow K^+ l^- \nu$. Accordingly, we describe the
measurement as theoretically clean: the WS pair $(\pi_s^+\ell^-)$ is
unambiguously a mixing signature. However, the time distribution of
the mixing signal is quadratic in the mixing variables. 
Therefore,  although the $(\Gamma t)^2
e^{-\Gamma t}$ term suits fixed-target experiments, the measurement is
relatively insensitive to small values of $x$ and $y$. Furthermore,
$x$ and $y$ are not measured separately. 
Finally, the
undetected neutrino complicates the measurement. 

At fixed-target
experiments, beginning with E791,
the technique is to reconstruct the semileptonic decay using
$m(\pi_{s} K^{+}\ell^{-})-m(K^{+}\ell^{-})$. 
The 
reconstructed peak has a large width 
due to the missing neutrino. However, the
detached vertex leads to a very clean signal in both $e$ and $\mu$
modes. The E791 result is~\cite{E791_LEPTON}
$R_{M}={\textstyle \frac{1}{2}}(x^{2}+y^{2})<0.5 \% \; {\rm at} \; 90\% \;{\rm
CL}$.
CLEO obtains $R_M< 0.87\%~@~90\%~\rm{CL})$~\cite{CLEO_LEPTON}
measured in the channels $D^0 \rightarrow K^+ \mu^- \bar{\nu_{\mu}}$ 
and $D^0 \rightarrow K^{*+} e^- \bar{\nu_{e}}$, respectively. 
The most restrictive measurement of $R_M$ comes from 
  FOCUS~\cite{FOCUS_LEPTON}  
in the mode $D^0 \rightarrow K^+ \mu^- \bar{\nu_{\mu}}$. 
\begin{equation}
R_M-\frac{1}{2}(x^{2}+y^{2})<0.12\% \; {\rm at} \; 90\% \;{\rm
CL}
\end{equation}
The bound on $R_M$ is a circle in the $x-y$ plane centered at the origin.
In Figure~\ref{allmix} this bound is shown as a circle with horizontal shading.

The future  outlook for searcher of charm mixing 
using semileptonic final states is as follows.
$B$ factory results are expected soon. The SuperBaBar working group
estimates that, with a 10~ab$^{-1}$ data sample,
the sensitivity would be $ R_{M} < 5 \times 10^{-4}$~\cite{Williams}. 
The technique can, in principle, also be employed by future
experiments at hadron machines because  the presence of the lepton
in the final state helps triggering. However, the technique 
has the greatest sensitivity at  $e^+ e^-$ machines operating 
at the  $\psi(3770)$. In all cases, 
if mixing is detected in
this channel and $y$ turns out to be  larger than or comparable to $x$, a
separate measurement of $y$ will be needed.

\subsubsection{HADRONIC FINAL STATES}
\label{hadr_fs}
A $D^0$ can produce a WS hadronic final state either
by undergoing a doubly Cabibbo-suppressed (DCS) decay or by first
oscillating into a $\bar{D^0}$ that subsequently undergoes a
Cabbibo-favored (CF) decay. The WS decay includes three components: one from the DCS decay, a second from
mixing, and a third from the interference between the first two.
Assuming $CP$ conservation and expanding the decay rate up to
$O(x^2$) and $O(y^2$) results in the following expression
for the time evolution of the hadronic WS decay rate:
\beq
r(t) = e^{-\Gamma\,t}\,\left[ R_D + \sqrt{R_D}\,y'\,(\Gamma t)
+ {\textstyle \frac{1}{2}}\,(x^{\prime 2} + y^{\prime 2})\,(\Gamma t)^2\right],
\label{mixrate}
\eeq
whereas the RS decays have simple exponential 
time dependence~$\propto \exp^{-\Gamma t}$. 
Here $R_D$ is defined by 
\beq
\frac{A_{f}}{\bar{A}_{\bar{f}}}\equiv-\sqrt{R_D}\,e^{-i\delta}~
\label{rddef}
\eeq
and gives the rate for the DCS component. The mixing contribution is
quadratic, and hence very small, but the interference term is
linear in $y^\prime$ and may result in a measurable deviation from
a pure exponential decay.

The specific case $f=K^+\pi^-$ has been studied. 
The parameters $x^\prime$ and
$y^\prime$ are related to the mixing parameters $x$ and $y$ by a
rotation~\cite{Bergmann}
\begin{equation}
x^{\prime} = x \cos \delta_{K\pi}+ y \sin \delta_{K\pi},
\hspace{1.6cm} y^{\prime}=y \cos \delta_{K\pi} - x \sin
\delta_{K\pi}.
\label{eq9GRO}
\end{equation}
A  strong-phase $\delta_{K \pi}$  exists between the DCS and CF
decay amplitudes~\cite{bsn,bp}. Although it vanishes in the SU(3) limit, this can be a badly broken symmetry.
In a measurement based on the time evolution of
the WS state, it is not possible to determine the phase. Since the
expression is quadratic in $x^\prime$, its sign is also not
determined. We discuss techniques to determine $\delta_{K\pi}$ 
in Section~\ref{strph_thr}.

To search for $CP$ violation in the time evolution of the WS 
hadronic state one applies Equation~\ref{mixrate} to $D^0$ and 
$\bar{D^0}$ separately.
One determines $\{R^{+}_{\rm WS}, x^{\prime
2}_{+}, y^{\prime}_{+}\}$ for $D^0$ candidates and $\{R^{-}_{\rm WS},
x^{\prime 2}_{-}, y^{\prime}_{-}\}$ for $\bar{D^0}$ candidates, 
where $R^{\pm}_{\rm WS}$ are the corresponding wrong-sign rates. The
separate $D^0$ and $\bar{D^0}$ results can be combined to form the
quantities
\begin{equation}
A_D = \frac{R^{+}_{D}-R^{-}_{D}}{R^{+}_{D}+R^{-}_{D}}; ~~\;\;\;
A_M=\frac{R^{+}_{M}-R^{-}_{M}}{R^{+}_{M}+R^{-}_{M}},
\label{meadows}
\end{equation}
where $R^{\pm}_{M} \equiv(x^{\prime 2}_\pm + y^{\prime 2}_\pm)/2$.
Here $A_M$ parameterizes $CP$ violation in the mixing amplitude and 
leads to 
\beq
\frac{q}{p} = (1+A_M)^{\frac{1}{2}}\,e^{i\phi},~
\label{amdef}
\eeq
where $\phi$ is responsible for $CP$ violation in the interference between
the DCS decay and mixing. 
Also, 
$A_D$ is related to $CP$ violation in the DCS decay amplitude 
and is defined by making the 
replacements $\sqrt{R_D}\to\sqrt{R_D}(1+A_D)$ for the case 
of initial 
$D^0$ and $\sqrt{R_D}\to\sqrt{R_D}/(1+A_D)$ for the $\bar{D}^0$.
In this way, the measurement of the time dependence is sensitive to
the combinations $y'\cos\phi\pm x'\sin\phi$ through
\begin{equation}
x^{\prime}_{\pm} = \sqrt[4]{\frac{1 \pm A_M}{1 \mp A_M}}(x^{\prime}
\cos \phi \pm y^{\prime} \sin \phi),\label{eq_4_pg4}
\end{equation}

\begin{equation}
y^{\prime}_\pm = \sqrt[4]{\frac{1 \pm A_M}{1 \mp A_M}}(y^{\prime}
\cos \phi \pm x^{\prime} \sin \phi).\label{eq_5_pg4}
\end{equation}
An offset in $\phi$ of $\pm \pi$ can be absorbed by a change in
sign of both $x^{\prime}$ and $y^{\prime}$, effectively swapping
the definition of the two physical $D^0$ states without any other
observable consequence.  In order to avoid this ambiguity, we use the
convention that $|\phi|< \pi/2$.
CLEO uses a different approximation \cite{CLEO_MIX_KPI} in its analysis, 
which is valid for $A_D, A_M \ll 1: R^\pm_D=(1 \pm
A_D)R_D, x^{\prime}_\pm=\sqrt{1 \pm A_M}(x^{\prime} \cos \phi \pm
y^{\prime} \sin \phi)$,~~$y^{\prime}_\pm= \sqrt{1 \pm A_M} (y\prime
\cos \phi \mp x^\prime \sin \phi)$.

The total time-integrated hadronic WS rate, assuming $CP$
conservation and normalizing to the total RS rate, is
\begin{equation}
R_{\rm WS} =\frac{\int \Gamma_{\rm WS}(t)~dt}{ \int\Gamma_{\rm RS}(t)~dt}=R_D +
\sqrt{R_D} y^\prime + 1/2(x^{\prime 2}+ y^{\prime 2}).
\label{mix_int}
\end{equation}
In this approximation, the mixing rate is $R_{M}=1/2(x^{\prime
2} + y^{\prime 2}) = 1/2 (x^2 + y^2)$. If there is no mixing in
the $D^0$ system, Equation~\ref{mix_int} reduces to
$R_{\rm WS}=R_{D}$. 

CLEO~\cite{CLEO_MIX_KPI} and BaBar~\cite{BABAR_MIX_KPI} 
have obtained results with 9.0~fb$^{-1}$ and 57.1~fb$^{-1}$ of data, respectively. 
A result from Belle should soon be
available~\cite{BELLE_MIX_KPI}. 
All three experiments have results for the
time-integrated WS decay rate $R_{\rm WS}$
\cite{CLEO_MIX_KPI,BABAR_MIX_KPI,BELLE_KPI_WS}.

WS candidate events of the types $D^0 \rightarrow K^+
\pi^-$ and $\bar D^0 \rightarrow K^- \pi^+$ are selected by
requiring the $\pi_s$ from the $D^{*}$ decay and the daughter $K$
of the $D^0$ to have identical charge (WS tag). To detect a
deviation from an exponential time distribution in WS events, a
likelihood fit to the distribution of the reconstructed proper
decay time $t$ is performed. The likelihood fit includes a signal
and a background component and models each as the convolution of a
decay-time distribution and a resolution function.

The WS sample is statistically limited, amounting to $\sim 1/300$
the size of the RS sample selected with the same criteria except
for the WS tag. The RS sample is used to constrain aspects in the
fit that are common to the two samples. CLEO and BaBar determine
the resolution functions of WS and RS signals and of the common
background types with the RS sample. In the WS sample, a
significant additional complication arises from the much lower
achievable purity, e.g., $\sim 50\%$ in the CLEO analysis.

CLEO and BaBar both fit using Equation~\ref{mix_int} and consider the
case with and without $CP$ violation. The results of both experiments
are consistent with the absence of mixing and  $CP$ violation.
We first consider the case with $CP$ conservation. 
CLEO finds one-dimensional limits of 
$(1/2)x^{\prime 2} < 0.038\%$ and $-5.2\% < y^\prime <0.2\%$ at
the 95\% CL. The systematic error is  $\pm 0.2\%$ $(\pm 0.3\%)$ for
$x^\prime$ $(y^\prime)$ dominated by knowledge of the background
shapes and acceptances.

CLEO's two-dimensional result is a contour on the $x^\prime$--$y^\prime$
plane that contains the true value of $x^\prime$ and $y^\prime$ with
$95\%$ confidence. The contour is constructed using a Bayesian approach.
Systematic errors are small but are not included. 
The contour is shown as the cross hatched region in
Figure~3, where the strong phase
shift $\delta_{K \pi}$
between the Cabibbo-favored and DCS decays
has been assumed to be zero so that the range in $x^\prime$ $(y^\prime)$
corresponds to the range of the allowed region in $x$  $(y)$.

In their fit, the BaBar collaboration allows $x^{\prime 2}$ to take
nonphysical negative values and finds that the most likely fit point
when $CP$ is conserved has a negative value of $x^{\prime 2}$.   BaBar observe a
correlation between $x^{\prime 2}$ and $y^\prime$. They present
their results as a contour on
the  $x^{\prime 2}$--$y^\prime$ plane (see \cite{BELLE_KPI_WS}), where the 
contour has been
constructed using a frequentist approach based on toy Monte Carlo and
systematic errors have been included.
BaBar made their contours available to the authors
of this review and we have redrawn the $x^{\prime2}$--$y$ contour
on the $x$--$y$ plane  by converting $x^{\prime2}$ to $x\prime$
and assuming  $\delta_{K \pi}$ to be zero. The BaBar redrawn contour is
displayed in Figure 3.  The numerical result is given in Table 4, where
the BaBar limits on $x^{\prime 2}$ and $y^\prime$ are obtained by
projecting their contour onto the corresponding axis. Also shown in
Figure~3, and numerically in Table 4, is a preliminary contour from
FOCUS \cite{paper20} derived from WS $D^0 \rightarrow K^+ \pi^-$.

The BaBar upper limit for $x^{\prime 2}$ is almost three times
larger than CLEO's but is based on a data sample six times larger.
However, CLEO and BaBar have used different techniques to obtain the $95\%$ CL
limits, and the treatment of the fit output parameter $x^\prime$ differs
as well, as can be seen in Figure~\ref{allmix}. The region allowed by the 
BaBar result is about a factor of two more restrictive than that from CLEO.
This is to be expected, since the measurement of $x~(y)$ using this technique
scales as $1/L^{1/4}$. For a discussion with an alternative point of view
see Reference~\cite{Grothe_review}.The sensitivity of this technique
to $x$ and $y$ will improve by a factor of two for each $B$ factory
with $500~{\rm fb}^{-1}$. 

A direct comparison of the CLEO and BaBar results is not possible
when $CP$ violation is allowed in the fit, because CLEO uses
as fit output parameters $x^\prime$, $y^\prime$, $R_D$ and $A_D$,
$A_M$,  and $\sin{\phi}$, 
whereas BaBar uses $A_D, $$x^{\prime 2}_+$, $y^{\prime}_+$, 
$R^+_D$, $R^+_M$ ($x^{\prime 2}_-$, and $y^{\prime}_-$, $R^-_D$, $R^-_M$)} 
for the $D^0$ ($\overline{D^0}$) case.
Fits that allow for CP violation lead to slightly less restrictive limits 
on $x^\prime$ and $y^\prime$ (se Table~\ref{tab:xy_comp}). In no case is evidence 
for CP violation found, i.e. BaBar finds $A_D$ and $R_M$ consistent with zero, and
CLEO finds $A_M$, $A_D$ and $\sin\phi$ consistent with zero.

%
%
%% COMP: \paragraph means fourth-level head
%
%
\paragraph{Extraction of the time-integrated WS decay rate
$R_{\rm WS}$:}
CLEO and BaBar measure $R_{\rm WS}$ by repeating the fits described in
the previous section with the assumption of no mixing in the
$D^{0}$ system, i.e., $x=y=0$ \cite{CLEO_MIX_KPI,BABAR_MIX_KPI}. 
CP violation is allowed, so the fit returns $R_{WS}=R_D$ and $A_D$. 
Belle uses a fit in the $M_D - \delta m$
plane to determine the time-integrated number of signal events in
the candidate samples \cite{BELLE_KPI_WS}. The ratio of the number
of signal events in the WS and RS candidate samples yields
$R_{\rm WS}$, where the systematic error is dominated by the
uncertainty on the background shapes used in the fit. 
Table~\ref{wrongsignhadronic} compares
results from CLEO, BaBar, and Belle to earlier  measurements of WS decays by
E791 \cite{E791_KPI_WS}, ALEPH \cite{ALEPH_KPI_WS}, and FOCUS
\cite{FOCUS_KPI_WS}. 

The results are in reasonable agreement. We compute the world average to be 
\beq
\langle R_{\rm WS}\rangle = (0.368\pm0.021)\%.
\label{rws_wa}
\eeq
The WS rate is at roughly the level expected in the
standard model (see Equation~\ref{def_b}). 
If mixing were present, the WS integrated rate would
differ from the standard-model expectation.

%
%
%% COMP: \paragraph indicates fourth-level heading
%
%

\paragraph{$D^0$ Decays to Multibody Final States:}
CLEO has also measured $R_{\rm WS}$ in the multibody channels
$D^0\rightarrow K^+\pi^-\pi^0$ and $D^0 \rightarrow
K^+\pi^-\pi^+\pi^-$ with the results $R_{\rm WS} =
(0.43^{+0.11}_{-0.10}\pm 0.07)\%$ and
$R_{\rm WS}=(0.41^{+0.12}_{-0.11}\pm 0.04)\%$,
respectively~\cite{CLEO_WS_KPIPI0,CLEO_WS_K3PI}. $R_{\rm WS}$
need not be the same for different decay modes, but
within the errors, $R_{\rm WS}$ is the same for all decay modes
measured so far.  With the large data samples from the $B$
factories, it may be possible to set $D^0$--$\bar{D^0}$ mixing limits
using combined Dalitz plot and proper time fits in multibody
modes. These modes may prove useful in searching for $CP$
violation and understanding strong-phase shifts~\cite{CHIANG}.

The decay $D^0\rightarrow K^{0}_{S} \pi^+\pi^-$ may be used to
measure $x$ and $y$ directly, since the strong-phase difference may
be extracted simultaneously in a time-dependent fit to the Dalitz
plot. This is possible because both the RS and WS decays in the
submode $D^0 \rightarrow K^{*\pm} \pi^{\mp}$ have the same final
state. Thus, one can fit for the phase difference directly. The
sign of $x$ can also be extracted from such a fit.
CLEO presented evidence for a WS amplitude and measured
the branching fraction relative to the RS mode to be
\begin{equation}
\frac{B (D^0 \rightarrow K^{*+} \pi^-)}{B(D^0 \rightarrow K^{*-}
\pi^+)}=(0.5\pm0.2^{+0.5}_{-0.1}\pm^{+0.4}_{-0.1})\% \label{eqsmithb}
\end{equation}
and the strong-phase difference between the RS and WS to be
$(189^\circ \pm 10 \pm 3^{+15}_{-5})^\circ$ using a
time-independent Dalitz plot fit~\cite{ASNER}. The last
uncertainty is due to the choice of resonances and model. 
No $CP$-violating effects were observed when the sample was separated into
$D^0$ and $\bar {D^0}$ subsamples. Results of a time-dependent fit
with limits on $x$, $y$, and $CP$ violation are expected soon from
CLEO, BaBar, and Belle. This channel may offer the greatest
sensitivity to $x$ at the large integrated luminosities already
collected by Belle and BaBar.

\subsubsection{THE MEASUREMENT OF  $y$}
\label{meas_y}
The mixing parameter $y$ can be determined by measuring the
lifetime difference between $D^0$ decays to $CP$-even and $CP$-odd
final states. 
\beq
y_{CP}= \frac{\Gamma(CP~{\rm even})-\Gamma(CP~{\rm odd})}
{\Gamma(CP~{\rm even})+\Gamma(CP~{\rm odd})}\simeq
\frac{\Gamma(D^0\to K^+ K^-)}{\Gamma(D^0\to K^-\pi^+)}-1,
\label{ycpdef}
\eeq
which then results in 
\beq
y_{CP}=y\,\cos\phi - x\sin\phi\,\left(A_M+A_{\rm prod}\right),
\label{ycpfin}
\eeq
where the production asymmetry is defined as
\beq
A_{\rm prod}\equiv\frac{N(D^0)-N(\bar{D}^0)}{N(D^0)+N(\bar{D}^0)}.
\eeq
It is assumed that $(A_M,A_{\rm prod})\ll 1$, where $A_M$ and 
$\phi$ are defined as in the previous section. 

Then $y_{CP}$ is determined from the slope of  the decay-time
distributions in samples of $D^0 \rightarrow K^- \pi^+$, which is
an equal mixture of $CP$-even and $CP$-odd final states, and $D^0
\rightarrow K^-K^+$ or $\pi^- \pi^+$, which are even final states. 
An unbinned maximum likelihood fit to the distribution of the
reconstructed proper decay time $t$  of the $D^0$ candidates at
$e^+e^-$ machines, and the reduced proper time at fixed-target
experiments, is performed. Because $y_{CP}$ is measured from the ratio of
lifetimes, many systematic effects cancel.

A fixed-target experiment, FOCUS, was the first to make this measurement
and found a value of $y_{CP}$ of about 3\% that was several standard
deviations from zero \cite{FOCUS_KPI_WS}. The $e^+ e^-$ experiments BaBar, Belle, and
CLEO have now all made this measurement with greater precision than
FOCUS.

The three experiments determine $y$ from unbinned maximum-likelihood
fits to the distribution of the reconstructed proper decay time, $t$, of
the $D^0$ candidates. Small biases may occur in the measured lifetimes;
these are at the 1\% level or below.  The precision with which they
are known is limited by the statistics of the simulation and is the
dominant source of systematic uncertainty for BaBar, whereas for Belle and CLEO  knowledge of the background contributions to the signal is the largest
source of systematic uncertainty.

The technique, resolution, and systematic errors at $\sqrt{s} = 10$ GeV are
quite different from those in fixed-target experiments. $D^0$ candidates are
selected by searching for pairs of tracks with opposite charge and
combined invariant mass near the expected $D^0$ mass. 
The interception point of the $D^0$ momentum
vector with the envelope of the interaction point (IP) provides
the production vertex of the $D^0$ candidate. The proper
decay time of a $D^0$ candidate is derived from its mass and
flight length.

BaBar and CLEO employ a $D^*$ tag. They refit each $\pi_s$ candidate track with 
the constraint that it coincide with the $D^0$
candidate production vertex. This reduces substantially the
mismeasurement of the $\pi_s$, momentum caused by multiple
scattering. Then $\delta m=m(D^{*+})-m(D^0)$ 
is required to be consistent with the
known value. Belle does not use a $D^*$ tag but  requires that the $D^0$
candidate flight path be consistent with originating at the IP.
All three experiments reject events with secondary charm
production from $B$-meson decays with a momentum cut. The analyses
by BaBar and Belle are for a subsample of the full data sets. 
Table~\ref{4pg9lifetimedifference} lists the
subsample sizes and the number of events.

Recently BaBar reported an improved analysis that allows for $CP$
violation. If $CP$ violation occurs, the $D^0 (\tau^+)$ and the
$\bar{D^0} (\tau^-)$ will have different lifetimes to decay to
$CP$-even states. 
The effective lifetimes can be combined to form
\begin{eqnarray}
Y &=& \frac{\tau^0}{\langle \tau \rangle} - 1,~~ \langle \tau \rangle = 
(\tau^+  + \tau^-)/2
\label{ynoncp}\\
\Delta Y &=& \frac{A_\tau \tau^0}{\langle \tau \rangle} ~~~~A_\tau = 
\frac{(\tau^+  - \tau^-)}{(\tau^+  + \tau^-)}.\label{deltay}
\end{eqnarray}
Both $Y$ and $\Delta Y$ are zero if mixing is absent. Otherwise, in
the limit of $CP$ conservation in the mixing amplitude (and 
little production asymmetry), $Y = y \cos \phi$ and
$\Delta Y = x \sin \phi$. All BaBar data (91~fb$^{-1})$ have been
used. Four independent samples were isolated, three tagged with a
$D^*$: $K^- \pi^+$ measures $\tau^0$; $K^- K^+$ measures
$\langle \tau \rangle$, and $A_{\tau}$; $\pi^+ \pi^-$ measures 
$\langle \tau \rangle$, and $A_{\tau}$; and
an untagged sample of $\approx 146,000$ $K^- K^+$ measures
$\langle \tau \rangle$. An unbinned likelihood fit to $m_{D^0},t,\sigma_t$
yields $\langle \tau \rangle, \tau_0$, and $A_{\tau}$. The statistical uncertainty is
small; for example, for $K^- \pi^+$, it is $\approx 0.9 $~fs,
corresponding to 0.5\% in $y$. Assuming the same value of $\phi$,
the modes $K^+ K^-$ and $\pi^+ \pi^-$ can be averaged. BaBar combines the modes to find
\begin{eqnarray}
Y = ( 0.8 \pm 0.4^{+0.5}_{-0.4})\%~
\end{eqnarray}
\begin{eqnarray}
\Delta Y = ( -0.8 \pm 0.6 \pm 0.2).\
\end{eqnarray}

This is the single most precise value of $Y$ and the first
measurement of $\Delta Y$.
\footnote{As this review was going to press, the Belle collaboration announced
the preliminary result  of a new analysis allowing for CP violation in mixing and 
interference.
Using $91~{\rm  fb}^{-1}$
of data \cite{Yabsley_2}. They find $y_{CP} = (1.15 \pm 0.69)\%$.
They also measure $ A_{\Gamma} = (-0.2 \pm 0.6 \pm 0.3)\%$,
where $A_\Gamma$ differs from $\Delta Y$ by a factor  $(1 + y _{CP})$.}  
Because $Y$ and  $\Delta Y$ are consistent
with zero, there is no evidence for mixing or $CP$ violation.

In the limit of CP conservation 
$Y=y $. We average $Y$ with
previous measurements of $y_{CP}$.  Figure~\ref{allycp} summarizes the 
experimental  situation: the FOCUS measurement is
high, whereas the recent BaBar and Belle results move the world average
closer to zero. The new measurement of $Y$ continues this trend.
Within the errors, the measured values are consistent. To compute
the world average, we exclude the first  BaBar measurement because the new
measurement of $Y$ supersedes it. We find
\begin{equation}
\langle y_{CP} \rangle = (0.8\pm 0.5)\%.
\end{equation}
The average is consistent with the standard-model expectation of a value of $
| y |$ 
close to zero. Figure~\ref{allmix}
shows the world-average value of $y_{CP}$ on the $x$--$y$ plane.

We note that $\Delta Y$ is not systematically limited and so the
precision with which it is determined can be expected to improve
by a factor of $\sqrt{5}$ with a 500-fb$^{-1}$ data sample at
BaBar or Belle.  With the $10~{\rm ab}^{-1}$ sample expected at
SuperBaBar, the collaboration estimates the
following reach: $|Y| < 0.1\%$ and $|\Delta Y| <
0.06\%$~\cite{Williams}.

Figure~\ref{allmix} combines the constraints on the $x$--$y$
plane for all measurements of mixing  from hadronic and semileptonic 
decays discussed so far. 
The
strong-phase difference is assumed to be zero for 
this comparison. 
It is seen that the BaBar measurement of $x^{\prime 2}-y^\prime$ is in 
reasonable agreement with the horizontal band corresponding to the world average of
$y$.
If $\delta = 40^{\circ}$ (the estimated maximum
in Reference~\cite{Falk}), the  elliptical constraints on $x$ and $y$ from
$D^0 \rightarrow K^- \pi^+$ would be rotated by $\delta$ in a
counterclockwise direction, 
resulting in greater overlap 
of the CLEO and FOCUS measurements with the world average value of $y$. 
These results would be even more consistent if $\delta>90^\circ$~\cite{Bergmann}.
The overlap of the BaBar measurement with the $y$ band is not significantly
improved by rotation.
The errors are too large to permit any strong conclusion at this time.
However, it is clearly very important to obtain a 
measurement of the strong
phase. 

\subsubsection{MIXING VIA QUANTUM COHERENCE AT THRESHOLD}
\label{qcoher}

At  the $\psi(3770)$ and
at the $\psi(4140)$, reactions $\psi(3770) \rightarrow D^0$--$\bar{D}^0$ 
and $\psi(4140) \rightarrow \gamma D^0$~$\bar{D}^0$ produce a $D^{0}$~$\bar D^{0}$ 
pair in a state that is quantum-mechanically coherent. This enables simple new 
methods to measure the $D^{0}-\bar D^{0}$ mixing parameters~\cite{JON}
in a way similar to that proposed in Reference~\cite{IKAROS}. 
Consider, for instance, $D^{0}\bar
{D}^{0}\rightarrow(K^{-}\pi^{+})(K^{-}\pi^{+})$. The initial
$D^{0} \bar {D}^{0}$ state is
\begin{equation}
|i > =\frac{1}{\sqrt 2}\{|D^{0}(k_{1}, t_{1})\bar
D^{0}(k_{2}, t_{2})\rbrace + \eta_C|D^{0}(k_{2}, t_{2})
\bar{D}^{0}(k_{1}, t_{1}) \rbrace \},
\end{equation}
where $\eta_C=\pm$ is the charge-conjugation eigenvalue of the $D^{0} \bar D^{0}$ 
pair.
At the $\psi(3770)$, $\eta_C=-1$, so the  system is in
the antisymmetric initial state. 
The case where the $D^0$ decay is Cabibbo-favored (CF) and the $\bar{D^0}$ decay
is DCS (e.g. $D^0 \rightarrow K^- \pi^+$ $\bar{D^0} \rightarrow K^-
\pi^+$) is indistinguishable from the the case where the
$\bar{D^0}$ decay is CF and the $D^0$ decay is DCS. Upon adding
the amplitudes, the DCS contribution cancels. 
Therefore, for
$D^{0} \bar D^{0}$ pairs produced from the decay of the
$\psi(3770)$, the time-dependent rate is
\begin{equation}
\Gamma(t) \propto \frac{\Gamma^2}{8}
e^{-\Gamma(t_{1}+t_{2})}|\frac{p}{q}^{2}|B(k_{1})|^{2}|B(k_{2})|^{2}
(x+y)^{2}(t_{1}-t_{2})^{2},
\label{eqyb18}
\end{equation}
where $B\equiv B_{K^{-}\pi^{+}}= A(D^0 \rightarrow K^- \pi^+)$,
i.e., the decay amplitude. Normalizing to $\Gamma [D^{0} \bar D^{0}
\rightarrow (K^{-} \pi^{+})(K^{+}\pi^{-})]$, which is given by
\begin{equation}
\Gamma(t) \propto
{\textstyle \frac{1}{2}}e^{-\Gamma(t_{1}+t_{2})}|B_{K^{-}\pi^{+}}(k_{1})|^{2}|B_{K^{+}\pi^{-}}
(k_{2})|^{2},
\label{eqyb20}
\end{equation}
implies for the time-integrated ratio
\begin{equation}%equation 14
R \bigg (
\frac{(K^{-}\pi^{+})(K^{-}\pi^{+})}{(K^{-}\pi^{+})(K^{+}\pi^{-})}
\bigg ) = \frac{x^2 + y^2}{2}  \bigg |\frac{p}{q} \bigg |^{2}
\frac{ |B_{K^{-}\pi^{+}}|^{2}}{|B_{K^{+}\pi^{-}} |^{2}}.
\label{yb21}
\end{equation}
This is similar to the case of semileptonic final states,
\begin{equation}
R \bigg (\frac{l^{\pm} l^{\pm}}{l^{\pm} l^{\mp}}
\bigg)=\frac{x^2 + y^2}{2}.
\label{eqyb22}
\end{equation}
We note that the DCS amplitude does not cancel for all decay
modes~\cite{IKAROS}. 
For the case where one final state is hadronic and the other
semileptonic, we have
\begin{equation}
R \left(\frac{l^{+}(K^{-}\pi^+)}{l^{+}(K^+\pi^-)}\right)=
%\left|\frac{\bar A_{K^{-}\rho^{+}}}{B_{K^{-}\rho^{+}}}\right|^{2} + %\frac{x^2 + y^2}{2}.
R_D + {\textstyle \frac{x^2 + y^2}{2}}.
\end{equation}
These unambiguous signatures for $D^0$--$\bar {D}^0$ mixing arise
because of  the quantum coherence of the initial state 
and are analogous to the situation for $B^0$--$\bar{B}^0$ pairs 
at the $\Upsilon (4S)$.

The measurement of $R_{M}$ can be performed unambiguously with
the decays $\psi^{\prime\prime} \rightarrow K^-\pi^+K^-\pi^+$ and
$\psi^{\prime\prime}\rightarrow \ell^\pm(KX)^\mp
\nu\ell^\mp(KX)^\mp \nu$.  The hadronic final state cannot be
produced from DCS decays. This final state is also very appealing experimentally, because it involves a two-body decay
of both charm mesons, with energetic charged particles in
the final state that form an overconstrained system. 
Particle identification is crucial in this measurement
because if 
both the kaon
and pion are misidentified in one of the two $D$-meson decays in
the event, it becomes impossible to discern whether mixing has occurred.

The number of RS events in the all-hadronic and hadronic-semileptonic 
channels combined produced in 3~fb$^{-1}$ is expected to be about 50,000, corresponding to
$\sqrt{R_{M}}\leq 1\%$ at 95\% CL. At threshold, the
sensitivity scales as $ x \propto L^{-1/2}$, whereas 
at a  $B$ factory, for the measurement of mixing through the time evolution of 
$D^0\to K^+\pi^-$, 
the scaling goes like  $ x \propto L^{-1/4}$. This
implies that a 3-fb$^{-1}$ data sample at charm threshold has
similar reach to a  $B$-factory data sample of 500 fb$^{-1}$. Because 
the charm-threshold measurement is background-free, it is
statistically limited rather than systematically limited. At BES III, where the data sample is expected to be 10 times greater, the limit will improve to $\sqrt{R_{M}}\leq 0.3\%$, but only if the particle
identification capabilities are adequate. 
If it were possible to obtain 500~fb$^{-1}$ at the
$\psi(3770)$, 
the limit would be $\sqrt{R_{M}}\leq 0.08\%$.

The investigation of mixing parameters at the $\psi(4140)$
provides valuable complementary constraints and a useful
consistency check. Note that the final-state $(K^-\pi^+)~(K^-\pi^+)$
and $(K^-\pi^+)~(\ell^+)$ from the $\eta_C=+1$ initial state have a decay
width proportional to $y^{\prime}$. Because $R_{M}$ is bigger than
$x$ or $y$, the determination of $y^{\prime}$ from $\eta_C = +1$ initial
states is comparable in sensitivity to measurements from the
higher-statistics $\eta_C =-1$ initial state.

\subsubsection{STRONG PHASES AT CHARM THRESHOLD}
\label{strph_thr}
We can also take advantage of the coherence of the $D$ mesons
produced at the $\psi^{\prime\prime}$ to extract the strong-phase
difference $\delta$ between the direct and DCS amplitudes that
appears in the time-dependent mixing measurements. Because the $CP$
properties of the final states produced in the decay of
the $\psi(3770)$ are anticorrelated, one $D$ state decaying into a
final state with definite $CP$ properties immediately identifies or
tags the $CP$ properties of the state ``on the other side.'' If one
state decays into, for example, $\pi^0 K_S$ with  $CP=-1$, the other
state is ``$CP$-tagged'' as being in the  $CP=+1$ state. This allows 
measurement of the branching ratio $B(D_{CP} \rightarrow K^-\pi^+)$
and $\cos \delta$. A triangle relation follows from the definition
of $D_{CP}$:
\begin{equation}
\sqrt{2}A(D_{CP} \rightarrow K^- \pi^+)=A(D^0 \rightarrow K^-
\pi^+) \pm A(\bar{D^0} \rightarrow K^- \pi^+). \label{pg32_25}
\end{equation}
This implies
\begin{equation}
1\pm2 \cos \delta \sqrt{R_D} = 2 \frac{B(D_\pm \rightarrow
K^-\pi^+)}{B(D^0 \rightarrow K^-\pi^+)}, \label{cosdel}
\end{equation}
where $R_D$ is the ratio of the DCS decay to the CF mode. We have
used the fact that $R_D \ll \sqrt{R_D}$ and neglected $CP$
violation in mixing, which could undermine the $CP$-tagging
procedure by splitting the $CP$-tagged state on one side into a
linear combination of $CP$-even and $CP$-odd states, thus requiring
time-dependent studies. Both effects, however, are negligible.  Now, if decays of both $D_+$ and $D_-$ are measured, $\cos
\delta$ can be obtained from the asymmetry,
\begin{equation}
A =\frac{B(D_+ \rightarrow K^- \pi^+)-B(D^-
\rightarrow K^- \pi^+)}{B(D^+ \rightarrow K^- \pi^+) + B(D^- \rightarrow K^- \pi^+)}
= 2 \sqrt{R_D} \cos\delta.
\end{equation}

The asymmetry $A$ is expected to be small. Thus we have
\begin{equation}
\Delta(\cos\delta) \approx \frac{1}{2\sqrt{R_D}\sqrt{N}},
\end{equation}
where $N$ is the total number of $CP$-tagged $K^-\pi^+$ from the  $C = -1$ 
initial state. At  CLEO-c, the total number $N$ is expected to be about 
32,000 in one year run of CLEO-c, 
leading to
an expected accuracy of about $\pm 0.05$  in $\cos\delta$.

Reference~\cite{gro_pg10_10} outlines an alternative  method to extract
$\delta_{K \pi}$ from a measurement of the rates of the DCS and CF
decays of the type $D \rightarrow K \pi$. This method requires the
determination of rate asymmetries in $D$ meson decays to $K_{L}
\pi$ and $K_{S} \pi$, and makes the assumption that the two $\Delta I = 1/2$ DCS amplitudes have equal phase. Belle~\cite{gro_pg10_11} finds
that the relevant measurements of $K_L$ and $K_S$ mesons are
possible with a statistical precision sufficient to constrain
$\delta_{K \pi}$, but their preliminary measurement is not yet
sensitive enough to do so:
\begin{equation}
A= \frac{\Gamma(D^0 \rightarrow K^{0}_{S}\pi^{0})- \Gamma(D^0
\rightarrow K^{0}_{L}\pi^0)} {\Gamma (D^0 \rightarrow
K^{0}_{S}\pi^{0})+ \Gamma(D^0 \rightarrow K^{0}_{L}\pi^0)} = 0.06
\pm0.05\pm 0.05. \label{eqsmithY}
\end{equation}
As the data sample grows, this measurement will become  interesting, but it will not be as precise as the determination of
$\delta_{K \pi}$ at threshold. Because the two measurements have
entirely different systematic errors, the $B$-factory measurement
will serve as a useful cross check.

\subsubsection{SEARCHES FOR $CP$ VIOLATION IN
THE $D$-MESON SYSTEM}

Searches for $CP$ violation in charm decays have reached accuracies
of several percent~\cite{yb48pg56,yb57pg56}. In order to
measure an asymmetry, one must know the flavor of the decaying meson. Charged $D$ 
mesons are self-tagging; the neutral $D$ meson requires the 
above-mentioned $D^{*}$ tag to determine whether the decaying 
particle was a $D^0$ or $\bar D^0$.

The fixed target experiments E791 and FOCUS have made some of the most sensitive 
searches for direct CP violation in the charm system.
A subtlety in the measurement is that there is a production asymmetry 
between $D^+$ and $D^-$ and between $D^0$ and
$\bar{D^0}$, and so these experiments normalize all asymmetries to
some known CF mode where there can be no $CP$
violation---for example, 
\begin{equation}
A_{CP}(KK\pi) =\frac{\eta(D^+)-\eta(D^-)}{\eta (D^+)+\eta (D^-)},
\end{equation}
where
\begin{equation}
\eta(D^\pm)=\frac{N(D^\pm \rightarrow K^\mp K^\pm
\pi^\pm)}{N(D^\pm \rightarrow K^\mp \pi^\pm \pi^\pm)}.
\end{equation}
Signals are isolated according to ($a$) the significance of the 
longitudinal separation between the decay (secondary) vertex and 
production (primary) vertex $(L/\sigma)$, where $L$ is the longitudinal 
separation and $\sigma$ is the calculated resolution in $L$; ($b$) the 
confidence level of the decay vertex fit; ($c$) the vector sum of the 
momenta from decay vertex tracks, which is required to point to the production 
vertex in the plane perpendicular to the beam; and ($d$) particle 
identification and momentum cuts that are specific to a given final state.

CDF recently presented their first search for $CP$ violation
in the charm system using the two-track hadronic
trigger~\cite{CDF_ACP}. Using 65 pb$^{-1}$ of data collected
during 2002, about 100,000 $D^0 \rightarrow K^- \pi^+$, $8,000$
$D^0 \rightarrow K^- K^+$ and $4,000$ $D^0 \rightarrow \pi^+
\pi^-$ candidates pass selection cuts including  $\delta m (D^*)$,
the impact parameter of the $D^0$, and the projected decay length
of the $D^0$. The already impressive statistics yield the most
sensitive searches for asymmetries in the following modes:
\begin{equation}
A_{CP} (D^0 \rightarrow K^+ K^-) = (2.0 \pm 1.7 \pm 0.6) \%,
\end{equation}
\begin{equation}
A_{CP} (D^0 \rightarrow \pi^+ \pi^-) = (3.0 \pm 1.9 \pm 0.6) \%.
\end{equation}
The dominant systematic error comes from the correction for the
charge asymmetry for  low-momentum tracks in the CDF tracking
system.

CLEO has made numerous searches for $CP$ violation in the charm
sector. There is no production asymmetry or appreciable detection
asymmetry in the $\Upsilon(4S)$ energy region, but statistics are
limited. 
Selected recent asymmetry measurements from all of these experiments 
are tabulated in
Tables~\ref{table:acp_dzero_summary} and  \ref{table:acp_dcharged_summary}  
and in Figures~\ref{fig:ACP_graph1}
and~\ref{fig:ACP_graph2}. We compute the following averages:
\begin{equation}
\langle A_{CP} (D^0 \rightarrow K^+ K^-)\rangle = (0.8 \pm 1.2 ) \%,
\end{equation}
\begin{equation}
\langle A_{CP} (D^0 \rightarrow \pi^+ \pi^-) \rangle  = (2.7 \pm 1.6) \%.
\end{equation}
There is no evidence for $CP$ violation at the current level of
sensitivity. 

An entirely different way
to search for $CP$ violation is to exploit quantum  
coherence at the $\psi^{\prime\prime}$.
The production process
\begin{equation}
e^+e^+ \rightarrow \psi^{\prime\prime} \rightarrow D^{0} \bar D^0
\end{equation}
produces an eigenstate of  $CP$+, in the first step, since the
$\psi^{\prime\prime}$ has $J^{PC}$ equal to $1^{--}$.
Consider the case where both the $D^0$  and the $\bar D^0$ decay
into $CP$ eigenstates. Then the decays
\begin{equation}
\psi^{\prime\prime} \rightarrow f^{i}_{+} f^{i}_{+} ~{\rm or} ~
f^{i}_{-} f^{i}_{-}
\end{equation}
are forbidden, where $f_+$ denotes a $CP+$ eigenstate and $f_-$
denotes a $CP-$  eigenstate. This is because
\begin{equation}
CP (f^{i}_{\pm}f^{i}_{\pm})=(-1)^\ell =-1
\end{equation}
for the $\ell=1$~$\psi^{\prime\prime}$.
Thus, observation of a final state such as $(K^+K^ )(\pi^+\pi^-)$ constitutes
evidence of $CP$ violation. Moreover, all pairs of $CP$ eigenstates, where both 
eigenstates are even or both are odd,  can be summed over for this measurement. This
provides a sensitive way to detect $CP$ violation in charm decays and
could become sensitive enough to see standard-model mechanisms.

Table~\ref{tab:tabYB18pg110} estimates the total number
of events that would be observed for maximal $CP$ violation in a one-year run at CLEO-c. 
The event samples are not large, but the measurement
is essentially background free. Moreover, this method is important because it has 
unique sensitivity to the quantum-mechanical phase in the amplitude.
This measurement can also be performed at higher energies, where
the final state $D^{*0}\bar D ^{*0}$ is produced. When either
$D^*$ decays into a ${\pi^0}$ and a $D^0$, the situation is the
same as above. When the decay is $D^{*0}\rightarrow \gamma D^0$,
the $CP$ parity is changed by a multiplicative factor of 1, and all
decays $f^{i}_{+} f^{i}_{-}$ violate $CP$~\cite{yb59pg56}.

The coherent nature of the wave function prevents measurements of
time-integrated mixing-induced $CP$ asymmetries from $CP$-odd $D^{0}
\bar D^{0}$ pairs produced at the $\psi$(3770). Thus, at this
energy, nonzero time-integrated asymmetries would be a
manifestation of direct $CP$ violation. The measurement of direct $CP$ 
violation in $D^0$ decays at the 
$\psi(3770)$ uses a different tagging technique from that at higher 
energies because the decay $D^{*+} \rightarrow D^0 \pi^+ $ is 
kinematically forbidden.  One $D^0$ meson in the event, the tag, is 
partially or fully reconstructed. One example of partial reconstruction 
is lepton flavor tagging, in which the lepton in the decay $D^0 
\rightarrow \ell^+ X$ is detected and the charge of the lepton 
determined. With the flavor of the tag established, the flavor of the 
other $D^0$ meson in the event is fixed.
Table \ref{tab:tabYB19pg110} summarizes the expected sensitivity for
direct $CP$ violation in a one-year run at the $\psi(3770)$ at CLEO-c.

$CP$-even initial states, such as those produced at the
$\psi(4140) \rightarrow \gamma D^{0}\bar D^{0}$, are amenable to
time-integrated $CP$ asymmetries that are nonvanishing and depend
only linearly on the mixing parameter $x$. In this case, there is a
smaller statistical accuracy,due to the smaller cross section,  
and the overall CLEO-c sensitivity is
of the order of 3\% using lepton flavor tags only.   Backgrounds
are expected to be modest and to have a negligible effect on the
measured asymmetries, in agreement with previous studies~\cite{yb32pg152,yb30pg152}.

The studies described here are only 
some examples of the many possible search strategies \cite{yb59pg56}. 
Dalitz plot analyses uncover interference effects that
are sensitive probes of $CP$-violating phases. Although some of
these measurements can be performed with comparable accuracy at
the  $B$ factories, the number of studies that can be performed at
threshold is very broad and includes some unique measurements
exploiting the quantum coherence of the initial state.

Current measurements of $CP$ violation in $D^0$ decay are at the
several-percent sensitivity level. The limits are considerably
worse in $D^+$ and $D_S$ decay. There are significant
opportunities to search for the effects of new physics via the
mechanism of $CP$ violation both at charm threshold and at the $B$
factories and CDF.

The most reliable way to compare the reach of future experiments
for $A_{CP}$ would be to compile a list of sensitivities estimated
by each experiment based on detailed simulations. In almost all
cases the simulations, or other detailed estimates, do not exist.
Instead we will use the simplistic benchmark of the
number of reconstructed $D^0 \rightarrow K^- \pi^+$ to estimate
sensitivities.  The $CP$ asymmetry scales like $1/\sqrt{N(D^0
\rightarrow K^- \pi^+)}$. To test this method, we note that the
experiments E687, E791, and CLEO II, which each reconstructed a
${\rm few}\times 10^4$ of these decays, had a similar $CP$ reach of
$\delta A_{CP} \sim 5\%$. Using this result as a normalization, one
expects $\delta A_{CP} \sim 3\%$ for FOCUS and $\delta A_{CP} \sim
2\%$ for CDF. Both are in agreement with the precision reported by the
experiments. We compute the sensitivities for future experiments
in Table~11. Experiments can be expected to probe $CP$
asymmetries at the $10^{-3}$ level during this decade. In the early part of the next decade, we can hope to see $CP$ asymmetries probed at the
$10^{-4}$ level.

\section{RARE CHARM DECAYS}
\label{sec_rare}
We have seen that charm processes tend to be affected by large nonperturbative  effects. However, for some modes, a window  exists in which theoretical predictions are   sufficiently under control to allow tests of the short-distance structure of the FCNC transition. 
This is, to some extent, the case in $c\to u\ell^+\ell^-$ modes, and therefore we concentrate on their potential. On the other hand, radiative charm decays, such as 
those mediated by $c\to u\gamma$, 
are largely dominated by long-distance physics. 
Their experimental accessibility presents an opportunity to study 
purely nonperturbative effects in radiative weak decays.
We first review the standard-model predictions for the leptonic, 
semileptonic, 
and radiative decays. Then we study the potential for new-physics signals
in $c\to u\ell^+\ell^-$. Finally, we survey present experimental knowledge and 
future prospects in rare charm decays.

\subsection{The Standard-Model Predictions}

The short-distance contributions to the $c\to u$ transitions
are  induced at one loop in the standard model.  
It is convenient to use an effective description with the 
$W$ boson and the  $b$ quark 
being integrated out as their  respective thresholds are reached in the renormalization group evolution~\cite{gw}.
The effective Hamiltonian is given by \cite{bghp,greub,bghp2}
\begin{eqnarray}
{\cal H}_{\rm eff} &=&-{4G_F\over\sqrt 2} \left[ \sum_{q=d,s,b} 
C_1^{(q)}(\mu)O_1^{(q)}(\mu) + C_2^{(q)}(\mu)O_2^{(q)}(\mu) \right.\nonumber\\ 
& &\left.+\sum_{i=3}^8 C_i(\mu)O_i(\mu)\right]  \,,\,\, m_b<\mu<M_W \nonumber\\
{\cal H}_{\rm eff} & = & -{4G_F\over\sqrt 2}\left[  \sum_{q=d,s} 
C_1^{(q)}(\mu)O_1^{(q)}(\mu) + C_2^{(q)}(\mu)O_2^{(q)}(\mu)\right.\nonumber\\ 
& &\left.+ \sum_{i=3}^8 C'_i(\mu)O'_i(\mu)\right]  \,,\,\, \mu<m_b\,, 
\label{heff}
\end{eqnarray}
with $\{O_i\}$ being the complete operator basis, 
$\{C_i\}$ the corresponding Wilson coefficients, and $\mu$ the renormalization
scale; the primed quantities
are those for which the  $b$ quark has been eliminated. 
In Equation~\ref{heff}, the Wilson coefficients contain the dependence
on the CKM matrix elements $V_{qq'}$. 
The CKM structure of these transitions  differs drastically from that of the analogous $B$-meson processes. 
The operators $O_1$ and $O_2$ are explicitly split into their 
CKM components,
\begin{equation}
O_1^{(q)}=(\bar{u}_L^{\alpha}\gamma_\mu q_L^{\beta})
(\bar{q}_L^{\beta}\gamma^\mu c_L^\alpha)\, , \qquad  
O_2^{(q)}=(\bar{u}_L^{\alpha}\gamma_\mu q_L^{\alpha})
(\bar{q}_L^{\beta}\gamma^\mu c_L^\beta)\, ,
\label{o1q}
\end{equation}
where $q=d,s,b$, and $\alpha$, $\beta$ are contracted color indices. 
The rest of the operator basis is defined in the standard way.
The QCD penguin operators are given by
\begin{eqnarray}
O_3&=&(\bar{u}_L^{\alpha}\gamma_\mu c_L^{\alpha})\sum_{q}(\bar{q}_L^\beta
\gamma^\mu q_L^\beta)\, , \qquad 
O_4 =(\bar{u}_L^{\alpha}\gamma_\mu c_L^{\beta})\sum_{q}(\bar{q}_L^\beta
\gamma^\mu q_L^\alpha)\, , \nonumber\\
O_5&=&(\bar{u}_L^{\alpha}\gamma_\mu c_L^{\alpha})\sum_{q}(\bar{q}_R^\beta
\gamma^\mu q_R^\beta)\, , \qquad 
O_6=(\bar{u}_L^{\alpha}\gamma_\mu c_L^{\beta})\sum_{q}(\bar{q}_R^\beta
\gamma^\mu q_R^\alpha). \label{qcdpen}
\end{eqnarray}
The electromagnetic and chromomagnetic dipole operators are
\beq
O_7 = \frac{e}{16\pi^2}m_c(\bar{u}_L\sigma_{\mu\nu}c_R)F^{\mu\nu}\, ,
\qquad 
O_8 = \frac{g_s}{16\pi^2}m_c(\bar{u}_L\sigma_{\mu\nu}
T^a c_R)G^{\mu\nu}_a\ ;
\label{gdipole}
\eeq
and finally the four-fermion operators coupling directly to the charged leptons
are
\beq
O_9=\frac{e^2}{16\pi^2} (\bar{u}_L\gamma_\mu c_L)( \bar{\ell}\gamma^\mu
\ell) \, , \qquad 
O_{10}=\frac{e^2}{16\pi^2} (\bar{u}_L\gamma_\mu c_L)( \bar{\ell}\gamma^\mu
\gamma_5\ell) .
\label{laxial}
\eeq
The matching conditions at $\mu=M_W$ for the Wilson coefficients
of the operators $O_{1-6}$ are 
\beq
C_1^{q}(M_W) = 0, \qquad C_{3-6}(M_W)=0, \qquad 
C_2^{q}(M_W)=-\lambda_q,
\label{mco16}
\eeq
with $\lambda_q=V^*_{cq}V_{uq}$. The corresponding conditions for the 
coefficients of the operators $O_{7-10}$ are 
\begin{eqnarray}
C_7(M_W)&=&-\frac{1}{2}\left\{ \lambda_s F_2(x_s) +\lambda_b F_2(x_b)
\right\},
\nonumber\\
C_8(M_W)&=&-\frac{1}{2}\left\{ \lambda_s D(x_s) +\lambda_b D(x_b)
\right\}, 
\nonumber\\
C^{(')}_9(M_W)&=&\sum_{i=s,(b)}\lambda_i 
\left[-\left(F_1(x_i)+2\bar{C}(x_i)\right)
+\frac{\bar{C}(x_i)}{2s_w^2}\right],
\nonumber\\
C^{(')}_{10}(M_W)&=&-\sum_{i=s,(b)} \lambda_i\frac{\bar{C}(x_i)}{2s_w^2}.
\label{c10mw}
\end{eqnarray}
In Equation~\ref{c10mw}, we define $x_i=m_i^2/M_W^2$; 
the functions $F_1(x)$, $F_2(x)$, and $\bar{C}(x)$ are 
those derived in Reference~\cite{IL81}, and 
the function $D(x)$ was defined in Reference~\cite{bghp}. 

To compute the $c\to u\ell^+\ell^-$ rate at leading order, 
operators in addition to $O_7$, $O_9$, and $O_{10}$ must contribute. 
Even in the absence of the strong interactions, the insertion of the 
operators $O_2^{(q)}$ in a loop would give a contribution 
sometimes referred to as leading-order mixing of $C_2$ with 
$C_9$. 
When the strong interactions are included, further mixing
of the four-quark operators with $O_{7-10}$ occurs. 
The effect of these QCD corrections in the renormalization group running from $M_W$ down to $\mu=m_c$ is particularly important in $C_7^{\rm eff}(m_c)$, 
the coefficient determining the $c\to u\gamma$ amplitude. As was shown in Reference~\cite{bghp}, the QCD-induced mixing with $O_2^{(q)}$
dominates $C_7^{\rm eff}(m_c)$. The fact that the main contribution
to the $c\to u\gamma$ amplitude comes from the insertion of four-quark 
operators inducing light-quark loops signals the presence of large long-distance effects. This was confirmed  \cite{bghp,greub} 
when these nonperturbative contributions were estimated and found to dominate the 
rate. Therefore, we must take into account 
effects of the strong interactions in $C_7^{\rm eff}(m_c)$. 
On the other hand, the operator $O_9$ mixes
with four-quark operators even in the absence of QCD 
corrections~\cite{buras}. 
Finally, the renormalization-group running does not affect $O_{10}$, i.e., $C_{10}(m_c)=
C_{10}(M_W)$.  
Thus, in order to estimate the $c\to u\ell^+\ell^-$ amplitude, it 
is a good approximation to consider the QCD effects only 
where they are dominant, namely in $C_7^{\rm eff}(m_c)$, whereas 
we expect these to be less dramatic in $C_9^{\rm eff}(m_c)$.

The leading-order mixing of $O_2^{(q)}$ with $O_9$ results in 
\begin{equation}
C^{('){\rm~eff}}_9=C_9(M_W)+\sum_{i=d,s,(b)}\lambda_i\left[
-\frac{2}{9}{\rm ~ln}\frac{m_i^2}{M_W^2} +\frac{8}{9}\frac{z_i^2}{\hat s}
-\frac{1}{9}\left(2+\frac{4z_i^2}{\hat s}\right)
\sqrt{\left| 1-\frac{4z_i^2}{\hat s}\right|}~{\cal T}(z_i) \right], 
\label{c9eff}
\end{equation}
where we have defined
\begin{equation}
{\cal T}(z)=\left\{
\begin{array}{cc}
2{\rm ~arctan}\left[\frac{1}{\sqrt{\frac{4z^2}{\hat s}-1}}\right] 
& ({\rm for~}\hat s < 4 z^2) \\ 
\\
{\rm ~ln}\left|\frac{1+\sqrt{1-\frac{4z^2}{\hat s}}}
{1-\sqrt{1-\frac{4z^2}{\hat s}}}\right| -i\pi~ & ({\rm for~}
\hat s > 4 z^2),
\end{array}
\right.
\label{deft}
\end{equation}
and $\hat s\equiv s/m_c^2$, $z_i\equiv m_i/m_c$.
The logarithmic dependence on the internal quark mass $m_i$ in the 
second term of Equation~\ref{c9eff} cancels against a similar term in 
the Inami-Lim function $F_1(x_i)$ entering in $C_9(M_W)$, 
leaving no spurious divergences in the $m_i\to 0$ limit.\footnote{
Fajfer et~al.~\cite{fajfer2}  do not take the mixing of $O_9$ with $O_2$ into account.
This results in a prediction for the short-distance components that is mainly given by these
logarithms.}

\subsubsection{  THE $c\to u\ell^+\ell^-$ DECAY RATES}
To compute the differential decay rate in terms of the Wilson coefficients, 
we use the two-loop QCD corrected value of $C_7^{\rm eff}(m_c)$ as obtained 
in Reference~\cite{greub}; we compute $C_9^{\rm eff}(m_c)$ from Equation~\ref{c9eff}  and
$C_{10}(m_c)=C_{10}(M_W)$ from Equation~\ref{c10mw}. The 
differential decay rate in the approximation of 
massless leptons is given by
\begin{eqnarray}
\frac{d\Gamma_{c\to u\ell^+\ell^-}}{d\hat s}&=& 
\tau_D~\frac{G_F^2\alpha^2m_c^6}{768\pi^5} ~(1-\hat s)^2 
\left[ 
\left(\left|C_9^{('){\rm~eff}}(m_c)\right|^2+\left|C_{10}\right|^2\right)
\left(1+2\hat s\right)\right.\nonumber\\
& &\left.+ 12 ~C_7^{\rm eff}(m_c){\rm ~Re}\left[C_9^{('){\rm~eff}}(m_c)\right]
+ 4 \left(1+\frac{2}{\hat s}\right)\left|C_7^{\rm eff}(m_c)\right|^2
\right],
\label{dbs}
\end{eqnarray}
where $\tau_D$ refers to the lifetime of either $D^{\pm}$ or $D^0$. 
We estimate the inclusive branching ratios for $m_c=1.5$~GeV, $m_s=0.15$~GeV, 
$m_b=4.8$~GeV and  $m_d=0$,
\beq
{\cal B}r_{D^+\to X_u^+ e^+e^-}^{\rm (sd)} \simeq 2\times10^{-8} \, ,
\qquad 
{\cal B}r_{D^0\to X_u^0 e^+e^-}^{\rm (sd)} \simeq 8\times10^{-9} \, .
\label{d0br}
\eeq
The dominant contributions to the 
rates in Equation~\ref{d0br} come from the leading-order mixing
of $O_9$ with the four-quark operators $O_2^{(q)}$, the second term in 
Equation~\ref{c9eff}.
When considering 
the contributions of various new-physics scenarios, one should remember that their magnitudes must be compared to the mixing of these operators. 
Shifts in the matching conditions for the 
Wilson coefficients $C_7$, $C_9$, and $C_{10}$, even when large, may
not be enough to give an observable deviation.

\subsubsection{  THE $c\to u\gamma$ RATE}
The short-distance $c\to u\gamma$ contribution to radiative charm decays
was first studied in detail by Burdman et~al.~\cite{bghp}, who found that the effects of the leading
logarithms on $C_7(mc)^{\rm eff.}$ enhanced the branching ratio
by several orders of magnitude. Even with such enhancement, the rates 
were too small. However, Greub et~al. noted~\cite{greub} that the 
leading logarithmic approximation was still affected by a fair amount of 
GIM suppression because the quark mass dependence on the 
resummed expressions was still mild. Going to two loops in the matrix 
elements of 
the operators in Equation~\ref{heff}, specifically in $O^{(q)}_2$, leads to a 
more substantial
mass dependence that in turn breaks GIM more efficiently. These two-loop 
contributions
dominate the short-distance radiative amplitude, giving, for 
instance~\cite{greub},
\begin{equation}
B^{(\rm sd)}(D^0\to X\gamma)\simeq 2.5\times10^{-8}.
\label{cug}
\end{equation}
Although this represents a very large enhancement even with respect  to the 
leading logarithmic approximation (about five orders of magnitude!), 
it is still 
small, especially when compared with the estimated size of long-distance 
contributions
(see below).  

\subsubsection{EXCLUSIVE SEMILEPTONIC DECAY MODES}
\label{exclusive}

The exclusive modes corresponding to $c\to u $ transitions are known to be 
dominated by long-distance dynamics. This is true for both the radiative and
the semileptonic decays. 
For the $D\to X\gamma$ exclusive modes, long-distance physics dominates
all observables. However, in $D\to X\ell^+\ell^-$, it is in principle possible
to escape the largest long-distance contributions by looking at regions of phase space
away from resonances. We now discuss in some detail the computation 
of $D\to\pi\ell^+\ell^-$ and $D\to\rho\ell^+\ell^-$ as presented in Reference~\cite{bghp2}. 
For completeness, the expectations in the exclusive radiative
and neutrino modes are surveyed at the end of this section.

As a crude first estimate of the contributions of long-distance physics, we 
can consider the resonance process $D\to X V\to X\ell^+\ell^-$, where 
$V=\phi,\rho,\omega$. 
We isolate contributions from this particular 
mechanism by integrating $d\Gamma/dq^2$ over each resonance peak 
associated with an exchanged vector or pseudoscalar meson.
The branching ratios thus 
obtained are in the 
${\cal O}(10^{-6})$ range~\cite{Singer:1996}.  

This result suggests that the long-distance contributions
overwhelm the short-distance physics and any new physics
that might be present. 
However, this is not always the case. 
A more thorough treatment requires looking at all the kinematically
available regions in $D\to X_u\ell^+\ell^-$, not just the resonance region.
The effect of these states can be 
thought of as a shift
in the short-distance coefficient $C_9^{\rm eff}$ in Equation~\ref{c9eff}, 
since $V\to\ell^+\ell^-$ selects a vector coupling to the leptons. 
This follows from Reference~\cite{lms}, which incorporates 
in a similar manner
the resonant contributions to $b\to q\ell^+\ell^-$ decays via a 
dispersion relation for $\ell^+\ell^-\to $~hadrons. This procedure
is manifestly gauge-invariant. 
The new contribution can be written via the replacement~\cite{lms} 
\begin{equation}
C_9^{\rm eff}\to C_9^{\rm eff} + \frac{3\pi}{\alpha^2}
\sum_{i} 
\kappa_i\frac{m_{V_i}\Gamma_{V_i\to\ell^+\ell^-}}{m_{V_i}^2-s-im_{V_i}
\Gamma_{V_i}},
\label{c9res}
\end{equation}
where the sum is over the various relevant resonances, $m_{V_i}$ and 
$\Gamma_{V_i}$ are the resonance mass and width, and the factor
$\kappa_i \sim {\cal O}(1)$ is a free parameter adjusted to fit 
the nonleptonic decays
$D\to X V_i$ when the $V_i$ are on shell. We obtain $\kappa_\phi\simeq 3.6$,
$\kappa_\rho\simeq 0.7$,  and $\kappa_\omega\simeq 3.1$. The latter result
comes from assuming $B(D^+ \rightarrow \pi^+\omega)=10^{-3}$, since a direct 
measurement is not available yet. 

%
%
%% COMP: \paragraph indicates a fourth-level heading
%
%
%\paragraph{\underline{$D^+\to\pi^+e^+e^-$}:} 
\paragraph{$D^+\to\pi^+e^+e^-$:}
The main long-distance contributions come from 
the $\phi$, $\rho$, and $\omega$ resonances. The $\eta$ and $\eta'$
effects are  negligible.
The dilepton mass distribution for this decay
takes the form~\cite{bghp2}
\begin{eqnarray}
 \frac{d\Gamma}{ds} = \frac{G_F^2\alpha^2}{192\pi^5}
|\vec{p}_\pi|^3\,|f_+(s)|^2\,
\left( \left|\frac{2m_c}{m_D} C_7^{\rm eff} + C_9^{\rm eff}\right|^2
+|C_{10}|^2\right),
\label{meedist}
\end{eqnarray} 
where $s=m_{ee}^2$ is the  square of the dilepton mass. 
Here we have used the heavy-quark spin-symmetry relations
that relate the matrix elements of $O_7$ to the ``semileptonic''
matrix elements of $O_9$ and $O_{10}$~\cite{iw90}. 
An additional form factor 
is formally still present, but its contribution to the 
decay rate is suppressed by $(m_\ell/m_D)^2$ and is neglected here.
Precise measurements of $D\to\pi\ell\nu$ will give us $f_+(q^2)$. 
In the meantime, we make use of the prediction
of  chiral perturbation theory for heavy hadrons 
(ChPTHH)~\cite{chpthh}, which 
at low recoil gives
\begin{equation}
f_+(s) = \frac{f_D}{f_\pi}\frac{g_{D^*D\pi}}{(1-s/M_{D^{*}}^2)}.
\label{fp}
\end{equation}
Here we use the recent CLEO measurement~\cite{cleo_g} 
$g_{D^*D\pi}=0.59\pm0.1\pm0.07$, and we take $f_D=200$~MeV. 
In Figure~\ref{pill}, we present this distribution as a function of the 
dilepton mass. 
The two narrow peaks are the $\phi$ and the $\omega$, which sits on top
of the broader $\rho$.
The total rate results in $B(D^+\to\pi^+e^+e^-)
\simeq 2\times10^{-6}$. Although  most of this branching ratio arises
from the intermediate $\pi^+\phi$ state, we can see from Figure~\ref{pill}
that new-physics effects as low as $10^{-7}$ can be observed as long as 
such sensitivity is achieved in the regions away from the $\omega$ and $\phi $
resonances, both at low and high dilepton mass squared.

Fajfer et~al.\ use a different approach to compute the long-distance effects in $D\to\pi\ell^+\ell^-$
\cite{fajfer2}. They estimate individual contributions to 
the amplitude by using a combination of 
vector meson dominance (VMD), factorization, ChPTHH, and hidden local symmetry.
In addition to the large inherent uncertainties associated with each of these methods, there is no clear prescription of how to 
fix the relative signs of the various contributions, nor---in contrast to the 
previous approach---the sign relative to the short-distance contribution.
Within these large uncertainties, the results of Reference~\cite{fajfer2} and 
Reference~\cite{bghp2} do not differ by much.

%
%
%% COMP: \paragraph means fourth-level heading
%
%
%\paragraph{\underline{$D^+\to\rho^+e^+e^-$}:} 
\paragraph{$D^+\to\rho^+e^+e^-$:} 
Once again, we follow closely the calculation of Burdman et~al.\ \cite{bghp2}.
Because fewer data are currently available on the 
$D\to V V'$ modes, we take the values of the $\kappa_i$ in 
Equation~\ref{c9res} from the fits to the $D^+\to\pi^+ V$ case studied above.
Again, once precise measurements of the $D\to\rho\ell\nu$ form factors
are available, heavy quark spin symmetry relations can be used to 
turn these into $D\to\rho\ell^+\ell^-$ form factors.
Lacking these at the moment, we use the extracted values from 
the $D\to K^*\ell\nu$ data~\cite{drhoff} and assume SU(3)
symmetry~\cite{testing}.
The total integrated branching ratio is 
$B(D^0\to\rho^0e^+e^-)=1.8\times 10^{-6}$ (i.e.,
$B(D^+\to\rho^+e^+e^-)=4.5
\times 10^{-6}$). As  Figure~\ref{rholl} shows, once 
again most of this rate 
comes from the resonance contributions.
However, there is also a region---in this 
case confined to low values of $m_{ee}$ owing to the kinematics---where sensitive measurements 
could test the standard-model short-distance structure of these transitions.

Fajfer et~al.\ \cite{fajfer1}  studied these modes, following the approach of 
Reference~\cite{fajfer2}. Their results~\cite{fajfer1}
show a large enhancement at low $m_{ee}$ when compared with Figure~\ref{rholl}. 
If this long-distance enhancement is present, it could dominate the low $m_{ee}$ 
region, rendering the $D\to V\ell^+\ell^-$ modes useless to test short-distance physics
at any level. Therefore, we examine this discrepancy carefully. 
A $1/m_{ee}^2$ enhancement can only appear as a result of 
nonfactorizable contributions. 
This is clear from References~\cite{gp95} and~\cite{soares}:  
the factorization amplitude for $D\to\rho V$, 
when combined with a gauge-invariant $(\gamma-V)$ mixing, leads to a null
contribution to $D\to V\ell^+\ell^-$. This is reflected also in the 
fact that the mixing of the operator $O_2$ with $O_7$ 
is nonfactorizable~\cite{soares}. 
A resonant contribution to $O_7$, leading to a $1/q^2$ 
behavior, is then proportional to $C_7^{\rm eff}$, which is 
mostly given by the 
$O_2$ mixing. In addition, when compared with the 
usual short-distance matrix element of $O_7$, this resonant contribution
will be further suppressed by the factor 
$g_V(q^2) A^{\rm nf}(q^2)~$,
where $g_V(q^2)$ is the $(\gamma-V)$ mixing form factor, and 
$A^{\rm nf}(q^2)$ parameterizes the nonfactorizable 
amplitude $\langle \rho V|O_7|D\rangle$, which is 
of ${\cal O}(\Lambda_{\rm QCD}/m_c)$~\cite{qcdfac}. 
Thus, even if we take the 
on-shell values for these quantities, the resonant contribution
to $O_7$ is likely to be below $10\%$ of the standard-model short-distance contribution.
The actual off-shell values at low $q^2$ far from the resonances are likely
to be even smaller. 
We then conclude that the $1/q^2$ enhancement is mostly 
given by the short-distance contribution. This is only noticeable
at extremely small values of the dilepton mass, so that it is
likely to be beyond the experimental sensitivity in the electron 
modes (due to Dalitz conversion), whereas in the muon modes 
it lies beyond the physical region.
On the other hand, the factorizable pieces contribute to the 
matrix elements of $O_9$, just as in Equation~\ref{c9res}, 
and give no enhancement at low values of $q^2$. 
We then conclude that there should be no very large long-distance contributions
at low $m_{ee}$. 

The $\rho$ modes  also contain angular information 
in the form of a forward-backward asymmetry for the lepton pair. 
Because this asymmetry results from the interference between 
the vector and the axial-vector couplings of the leptons, 
it is negligible in the standard model, since vector couplings due to vector mesons
overwhelm axial-vector couplings. This is true even away from the 
resonance region, since the coefficients $C^{(\prime)}_9{\rm eff}$ and 
$C_7^{\rm eff}$ get large enhancements due to mixing with 
$O_2$ and QCD corrections, whereas $C_{10}$---the axial-vector coupling---is not affected by any of these, which results in a very small interference.

For both the $\pi$ and $\rho$ modes, 
the sensitivity to new-physics effects is reserved 
for large ${\cal O}(1)$ enhancements
because the long-distance contributions are still important even 
away from the resonances. Table~\ref{rpv_table} summarizes theoretical predictions from Reference~\cite{bghp2}. In addition, some modes are
almost exclusively long-distance physics. Examples are 
$D^0\to \bar{K}^{0(*)}\ell^+\ell^-$ and the radiative $D^0\to \bar{K}^{0(*)}\gamma$,
dominated by $W$ exchange diagrams, as well as 
$D^\pm\to \bar{K}^{\pm(*)}\ell^+\ell^-$ and the radiative $D^\pm\to \bar{K}^{\pm(*)}
\gamma$, which contain both $W$ annihilation and exchange. The measurements of these modes,
although not directly constraining new physics, will help us understand
long-distance physics. This may prove crucial to test the short-distance physics
in the $\pi$ and $\rho$ modes.

%
%
%% COMP: \paragraph means fourth-level heading
%
%
%\paragraph{\underline{Exclusive Radiative Decays:}}
\paragraph{Exclusive Radiative Decays:}
Exclusive decays mediated by the $c \to u \gamma$ transition 
are expected to be plagued by large
hadronic uncertainties. As mentioned, the large mixing of the $O_7$ operator
with the four-quark operators, especially $O_2$, and the propagation of 
light quarks 
in the loops indicate the presence of potentially large 
nonperturbative effects.  These are not calculable from first principles nor
in a controlled
approximation (other than lattice gauge theory). Moreover, even if lattice 
computations
of these effects become available, they typically overwhelm the standard-model short-distance contributions. Thus, modes such as $D\to\rho\gamma$ are not expected to 
be a probe of the short-distance structure of the standard model to the extent
$B\to K^*\gamma$ can be if the transition form factor is known precisely. 

On the other hand, one can try to estimate the size of the long-distance 
contributions
and therefore the branching fractions of these modes. This is interesting in its own 
right;
experimental observation of these modes will give us guidance in our otherwise 
limited understanding of these nonperturbative effects. 

Several attempts at estimating the long-distance contributions 
have been made \cite{bghp,ksw,fajld1,fajld2,others}. 
An example is the decay $D^0\to\rho^0\gamma$.
We can identify two types of long-distance contributions: pole-mediated
and   vector-meson dominance (VMD) transitions. 
Pole contributions can be thought of as driven by ``annihilation'' diagrams.
The effective weak Hamiltonian underlying these transitions is 
\begin{equation}
H_w = -\frac{4G_F}{\sqrt{2}}\,\left[a_1(\bar{u}_L\gamma_\mu d'_L)(\bar{s'}_L
\gamma^\mu c_L)
     + a_2\,(\bar{s'}_L\gamma_\mu d'_L)(\bar{u}_L\gamma^\mu c_L)\right],
\label{hwa}
\end{equation} 
with $d'\equiv V_{ud}d+V_{us} s$ and $s'\equiv V_{cs}s+V_{cd} d$. 
There are two pole diagrams giving~\cite{bghp}
\begin{eqnarray}
A(D^0\to\rho^0)_{\rm pole} &=& \frac{g_{\rho\pi\gamma}}{m_D^2 - m_\pi^2}
\langle\pi^0|H_w|D^0\rangle \nonumber\\
 && + \frac{g_{D^{*0} D^0\gamma}}{m_D^2 - m_{D^*}^2}
\langle\rho^0|H_w|D^{*0}\rangle.
\label{polams}
\end{eqnarray}
Here, $H_w$ refers to the four-fermion weak  Hamiltonian governing nonleptonic
weak decays. For instance, 
\begin{equation}
\langle\pi^0|H_w|D^0\rangle = -\frac{G_F}{\sqrt{2}}a_2(m_c)V_{ud}^*V_{cd}
\frac{f_{\pi}}{\sqrt{2}}f_D m_D^2,
\label{d2rho} 
\end{equation}
where $a2(m_c)$ is determined in $D$ nonleptonic decays. 
These two pole diagrams may cancel rather efficiently. 
Since  the details are extremely 
model-dependent, we could ask what  branching ratio would be implied by 
one of them alone. This typically leads to~\cite{bghp}
$B(D^0\to\rho^0\gamma)_{\rm pole}\leq{\rm few}\times10^{-7}$.  
However, QCD sum rules may be used~\cite{ksw} to compute
the annihilation contributions, and these are found to give 
$B(D^0\to\rho^0\gamma) \simeq{\rm few}\times10^{-6}$.

On the other hand, VMD contributions come from considering the 
nonleptonic intermediate states. In this case, this corresponds to 
$D^0\to\rho^0 V\to\rho^0\gamma$, where the neutral vector boson $V$ turns
into an on-shell photon. Various methods have been used to compute the nonleptonic 
and $V\to\rho$ amplitudes~\cite{bghp,fajld1,fajld2}.  
A common assumption to estimate the VMD amplitude has been that of 
factorization~\cite{factor}. However, the contribution of 
the factorized nonleptonic amplitude vanishes when the photon is on-shell
%\cite{vmdzero}.
\cite{gp95,soares}.
This is  a consequence of gauge invariance and is related to the 
fact that the mixing of four-quark operators with the photon penguin operator $O_7$
vanishes unless nonfactorizable gluons are exchanged.
Thus, nonfactorizable contributions to the nonleptonic amplitude 
constitute the leading effect in the VMD amplitude. It is therefore possible that
the VMD contributions to weak radiative decays of charm mesons are overestimated.
At the same time, it is possible that the charm quark is not heavy enough 
for the nonfactorizable effects to be suppressed. The suppression is formally 
$O(\Lambda/m_c)$, with $\Lambda$ a typical scale of strong interactions. 
We conclude that uncertainties in these modes are still very large. 
Table~\ref{drad} summarizes predictions for these decays in different model 
computations. 
The Belle collaboration recently announced a measurement of 
$D^0\to\phi\gamma$, with a branching ratio of \cite{Yabsley} 
$B(D^0 \rightarrow \phi \gamma) = (2.60^{+0.70 + 0.15}_{-0.61 -0.17})
\times 10^{-5}$. As one can see from Table~11, this is consistent
with the upper end of the predictions in 
%Burdman et al. 1995, 
Reference \cite{bghp}, which were 
obtained by making use of VMD plus the data on the relevant nonleptonic 
decay in addition to the pole contributions. If this trend is confirmed in this
as well as other modes, this might point in the direction of large 
nonfactorizable contributions. 
Experimental bounds are closing in on some of these predictions and will 
undoubtedly shed light on the size of these long-distance effects.

\subsubsection{OTHER DECAY MODES}
Our discussion has neglected many decay modes to focus on those that can potentially test the short-distance structure of the 
standard model and are reachable in present or planned experiments. 
Here is a sample of other modes  studied in the literature. 

%
%
%% COMP: \paragraph means fourth-level heading
%
%
%\paragraph{\underline{$D^0\to\gamma\gamma$}:}
\paragraph{$D^0\to\gamma\gamma$:}
The standard-model short-distance contributions can be obtained from the two-loop 
$c\to u\gamma$ amplitude. This results in~\cite{bghp2}
$B^{\rm SD}(D^0\to\gamma\gamma)\simeq 3\times10^{-11}$. 
There are several types of long-range effects. Fajfer et~al. ~\cite{fajfergg} 
estimate these effects using ChPTHH to one loop. This gives
$B^{\rm LD}(D^0\to\gamma\gamma)\simeq (1\pm0.5)\times10^{-8}$. 
Burdman et~al.\ ~\cite{bghp2} consider various long-distance effects and obtain similar results. In this case, the main contributions are 
found to come from VMD and the $K^+ K^-$ unitarity contribution. 

%
%
%% COMP: \paragraph means fourth-level heading
%
%
%\paragraph{\underline{$D^0\to\ell^+\ell^-$}:}
\paragraph{$D^0\to\ell^+\ell^-$:}
The short-distance contributions to this mode are also extremely suppressed, 
not only by helicity
but also by the quark masses in the loop. Unlike in $c\to u\gamma$, the mixing with $O_2$ 
does not help. In Reference~\cite{bghp2}, the short distance is estimated at
$B^{\rm SD}(D^0\ell^+\ell^-)\simlt 10^{-18}$. The most important source of long-distance effects is the two-photon unitary contribution, which 
gives
\beq
B^{\rm LD}(D^0\to\ell^+\ell^-)\simeq 3\times10^{-5}~B(D^0\to\gamma\gamma).
\label{ldll}
\eeq
%
%
%% COMP: \paragraph means fourth-level heading
%
%
%\paragraph{\underline{$D\to X\nu\bar\nu$}:}
\paragraph{$D\to X\nu\bar\nu$:}
Short- and long-distance contributions to $c\to u\nu\bar\nu$ processes in the standard model
are extremely small, typically resulting in~\cite{bghp2} 
$B(c\to u\nu\bar\nu)\simlt 10^{-15}$.

\subsection{Rare Charm Decays and New Physics}
\label{sec_new}
As mentioned, charm-changing neutral-current processes such as $D^0$--$\bar{D}^0$ mixing and 
rare charm decays complement the constraints on extensions of the standard model obtained from processes initiated by down quarks, such as kaon and $B$-meson transitions.
Although we have seen that bounds on $\Delta m_D$ are quite constraining in a variety of models, new physics may still show itself
in rare charm decays. 
We mainly review the potential for signals from  supersymmetric
theories (with and without
$R$-parity conservation) and from new strong dynamics at the TeV scale. We briefly comment 
on the sensitivity to other new physics, such as theories with extra dimensions and 
extended gauge and matter sectors.  

\subsubsection{THE MINIMAL SUPERSYMMETRIC STANDARD MODEL}
\label{sec_mssm}
The MSSM adds to the standard-model description of loop-mediated processes contributions due to
gluino-squark exchange, chargino/neutralino-squark
exchange, and charged Higgs-quark exchange. This last contribution
carries the same CKM structure as in the standard-model
loop diagram and is proportional to the internal and external quark
masses; thus, its effects in rare charm transitions are small and
we neglect it here.  The gluino-squark contribution proceeds via flavor-diagonal vertices proportional to the strong coupling constant and in
principle dominates the CKM-suppressed, weak-scale strength
chargino/neutralino-squark contributions.  We therefore consider only the
case of gluino-squark exchange here as an estimate of the potential size
of SUSY effects in rare charm decays.  

A typical squark-gluino contribution is depicted in 
Figure~\ref{mssm}. 
The corresponding effects in the $c\to u$ transitions were studied for $D\to X_u\gamma$~\cite{wyler}  and for 
$D\to X_u\ell^+\ell^{-}$~\cite{bghp2}.  
Within the context of the mass insertion approximation \cite{susyfcnc}, the effects
are included in the Wilson 
coefficients corresponding to the decay $D\to X\ell^+\ell^-$ via
\begin{equation}
C_i = C_i^{\rm SM} + C_i^{\tilde g}\,,
\end{equation}
for $i=7,9,10$.
Allowing for only one insertion, the explicit contributions 
from the gluino-squark diagrams are~\cite{lunghi,wyler,bghp2}
\begin{equation}
C_7^{\tilde g} = -\frac{8}{9}\,\frac{\sqrt 2}{G_FM_{\tilde q}^2}
\pi\alpha_s\,\left\{ (\delta^u_{12})_{LL}\frac{P_{132}(u)}{4} 
 + (\delta^u_{12})_{LR}P_{122}(u)\frac{M_{\tilde g}}{m_c}
\right\},
\label{c7g}
\end{equation}
and 
\begin{equation}
C_9^{\tilde g} = -\frac{8}{27}\,\frac{\sqrt 2}{G_FM_{\tilde q}^2}
\pi\alpha_s\,(\delta^u_{12})_{LL}P_{042}(u),
\label{c9g}
\end{equation}
with the contribution to $C_{10}$ vanishing at this order because of the
helicity structure.

If we allow for two mass insertions, there is a contribution to 
$C_{9,10}$ given by
\begin{equation}
C_{10}^{\tilde g} = -\frac{1}{9}\,\frac{\alpha_s}{\alpha}\,
(\delta^u_{22})_{LR}  (\delta^u_{12})_{LR} P_{032}(u) = -{C_9\over
1-4\sin^2\theta_W}.
\label{c10g}
\end{equation}
Here, 
$u=M_{\tilde g}^2/M_{\tilde q}^2$
and the functions
$P_{ijk}(u)$ are defined as 
\begin{equation}
P_{ijk}(u) \equiv \int_{0}^{1} dx \ \frac{x^i(1-x)^j}{(1-x+ux)^k}.
\label{pdef} 
\end{equation} 
In addition, the operator basis can be extended by the ``wrong chirality'' 
operators $\hat O_7$, $\hat O_9$, and $\hat O_{10}$, obtained  by switching the 
quark chiralities in Equations~\ref{gdipole} and \ref{laxial}. The 
gluino-squark contributions to the corresponding Wilson coefficients
are
\begin{eqnarray}
\hat C_7^{\tilde g} &=& -\frac{8}{9}\,\frac{\sqrt 2}{G_FM_{\tilde q}^2}
\pi\alpha_s\,\left\{ (\delta^u_{12})_{RR}\frac{P_{132}(u)}{4} 
 + (\delta^u_{12})_{LR}P_{122}(u)\frac{M_{\tilde g}}{m_c}
\right\},
\label{c7pg}\\
\hat C_9^{\tilde g} &=& -\frac{8}{27}\,\frac{\sqrt 2}{G_F M_{\tilde q}^2}
\pi\alpha_s\,(\delta^u_{12})_{RR}P_{042}(u)-(1-4\sin^2\theta_W)\hat 
C^{\tilde g}_{10},\nonumber\\
\hat C_{10}^{\tilde g} &=& -\frac{1}{9}\,\frac{\alpha_s}{\alpha}\,
(\delta^u_{22})_{LR}  (\delta^u_{12})_{LR} P_{032}(u), \nonumber
\end{eqnarray}
where the expression for $\hat C_{10}^{\tilde g}$ is again obtained with 
a double insertion. 

As noted in References~\cite{lunghi} and \cite{wyler}, in both $C_7^{\tilde g}$ 
and $\hat C_7^{\tilde g}$ the term in which the squark chirality labels 
are mixed introduces the enhancement factor 
$M_{\tilde g}/m_c$. In the standard model, the chirality flip that appears in 
$O_7$ occurs by a flip of one external quark line, resulting in
a factor of $m_c$
included in the operator's definition.\footnote{The 
$m_u$ term, proportional to the ($1-\gamma_5$) in the operator, is neglected.}
However, in the gluino-squark diagram, the insertion of 
$(\delta^u_{12})_{RL}$ forces the chirality flip to take place 
in the gluino line, thus introducing a $M_{\tilde g}$ factor instead of
$m_c$.  This yields a significant enhancement in the short-distance contributions to the process
$D\to X_u\gamma$~\cite{wyler}, which is unfortunately obscured by the 
large long-range effects. This is not the case in 
$c\to u\ell^+\ell^-$ processes. 

In order to estimate the effects in $c\to u\ell^+\ell^-$ transitions from
the gluino contributions, we make use of the bounds of Table~\ref{numssm}.
Figures~\ref{pill_mssm} and \ref{rholl_mssm} show the 
dilepton mass distribution as a function of the dilepton mass for 
$D^+\to\pi^+ e^+ e^-$ and $D^0\to\rho^0 e^+ e^-$, respectively (from Reference~\cite{bghp2}). 
Four sample cases are considered there \cite{bghp2}: 
 (I) $M_{\tilde g}=M_{\tilde q}=250$~GeV, 
 (II) $M_{\tilde g}=2\,M_{\tilde q}=500$~GeV, 
 (III) $M_{\tilde g}= M_{\tilde q}=1000$~GeV, and 
 (IV) $M_{\tilde g}=(1/2)\,M_{\tilde q}=250$~GeV.

We first examine the $D^+\to\pi^+ e^+ e^-$ case.
Although the net effect is relatively small in the integrated rate (an increase
$\simeq 20\%$
or smaller), the enhancement due to the SUSY contributions is most conspicuous
away from the vector resonances, particularly for low dilepton masses. 
Experiments sensitive to the dilepton mass distribution 
at the branching fraction of $10^{-7}-10^{-8}$ can detect these SUSY contributions.  
However, the decays to a  vector meson, such  as 
$D\to\rho e^+ e^-$, are more sensitive to the gluino exchange, 
as  Figure~\ref{rholl_mssm} shows.

The effect is quite pronounced and  lies almost entirely
in the low $m_{ee}$ region. This is mostly
due to the contributions of 
$(\delta^u_{12})_{RL}$ to $C_7$ and $C_7'$, 
which contain the $M_{\tilde g}/m_c$ enhancement as advertised
above. This effect is intensified at low $q^2=m_{ee}^2$ owing to the 
photon propagator (see, e.g., Equation~\ref{dbs} for the inclusive decays). 
This low-$q^2$ enhancement of the $O_7$ contribution is present in exclusive
modes with vector mesons, such as $D\to\rho\ell^+\ell^-$, but not in modes
with pseudoscalars, such as $D\to\pi\ell^+\ell^-$, since gauge invariance
forces a cancellation of the $1/q^2$ factor in the latter case (see, e.g., 
Equation~\ref{meedist}).
This is   apparent from a comparison of the low dilepton mass regions
in Figures~\ref{pill_mssm} and~\ref{rholl_mssm}. 
Thus we see that rare charm decays are indeed sensitive to a generic
extension of the standard model such as the MSSM. This is particularly true of
the $D\to\rho\ell^+\ell^-$ modes. 

\subsubsection{SUPERSYMMETRY WITH $R$-PARITY VIOLATION}
\label{sec_rpv}
Imposing $R$-parity conservation in the MSSM  prohibits 
baryon- and lepton-number--violating terms in the 
superpotential.  However, other symmetries can be invoked to prohibit rapid 
proton decay, such as baryon parity or lepton parity (see, e.g., \cite{ross}), and hence
allow for $R$-parity
violation.
The $R$-parity--violating superpotential 
can be written as\footnote{We ignore bilinear terms that are 
not relevant to our discussion of FCNC effects.} 
\begin{equation}
{\cal W}_{R_p} =\epsilon_{ab}\left\{ 
{\textstyle \frac{1}{2}}\lambda_{ijk}L^a_iL^b_j\bar{E}_k
+\lambda'_{ijk}L_i^aQ^b_j\bar{D}_k 
+{\textstyle \frac{1}{2}}\epsilon_{\alpha\beta\gamma}\lambda^{''}_{ijk}\bar{U}^\alpha_i
\bar{D}^\beta_j\bar{D}^\gamma_k \right\},
\label{rpv}
\end{equation}
where $L$, $Q$, $\bar E$, $\bar U$, and $\bar D$ are the chiral superfields
in the MSSM. The  SU(3) color indices are denoted by 
$\alpha,\beta,\gamma=1,2,3$, the 
SU(2)$_L$ indices by $a,b=1,2$, and the generation indices by $i,j,k=1,2,3$. 
The fields in Equation~\ref{rpv} are in the weak basis.  

The $\lambda'_{ijk}$
term is the relevant one for the rare charm decays we consider here
because it can give rise to tree-level squark-exchange contributions to decay channels such as $D\to X\ell^+\ell^-$, 
$D\to\ell^+\ell^-$, 
as well as the lepton-flavor--violating $D\to X\mu^+e^-$ and 
$D\to\mu^+ e^-$ modes.  Before considering the FCNC effects in $D$ 
decays, we need to rotate the fields to the mass basis.
This leads to 
\begin{equation}
{\cal W}_{R_p}= \tilde{\lambda'}_{ijk}\left[N_i V_{jl} D_l
-E_iU_j\right] \bar{D}_k+\cdots \, ,
\label{mbasis}
\end{equation}
where $V$ is the CKM matrix and we define
\begin{equation}
\tilde{\lambda'}_{ijk}\equiv \lambda'_{irs} {\cal U}^L_{rj} 
{\cal D}^{*R}_{sk}.
\label{newlambda}
\end{equation}
Here, ${\cal U}^L$ and ${\cal D}^R$ are the matrices used to rotate the 
left-handed up- and right-handed down-quark fields to the mass basis. 
Written in terms of component fields, this interaction now reads
\begin{eqnarray}
{\cal W}_{\lambda'}&=&\tilde{\lambda}'_{ijk} \left\{V_{jl}[ 
\tilde{\nu}_L^i\bar{d}_R^kd_L^l + \tilde{d}_L^l\bar{d}_R^k\nu_L^i 
+ (\tilde{d}_R^k)^*(\bar{\nu_L^i})^cd_L^l]\right.\nonumber\\
& &\left.
-\tilde{e}_L^i\bar{d}_R^ku_L^j - \tilde{u}_L^j\bar{d}_R^ke_L^i 
-(\tilde{d}_R^k)^*(\bar{e_L^i})^cu_L^j \right\}.
\label{incomps}
\end{eqnarray}
The last term in Equation~\ref{incomps} can give rise to the processes
$c\to u\ell\ell^{(')}$ at tree level via the exchange of a down squark.
This leads to effects that are proportional to 
$\tilde{\lambda}'_{i2k}\tilde{\lambda}'_{i1k}$ with $i=1,2$ (owing to
kinematical restrictions).

Constraints on these coefficients have been derived  (for a recent review, see \cite{rpv1}).
For instance, Agashe \& Graesser~\cite{agashe}
obtain tight bounds from $K^+\to\pi^+\nu\bar\nu$ by assuming that only one $R$-parity--violating 
coupling satisfies $\tilde{\lambda}'_{ijk}\neq 0$. We update this bound 
by using the latest experimental result~\cite{k2pinunubar} 
$B(K^+ \to\pi^+\nu\bar\nu)=(1.57^{+ 1.75}_{- 0.82})\times10^{-10}$, 
which yields
$\tilde{\lambda}'_{ijk} <0.005$.  However, this bound can be avoided
in the single-coupling scheme~\cite{agashe}, where only one $R$-parity--violating coupling is taken to be nonzero in the weak basis. 
In this case, it is possible that flavor rotations may restrict the
$R$-parity-breaking--induced  flavor violation to be present in either the
charge $-\frac{1}{3}$ or $+\frac{2}{3}$ quark sectors, but not both.
Then large effects are possible in the up sector 
for observables such as $D^0$--$\bar{D^0}$ mixing and rare decays without
affecting the down-quark sector.

Agashe \& Graesser~\cite{agashe}
obtain a rather loose constraint on the $R$-parity--breaking couplings from 
$D^0$ mixing, which could result in large effects in $c\to u\ell\ell^{(')}$ 
decays. Here, we take a conservative approach and use more 
model-independent bounds. The constraints on the $R$-parity--breaking couplings
for the processes of interest here
are collected in Table~\ref{rpvbounds} from Reference~\cite{rpv1}.
The charged-current universality bounds assume three
generations. The $\pi$-decay constraint is given by the quantity
$R_\pi=\Gamma(\pi\to e\nu)/\Gamma(\pi\to\mu\nu)$. The limits obtained from
$D\to K\ell\nu$ were first obtained in Reference~\cite{rpv2}.

We first consider the contributions to $c\to u\ell^+\ell^-$. 
The tree-level exchange of down squarks results in 
the effective interaction~\cite{bghp2}
\begin{equation}
\delta {\cal H}_{\rm eff} = -\frac{\tilde{\lambda}'_{i2k}\tilde{\lambda}'_{i1k}}
{m^2_{\tilde{d}^k_R}}\, (\bar{\ell}_L)^cc_L\,\,\bar{u}_L(\ell_L)^c,
\label{dheff1}
\end{equation}  
which, after Fierz rearrangement, gives
\begin{equation}
\delta {\cal H}_{\rm eff} =
-\frac{\tilde{\lambda}'_{i2k}\tilde{\lambda}'_{i1k}}
{2 m^2_{\tilde{d}^k_R}}\,(\bar u_L\gamma_\mu c_L)(\bar\ell_L\gamma^\mu\ell_L).
\label{dheff2}
\end{equation}
This corresponds to contributions at the  high-energy scale to the Wilson coefficients $C_9$ and 
$C_{10}$ given by 
\begin{equation}
\delta C_9=-\delta C_{10} = \frac{\sin^2\theta_W}{2\alpha^2}
\left(\frac{M_W}{m_{\tilde{d}^k_R}}\right)^2\,
\tilde{\lambda}'_{i2k}\tilde{\lambda}'_{i1k}.
\label{dcs}
\end{equation}

If we now specify $\ell=e$ and use the bounds from Table~\ref{rpvbounds}, 
we arrive at the constraint
\begin{equation}
\delta C_9^e=-\delta C_{10}^e \le 1.10\,\left(\frac{\tilde{\lambda}'_{12k}}
{0.04}\right)\,\left(\frac{\tilde{\lambda}'_{11k}}
{0.02}\right). 
\label{dce}
\end{equation}
Notice that these are independent of the squark mass, which cancels.
Taking this upper limit on
the Wilson coefficients results in the 
dotted-dashed lines of Figures~\ref{pill} and~\ref{rholl} corresponding to 
$D^+\to\pi^+e^+e^-$ and $D^0\to\rho^0e^+e^-$, respectively. 
The effect in these rates is small, of order $10\%$ at most, 
whereas the experimental bounds are a factor of $20$ above 
this level in the most restrictive case (given by the pion mode).
On the other hand,   
for $\ell=\mu$, we obtain 
\begin{equation}
\delta C_9^\mu=-\delta C_{10}^\mu \le 17.4\,\left(\frac{\tilde{\lambda}'_{22k}}
{0.21}\right)\,\left(\frac{\tilde{\lambda}'_{21k}}
{0.06}\right).
\label{dcmu}
\end{equation}
When the existing constraints on $\tilde{\lambda}'_{1jk}$ are taken into account, 
$R$-parity violation results in small deviations in $D\to (\pi,\rho) e^+ e^-$. However, the 
bounds on $\tilde{\lambda}'_{2jk}$ are loose and lead to very large effects in the
$\ell=\mu$ modes. In fact, the allowed values from other observables saturate 
the current experimental limits for $D\to\pi\mu^+\mu^-$ and $D\to\rho\mu^+\mu^-$,
resulting in the bound~\cite{bghp2}
\begin{equation}
\tilde{\lambda}'_{22k}\,\tilde{\lambda}'_{21k} < 0.004.
\label{rpvbound}
\end{equation}
These large effects would be clearly observable away from the resonances. 

In addition, the angular information in $D\to\rho\mu^+\mu^-$ 
can be used to  confirm the new-physics origin of the large deviations.
If we define the forward-backward asymmetry for leptons as 
\begin{equation}
A_{FB}(q^2)=\frac{
\int_{0}^{1}\frac{d^2\Gamma}{dxdq^2} dx -
\int_{-1}^{0}\frac{d^2\Gamma}{dxdq^2} dx
}
{\frac{d\Gamma}{dq^2}} \ \  ,
\label{afbdef}
\end{equation}
where $x\equiv\cos\theta$, with $\theta$ being the angle between the $\ell^+$ 
and the decaying $D$ meson in the $\ell^+\ell^-$ rest  frame 
(as mentioned in Section~\ref{exclusive}), the standard-model prediction for this 
quantity is nearly zero. This is because 
the rate comes almost entirely from the long-distance contributions to the 
lepton vector coupling (e.g., $O_9$), making the interference with 
the axial-vector couplings negligible. 
This can be seen 
by inspecting the numerator of the 
asymmetry~\cite{testing,rareas}
\begin{equation}
A_{FB}(q^2) \sim  4\;m_D\;k\;C_{10}
\left\{ C_9^{\rm eff}\;g\;f +\frac{m_c}{q^2}C_7^{\rm eff}
\;\left(f\;G - g\;F\right) \right\},
\label{numerator}
\end{equation}
where $k$ is the vector meson three-momentum in the $D$ rest frame, and 
$f$, $g$, $F$, and $G$ are form factors.
Because the standard-model amplitude is dominated by the long-distance vector intermediate 
states, we have $C_9^{\rm eff}\gg C_{10}$. New-physics contributions that make 
$C_{10}\simeq C_9^{\rm eff}$ will hence 
generate a sizeable asymmetry. This is actually the case in $R$-parity-violating SUSY.
Burdman et~al.\ \cite{bghp2} show that the
asymmetry in $D\to\rho\mu^+\mu^-$ is predicted to be quite large for the allowed values
of the $R$-parity--violating parameters. 

The effective Wilson coefficients of Equation~\ref{dcs}
also lead to a contribution to the two-body decay
$D^0\to\mu^+\mu^-$.
The $R$-parity--violating contribution to the branching
ratio then reads
\begin{equation}
B^{\not R_p}(D^0\to\mu^+\mu^-)= 
\tau_{D^0}\,f_D^2\,m_\mu^2\,m_D\,\sqrt{1-\frac{4m_\mu^2}{m_D^2}}
\;\frac{\left(\tilde{\lambda}'_{22k}\tilde{\lambda}'_{21k}\right)^2}
{64\pi\,m_{\tilde{d}_k}^4}.
\label{rpv_d2mu}
\end{equation}
Applying the bound in Equation~\ref{rpvbound} gives the constraint\footnote{In Reference~\cite{bghp2}, this expression (Equation~86) 
was incorrectly given. Also, the branching ratio stated there did not reflect the bound from Equation~122, but the less restrictive bounds to the individual couplings.} 
\begin{equation}
B^{\not R_p}(D^0\to\mu^+\mu^-) < 3.5\times 10^{-7}\;
\left(\frac{\tilde{\lambda}'_{22k} \tilde{\lambda}'_21k}{0.004}\right)^2.
\label{d2mu_bound} 
\end{equation}
Thus, $R$-parity violation could give an effect that could soon be probed in these modes. 

Next, we consider the products of $R$-parity--violating couplings 
that lead to lepton flavor violation.
For instance, the products $\tilde{\lambda}'_{11k}\tilde{\lambda}'_{22k}$ and 
$\tilde{\lambda}'_{21k}\tilde{\lambda}'_{12k}$ will give rise to
$D^+\to\pi^+\mu^+ e^-$.
This leads to 
\begin{equation}
\delta C_9^{\mu e}=-\delta C_{10}^{\mu e}=
4.6\times\left\{ \left(\frac{\tilde{\lambda}'_{11k}}{0.02}\right)
\left(\frac{\tilde{\lambda}'_{22k}}{0.21}\right)
 + 
\left(\frac{\tilde{\lambda}'_{21k}}{0.06}\right)
\left(\frac{\tilde{\lambda}'_{12k}}{0.04}\right)
\right\},
\label{d2pimue}
\end{equation}
which results in $B^{\not R_p}(D^+\to\pi^+\mu^+ e^-)
< 3\times10^{-5}$.
Here again, experiment is on the verge of being sensitive to $R$-parity--violating effects.  Similarly, for the corresponding 
two-body decay we have
\begin{equation}
 B^{\not R_p}(D^0\to\mu^+e^-) <0.5 \times10^{-6}\times
\left\{ \left(\frac{\tilde{\lambda}'_{11k}}{0.02}\right)
\left(\frac{\tilde{\lambda}'_{22k}}{0.21}\right)
 + 
\left(\frac{\tilde{\lambda}'_{21k}}{0.06}\right)
\left(\frac{\tilde{\lambda}'_{12k}}{0.04}\right)
\right\}.
\label{d2mue_2body}
\end{equation}

Last, we consider the $R$-parity--violation contributions to $D^\pm\to e^\pm\nu$ 
and $D_s^\pm\to e^\pm\nu$. These decay modes
receive large enhancements mostly from the $s$-channel exchange of sleptons \cite{akeroyd}.
Unlike the $t$-channel squark exchange discussed above, this results in 
amplitudes that are unsuppressed by helicity. 
The largest lepton-number violation occurs in $D^\pm\to e^\pm\nu_\tau$ through the 
product $|\lambda_{231}^*\tilde{\lambda'}_{221}|$, and 
in $D_s^\pm\to\ e^\pm\nu_\tau$ through $ |\lambda_{231}^*\tilde{\lambda'}_{222}|$.
Here $\lambda_{ijk}$ is the strength of the cubic lepton superfield interactions
in the $R$-parity--violating superpotential term $\epsilon_{ab}L^a_iL^b_j\bar{E}_k$.
When current bounds are taken into account, it is found that $R$-parity violation
could result in~\cite{akeroyd}  
$B^{\not R_p}(D^\pm\to e^\pm\nu)\simeq 10^{-4}$, several orders of 
magnitude above the standard-model prediction. Even more dramatically, we could have 
$B^{\not R_p}(D_s^\pm\to e^\pm\nu)\simeq 5\times10^{-3}$. 
Table~\ref{rpv_table} summarizes the predictions from Reference~\cite{bghp2} 
in both the standard model and $R$-parity--violating SUSY.

\subsubsection{STRONG DYNAMICS}
\label{strograre}

%
%
%% COMP: \paragraph means fourth-level heading
%
%

\paragraph{Technicolor Models:}
In standard technicolor theories, both fermions and 
technifermions transform under the new gauge interaction of
extended technicolor (ETC). 
As we saw in Section~\ref{ddb_strong}, this leads to the
presence of four-quark operators coming from the diagonal ETC generators
and characterized by a mass-scale $M$ bounded by \ddb~mixing to be
greater than 100~TeV or so.
However, additional operators are generated at low energies that are not
suppressed by $M$.
The condensation of technifermions leading to electroweak symmetry breaking
leads to fermion mass terms of the form 
\begin{equation}
m_q\simeq \frac{g^2_{\rm ETC}}{M_{\rm ETC}^2} \langle \bar T
T\rangle_{\rm TC}.
\end{equation}

Operators arising from the technifermion interactions 
have been shown~\cite{css} to give rise to FCNC 
involving the $Z$ boson,  
\begin{eqnarray}
\xi^2\frac{m_c}{8\pi v}\frac{e}{ \sin2\theta_W} {\cal U}^{cu}_{L} 
Z^\mu~(\bar u_L \gamma_\mu c_L) \quad {\rm and} \quad 
\xi^2\frac{m_t}{8\pi v}\frac{e}{\sin2\theta_W} 
{\cal U}^{tu}_{L} {\cal U}^{tc*}_{L}
Z^\mu~(\bar u_L \gamma_\mu c_L) , 
\label{etc2}
\end{eqnarray}
where ${\cal U}_L$ is the unitary matrix rotating left-handed up-type 
quark fields into their mass basis and 
$\xi$ is a model-dependent quantity of 
${\cal O}(1)$.
The induced flavor-conserving $Z$ coupling was first studied in 
Reference~\cite{css}, and flavor-changing effects in $B$ decays have been 
examined in References~\cite{rs} and \cite{gns}.  

The flavor-changing vertices in Equation~\ref{etc2} induce contributions 
to $c \to u\ell^+\ell^-$. These appear mostly as a shift in the 
Wilson coefficient $C_{10}(M_W)$, 
\begin{equation}
  \delta C_{10}\simeq {\cal U}_{cu}^L\,\frac{m_c}{2v}\, 
\frac{\sin^2\theta_W}{\alpha}\simeq 0.02,
\label{dc10}
\end{equation}
where we assume ${\cal U}^{cu}_L\simeq \lambda\simeq 0.22$ 
(i.e., one power of the Cabibbo
angle) and we take $m_c=1.4~$GeV. Although this represents 
a very large enhancement with respect to the 
standard-model value of $C_{10}(M_W)$, it does not translate into a large
deviation in the branching ratio.
As we have seen, these are dominated by the mixing of 
the operator $O_2$ with $O_9$, leading 
to a very large value of $C_9^{\rm eff}$. The contribution in 
Equation~\ref{dc10} represents only 
a few-percent effect in the branching ratio with respect to the standard model.
However, the effect is quite large in the region of low dilepton mass.

Furthermore, the interaction in Equation~\ref{etc2} can also 
mediate $D^0\to\mu^+\mu^-$. The corresponding
amplitude is 
\begin{equation}
  {\cal A}_{D^0\mu^+\mu^-}\simeq {\cal U}_{cu}^L\, \frac{m_c}{2\pi v}\, 
\frac{G_F}{\sqrt{2}}\,\sin^2\theta_W
  f_D\,m_\mu \,.
  \label{d2mm}
\end{equation}
This results in the branching 
ratio $B^{\rm ETC}(D^0\to\mu^+\mu^-)\simeq 0.6\times10^{-10}$,
which, although still small, 
is  not only several orders of magnitude larger than the standard-model short-distance
contribution but also more than two orders of magnitude larger than 
the long-distance estimates.

Finally, the FCNC vertices of the $Z$ boson in Equation~\ref{etc2}
also give large contributions to $c\to u\nu\bar\nu$. The enhancement is 
considerable and results in the branching ratio
\begin{equation}
B^{\rm ETC}_{D^+\to X_u\nu\bar\nu}
\simeq \xi^4\,\left(\frac{{\cal U}_L^{cu}}{0.2}\right)^2
\,2\times10^{-9}.
\label{etcnn}
\end{equation}

%
%
%% COMP: \paragraph means fourth-level heading
%
%
\paragraph{Topcolor:}
As discussed in Section~\ref{ddb_strong}, the topcolor interactions
must be nonuniversal, and so they mediate FCNCs 
at tree level. 
After the rotations of quark fields defined in Equation~\ref{rot2mass}, 
the exchange of top-gluons generates four-fermion couplings of the form
\beq
\frac{4\pi\alpha_s\cot\theta^2}{M^2_G} \, {\cal U}^{tc*}{\cal U}^{tu}\,
(\bar u\gamma_\mu T^a t)
(\bar t\gamma^\mu T^a c)
\label{topc}
\eeq
where ${\cal U}^{ij}={\cal U}^{ij}_L+{\cal U}^{ij}_R$ 
and $M$ is the 
mass of the exchanged color-octet gauge boson.  
The one-loop insertion of this operator results in contributions to the 
operators ${\cal O}_9$ and ${\cal O}_{10}$ in $c\to u\ell^+\ell^-$
as well as in the purely leptonic decays. These could lead to large
effects when compared to the standard model
\beq
\delta C_{10}\simeq 2\,\delta C_9\simeq 
0.01\times\left(\frac{{\cal U}^{tc*}{\cal U}^{tu}}{\sin^5\theta_c}\right)
\,\left(\frac{1{\rm ~TeV}}{M_G}\right)^2,
\label{csintopc}
\eeq
where the effects are rather modest unless the quark 
rotation matrices are larger than expected. This would not be unnatural, 
for instance, for ${\cal U}_R$, since the rotations of right-handed quarks
are not related to any known observable in the standard model. 

\subsubsection{OTHER NEW-PHYSICS SCENARIOS}
Extensions of the standard model leading to effects in rare charm decays also tend
to result in large contributions to \ddb~mixing. 
Burdman et~al.\ have evaluated a long list of these scenarios in detail~\cite{bghp2}. 
Generically, the effects are either negligible or amount to $O(1)$ enhancement
over the standard-model short-distance contributions. 

As mentioned in Section~\ref{ddb_othernp}, compact extra 
dimensions
may lead not only to massive scalar and fermionic states but also to nonuniversal couplings of the standard-model fermions to Kaluza-Klein 
excitations of gauge bosons 
that may induce flavor-violating loop effects. 
In general, the largest effects in rare charm decays are associated with 
massive neutral gauge bosons (e.g., KK excitation of a $Z'$), which 
generate a FCNC current in the up-quark sector and then decay into   
either charged leptons or neutrinos. With masses starting around the TeV scale, 
these states could lead to $O(1)$ enhancements in $c\to u\ell^+\ell^-$ 
modes, when compared to the standard-model short-distance predictions. 
Thus, in the charged-lepton modes this can translate into observable effects in the 
low $m_{\ell\ell}$ window. The enhancement in the $c\to u \nu\bar\nu$ modes
could be several orders of magnitude above the standard-model predictions, although they may be 
very difficult to observe. 

Many other new-physics scenarios involving additional matter and/or gauge fields
would lead to contributions to flavor physics and in particular to rare charm decays.
Most contributions that are potentially large tend to be 
constrained by \ddb~mixing. Such is the case, for instance, with extra down-type
quarks and extra gauge bosons. 

\subsection{Experimental Bounds and Prospects}

Experiments that can measure rare $D$
decay branching ratios at the level of a part per million will start to
confront models of new physics in an interesting way. Although
current experimental limits are, in most cases, well above this
level of sensitivity, the arsenal of experiments
opening up from CDF to the $B$ factories and charm factory may allow
observation of rare decays within the decade.  

We first review the present experimental status of charm rare decays. 
We then provide 
estimates of the reach of the currently running experiments CLEO-c and 
BaBar/Belle made by the CLEO-c collaboration, estimate the rare-decay 
reach of future facilities, and finally
briefly consider the experimental dilepton mass resolution in decays of 
the type  $D \rightarrow  \pi e^+ e^-$. 

Fixed-target experiments have traditionally had the best rare-decay limits for low-multiplicity final states involving charged 
particles.  In fixed-target experiments, rare decays are isolated using 
the same procedure used to isolate nonrare charm decays.  The procedure 
is described in Section 2.4.6 of this review.

Table~\ref{tab:Will} and Figure~\ref{fig:Will}
show some of the best recent results and compare them to theory.
At CDF, the search strategy for rare charm decays is still being refined
but it will be similar to that developed for rare $B$ decays such as
$B^+\to K^+\mu^+\mu^-$~\cite{cdf_b2kmm}.
Recently CDF has produced the best
upper limit for $D^0 \rightarrow \mu^+ \mu^-$~\cite{CDF_ACP}.

At $e^+e^-$ $B$ factories operating at the $\Upsilon(4S)$, the background to
rare charm decays is very large, and so far searches have been performed
only for $D^0$ modes where the $D^*$ tag can be used to remove the
significant combinatoric background. 
Using this technique, the first observation of a FCNC mode in the charm
system has recently been made by BELLE \cite{Yabsley}, 
as shown in Figure~\ref{d2figamma}. 
The measured branching ratio is 
$B(D^0\to\phi\gamma)=(2.60^{+0.70+0.15}_{-0.61-0.17})\times 10^{-5}$.

Tables~\ref{tab:YB20pg112} and ~\ref{tab:YB20pg112b}  list a
collection of rare $D^0$ nodes. The second column   
is the best limit currently available from an
$e^+e^-$ experiment running at or near the $\Upsilon(4S)$
resonance. The  third column shows cases where the
best limit was not obtained on the $\Upsilon(4S)$. 
The next-to-last column estimates the sensitivity of an $e^+e^-$
experiment with 400 fb$^{-1}$ of integrated luminosity at or near
the $\Upsilon(4S)$, based on the efficiencies and background
levels found by CLEO, and the final column shows the projected
CLEO-c limits based on Monte Carlo estimates of efficiency and
tagging rates, and assuming an integrated luminosity of 3 fb$^{-1}$ on the $\psi$(3770) with a 10\% tagging probability. This
amount of data could be accumulated in one year. At the $\psi(3770)$ 
the assumption is made for all modes
that the signal is background-free.

Tables ~\ref{tab:YB21pg113} and ~\ref{tab:YB21pg113b} show the
current experimental status as well as projected sensitivities for
a collection of rare $D^+$ decays. In these cases, there are no
recent limits available from $e^+e^-$ experiments, so only the
current world best measurement and the projected CLEO-c
sensitivity [again based on 3 fb$^{-1}$ of integrated luminosity
on the $\psi$(3770)] are shown.

If tagging is not needed, the limits at CLEO-c  will improve by a
factor of ten.  Although these estimated limits are quite rough,
we can draw some general conclusions from them. In general,
CLEO-c and the $B$-factory experiments have complementary strengths.
For some final states, CDF will have the greatest sensitivity.\footnote{We note that there is no commonly agreed prescription for calculating 
expected experimental upper limits. For an alternative, and more 
optimistic, estimate of the sensitivity to rare charm decays at a 
$B$ factory, see Reference \cite{Williams}.
}

The most reliable way to compare the reach of future experiments
for rare decays would be to compile a list of sensitivities
estimated by each experiment based on detailed simulations.
However, as with $\delta A_{CP}$, in almost all cases
the  simulations or other detailed estimates are nonexistent. 
Instead, to compare the reach of future experiments for rare
decays we again use the number of  $D^0 \rightarrow K^- \pi^+$
reconstructed by each experiment. The mode we choose for the
extrapolation for hadron experiments is  $D^+ \rightarrow \pi^+
\mu^+ \mu^-$. The normalization is provided by the E791 search for
that mode, which found $B(D^+ \rightarrow \pi^+ \mu^+ \mu^-) <
5.2 \times 10^{-5}$ at 90\%~CL. Because background is present, we
scale this result by $1/\sqrt{N(D^0 \rightarrow K^- \pi^+)}$. For
the $e^+e^-$ experiments, the mode we choose for the extrapolation
is $D^0 \rightarrow \pi^0 e^+ e^-$. The normalization for machines
operating in  the $B$-meson threshold region is the result of the
CLEO II search, which found
 $B(D^0 \rightarrow \pi^0 e^+
e^-) < 4.5 \times 10^{-5}$ at 90\%~CL. We scale this by $1/\sqrt
{N(D^0 \rightarrow K^- \pi^+)}$ in order 
to obtain upper limits for the  $B$-factory and super-$B$-factory experiments. 
For the charm factories,
reconstructing the second charm meson in the event ensures a
background-free situation and the scaling is  $1/L$ using the
CLEO-c estimated sensitivity as the normalization. The results of
this exercise are in Table~\ref{REACH}.

To estimate the
number of events each experiment will observe, we use 
$B(D^+ \rightarrow \pi^+\mu^+ \mu^- ) \sim 6.2 \times 10^{-6}$ 
and $B(D^0 \rightarrow \pi^0 e^+e^- ) \sim 8 \times 10^{-7}$. 
As can be seen from the results in Table \ref{REACH}, sensitivity 
to interesting physics in these modes can be achieved by BES~III, 
BTeV, or the ``super'' machines.

For the case of the dilepton modes $D\to\pi\ell^+\ell^-$ and $D\to\rho\ell^+\ell^-$, 
as emphasized in Section~\ref{exclusive}, the sensitivity to short-distance
physics, whether it is that of the standard model or one of its extensions, is in the region of low
dilepton mass away from the resonances.
Figure~14 shows our Monte Carlo simulation of the experimental dilepton 
mass resolution in the decay $D^+ \rightarrow \pi^+ e^+ e^-$ for $D$ 
mesons produced  at
the $\Upsilon(4S)$ and  for $D$ mesons produced at threshold. The 
dilepton mass resolution is excellent in both cases. The dilepton mass 
resolution will also be very good at hadron facilities. In all cases, the 
dilepton resolution is expected to be adequate to sort between the 
standard-model and new-physics predictions for this class of decays.

\section{SUMMARY}
$D^0$--$\bar{D}^0$ mixing and rare charm decays will be 
further probed by experiment in the coming years. 
We have examined the potential of these measurements to test
the standard model and its extensions. In the case of
$D^0$--$\bar{D}^0$ mixing, the irreducible background from 
long-distance physics makes it difficult for a future 
positive observation to be attributed to short-distance physics.
However, tight limits on $\Delta m_D$ result in very important complementary
constraints on physics beyond the standard model. 
On the other hand, although they are also generally 
affected by large theoretical uncertainties, we have seen that 
some rare charm decay modes can be used to probe short-distance physics. 
Such is the case with $c\to u\ell^+\ell^-$ decay modes. 
Sensitivity to the nonresonant regions in these modes could be achieved 
by experiments in the near future. In sum, these charm physics observables, 
when considered together with the results from the various $B$ and kaon
physics experiments, constitute a necessary low-energy complement
to direct searches for new physics at high energies.

\vspace*{1cm}
{\bf Acknowledgments}
G.B.\ thanks JoAnne Hewett, Eugene Golowich, and Sandip Pakvasa for 
collaborations on the subject of this review. I.S. 
thanks many experimental colleagues from the BaBar, BELLE, BESII, BTeV, 
CLEO and FOCUS collaborations for helpful discussions 
and for making results available for this review
before they were in the public domain. He also thanks Francesca Shipsey
for sharing her unique perspective on particle physics. 
Many thanks go to 
Brandon McCarty, Victor Pavlunin, Alex Smith and 
Seunghee Son for providing technical support.
The authors also want to thank Will Johns, Zoltan Ligeti, and Alexey Petrov 
for helpful conversations, suggestions and careful reading of the manuscript. 
The  work of G.B.\ was supported by the Director, Office of Science, 
Office of High Energy and Nuclear Physics of the US Department of 
Energy (DOE) under contract DE-AC0376SF00098, and that of I.S.\ by DOE contract 
DE-FG02-91ER40681.

\clearpage

\begin{figure}
\centerline{
\psfig{figure=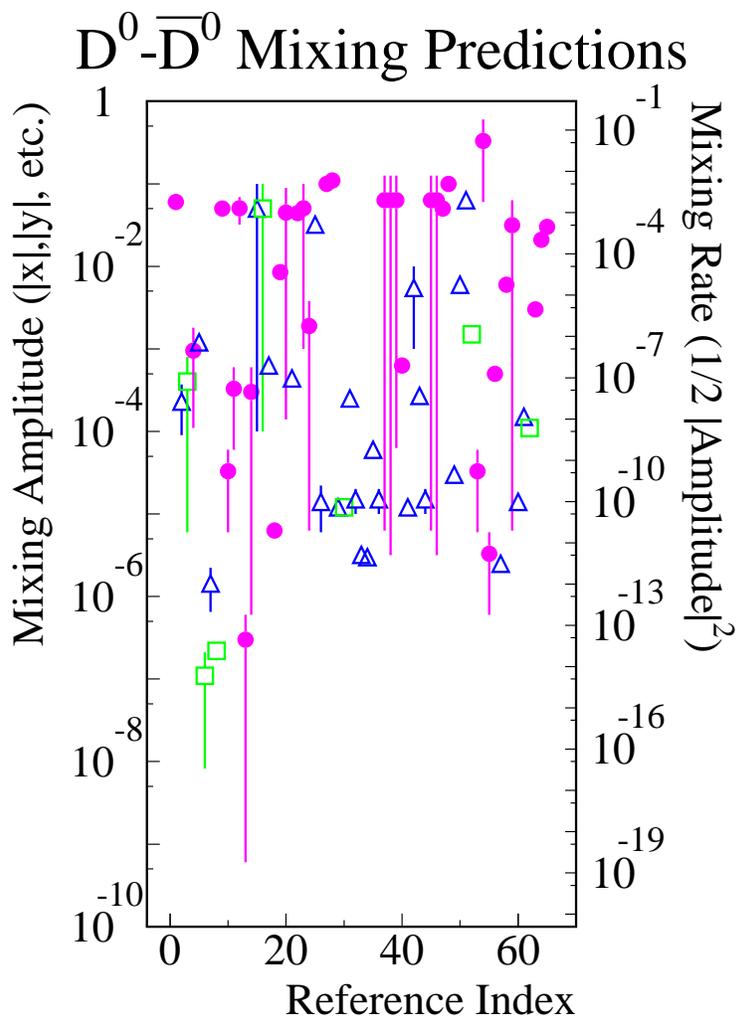,height=5.6in,angle=0}}
\caption{Predictions for \ddb ~mixing in the standard model and in some new-physics 
scenarios. From Reference~\cite{hnelson}. $\triangle$: Standard-model predictions for $x$. 
$\Box$: 
standard-model predictions
for $y$. $\bullet$:~New-physics 
predictions for $x$. The horizontal axis corresponds to 
the references in Reference~\cite{hnelson}.}
\label{hnelson}
\end{figure}

\begin{table}
\caption
{Bounds on $(\delta^u_{12})_{LL}$, $(\delta^u_{12})_{LR}$ from 
$D^0$--$\bar{D^0}$
mixing~\cite{wyler} (neglecting the strong phase).
%%
%% COMP: Please turn the rest of the caption into a footnote to the table.
%%
All constraints should be 
multiplied by $(M_{\tilde q}/500{~\rm GeV})$.
}
\label{numssm}
\begin{center}
\begin{tabular}{@{}|c|c|c|@{}} \hline\hline
$M^2_{\tilde g}/M^2_{\tilde q}$ & $(\delta^u_{12})_{LL}$ & 
$(\delta^u_{12})_{LR}$ 
\\ \hline
%& & \\
$0.3$ & $0.03$ & $0.04$ \\
$1.0$ & $0.06$ & $0.02$ \\
$4.0$ & $0.14$ & $0.02$ \\
\hline
\end{tabular}
\end{center}
\end{table}

\begin{table}
\begin{center}
\caption{A comparison of 
 the LEP
experiments~\cite{ALEPH}, CDF~\cite{CDF_trigger},
E791~\cite{E791_detector}, FOCUS ~\cite{FOCUS_detector},CLEO
~\cite{CLEO_detector},  BaBar ~\cite{BABAR_detector}, and
Belle~\cite{Belle_detector}.
Each experiment has performed
searches for charm mixing, rare decays, and $CP$
violation. $K^- \pi^+$ is the number of reconstructed $D^0
\rightarrow K^- \pi^+$ used in published measurements. The proper
time resolution for charm hadrons is denoted in $\sigma_t$.
}
\vspace*{0.3cm}
\label{tab:summary}
\begin{tabular}{|c|c|c|c|c|c|c|c|}
\hline & \multicolumn{2}{|c|}{Fixed Target}&
\multicolumn{4}{|c|}{$e^+e^-$} & $p \bar p$\\
\hline
&E791&FOCUS&LEP&CLEO&\multicolumn{2}{|c|}{ BaBar/Belle}&CDF\\
\hline Beam& Hadron &Photon& $e^+e^- \rightarrow Z^0$
&\multicolumn{3}{|c|}{
$e^+e^-$}&$p \bar p$\\
\hline
$K^- \pi^+$&$\sim 2 \times 10^4$&$\sim 2 \times 10^5$ & $\sim 10^4$ /expt. & 
$\sim 2 \times 10^5$ &\multicolumn{2}{|c|}{$\sim 10^6$}&$ \sim 5 \times 10^5$\\
\hline $\sigma_t$ & $\sim 40$ fs & $\sim 40$ fs  & $\sim 100$ fs & $\sim 140$ fs
&\multicolumn{2}{|c|}{$\sim 160$ fs} & $\sim 50$ fs\\
\hline
\end{tabular}
\end{center}
\end{table}

\begin{table}
\begin{center}
\caption{Expected charm data sets for existing experiments and
proposed facilities.
``Current'' is the size of the accumulated data
set used in published physics analyses. ``Full'' is the size of the
proposed complete data test,  or the quantity of data collected
per $10^7$~s as indicated. Super Charm is not a proposed
facility but is included for comparison. $K^+\pi^-$ is the number
of reconstructed $D^0 \rightarrow K^- \pi^+$ in the full
data set.
}
\vspace*{0.3cm}
\label{tab:summaryfuture}
\begin{tabular}{|c|c|c|c|}
\hline
Experiment &    Current &       Full &   $K^-\pi^+$ \\
\hline
BaBar &   91 fb$^{-1}$ &  500 fb$^{-1}$ & $6.6 \times 10^6$ \\
\hline
Belle &  46.2 fb$^{-1}$ & 500 fb$^{-1}$   & $6.6 \times 10^6$ \\
\hline CDF (Run II)    & 65 pb$^{-1}$ & 4.4 fb$^{-1}$  & $ 30 \times 10^6$ \\
\hline CLEO-c & -& 3 fb$^{-1}$ &  $5.5 \times 10^5$ \\
\hline BESIII & -& 30 fb$^{-1}$ &  $5.5 \times 10^6$ \\
\hline Super Charm & -& 500 fb$^{-1}$ &  $9.2 \times 10^8 / 10^7$ s \\
\hline SuperKEKB &  - & 2 ab$^{-1}$ & $ 2.5 \times 10^7 / 10^7$ s \\
\hline SuperBaBar & - & 10 ab$^{-1}$ & $ 1.3 \times 10^8 / 10^7$ s \\
\hline BTeV & - &  &  $\sim 6 \times  10^8 / 10^7$ s \\
\hline
\end{tabular}
\end{center}
\end{table}

\begin{table}
\caption{\small Comparison of the 95\%-CL limits for the fit
output parameters of BaBar, CLEO, and FOCUS when $CP$ conservation is
assumed in the fit, and when CP violation is allowed 
(denoted by CPV). The FOCUS entries are one dimensional limits.
%%
%% COMP: Please turn the rest of the caption into a footnote to the table.
%%
BaBar's limits include systematic
uncertainties and were obtained in a fit that allowed $x^{\prime 2} < 0$.
}
\centerline{\footnotesize } \centerline{
\begin{tabular}{|c||c| c| c| c| }
\hline & $R_D$ & $y^\prime$ & $x^\prime$ & $x^{\prime 2}$ \\
\hline
CLEO ~[\%]&(0.24, 0.69) & ($(-5.2, 0.2$) & ($-2.8, 2.8$)  & $<0.0076$\\
BaBar~[\%]&(0.24, 0.49) & ($(-2.7, 2.2$) & --  & $<0.20$\\
FOCUS [\%]&(0.43, 1.78) & ($-12.4, -0.6$) & ($-3.9, 3.9$) & -\\
CLEO (CPV)~[\%]&(0.24, 0.71) & ($-5.8, 1$) & ($-2.9, 2.9$) & $<0.082$  \\
BaBar (CPV)~[\%]&(0.23, 0.52) & ($-5.6, 3.9$) & -  & $<0.22$  \\
\hline
\end{tabular}}
\label{tab:xy_comp}
\end{table}

\begin{figure}
\centerline{
\psfig{figure=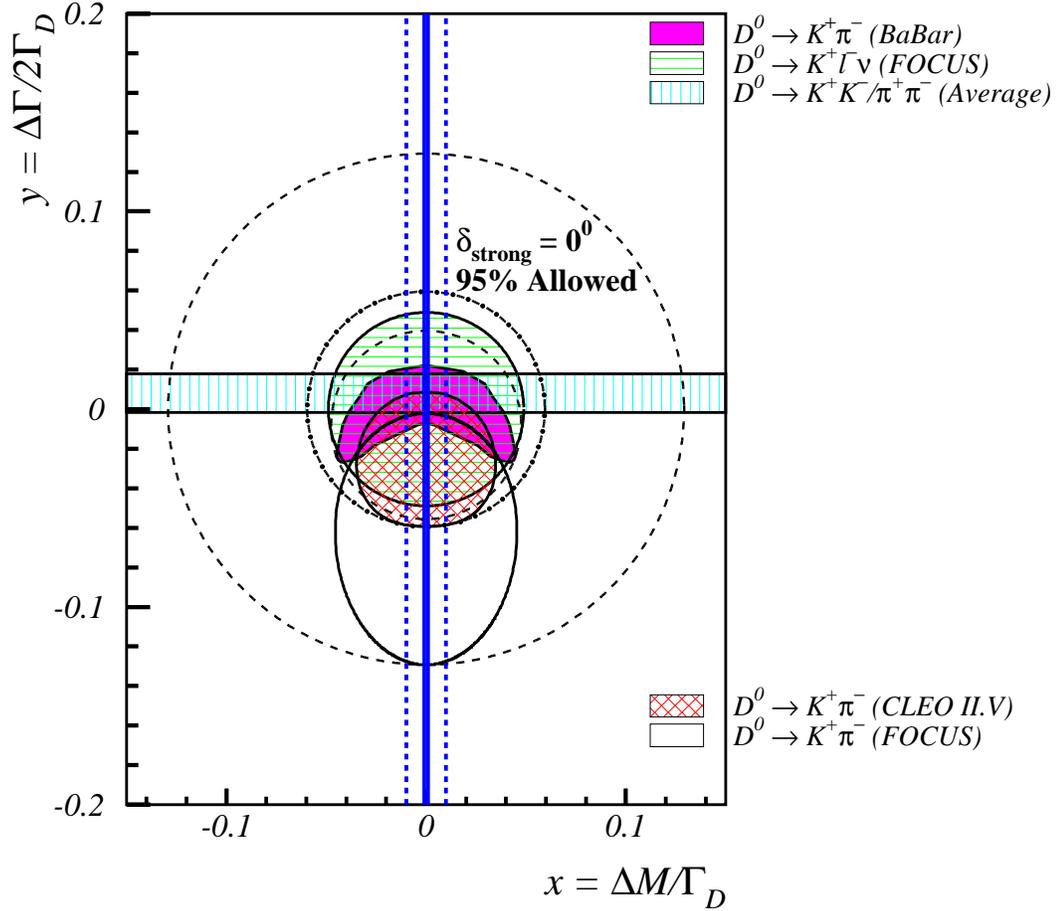,height=5.6in,angle=0}}
\caption{ Our present knowledge of $D^0$--$\bar{D^0}$ mixing. The
solid vertical lines indicate a ``typical'' standard-model
prediction for $x$. The dashed vertical lines indicate the upper
range of non-standard-model predictions for $x$. 
The horizontal band is the world average $95\%$ CL limit in $y$. The circle 
with horizontal shading is the $95\%$
CL limit in $(x,y)$.
The strong-phase
shift $\delta_{K \pi}$ between the Cabibbo-favored and doubly Cabibbo-suppressed decays
is assumed to be zero in plotting the $D^0 \rightarrow K^+ \pi^-$
results, where in each case $CP$ conservation is assumed.
For the CLEO and FOCUS measurements, the statistical error is included; for
the BaBar measurement, both the statistical and systematic errors are
included.
The strong phase shift is expected to be close to
zero, but until it is actually measured, the allowed region from the
$D^0 \rightarrow K^+ \pi^-$ measurements must be expanded to
include the area swept out by rotating these regions about the
origin. 
The three circles (small radius dashed, dot-dashed, and large radius dashed) 
are $2\pi$ rotations
of the BABAR, CLEO, and FOCUS regions, respectively. }
\label{allmix}
\end{figure}

\begin{table}
\begin{center}
\caption{ Measurements of the time-integrated rate for the 
wrong-sign decay $D^0\to K^+\pi^-$, normalized to the right sign decay
$D^0\to K^-\pi^+$}
\label{wrongsignhadronic}
\vspace*{0.3cm}
\begin{tabular}{|c|c|c|c|c|}
\hline
&$K^-\pi^+$&$K^+\pi^-$&$R_{\rm ws}[\%]$&$A_D[\%]$\\
\hline
E791~\cite{E791_KPI_WS}&5.6K&not quoted&$0.68^{+0.34}_{-0.33}\pm0.07$
&-\\
\hline
ALEPH~\cite{ALEPH_KPI_WS}&  1038 & 19 & $1.84 \pm 0.59 \pm 0.34$ &-\\
\hline
FOCUS~\cite{FOCUS_KPI_WS}&37K&150&$0.404\pm0.085\pm0.025$&-\\
\hline
CLEO~\cite{CLEO_MIX_KPI}&13.5K&45&$0.332^{+0.063}_{-0.065}\pm0.040$
&$-2^{+19}_{-20}\pm1$\\
\hline
Belle~\cite{BELLE_MIX_KPI}& 83K & 317 &$0.372 \pm 0.025^{+0.009}_{-0.014}$&-\\
\hline
BaBar~\cite{BABAR_MIX_KPI}&120K&430&$0.357\pm0.022\pm0.027$
&$9.5\pm 6.1 \pm 8.3$\\
\hline
Average&&&$0.368 \pm 0.021$&\\
\hline
\end{tabular}
\end{center}
\end{table}

\begin{table}%4 pg 9
\begin{center}
\caption{Current $D^0$ lifetime difference
measurements. BaBar (New) is a measurement of 
$y\cos\phi$ and supersedes BaBar. Only the former is used to compute 
the world average value of $y_{CP}$.} \label{4pg9lifetimedifference}
\vspace*{0.3cm}
\begin{tabular}{|c|c|c|c|c|c|}
\hline
&$K^- \pi^+$&$K^-K^+$&$\pi^-\pi^+$&$y_{CP}$&ref.\\
\hline
E791&35K&3.2K&$-$&$(0.8\pm2.9\pm1.0)\%$&~\cite{tb4_7}\\
\hline
FOCUS&120K&10K&$-$&$(3.4\pm1.4\pm0.7)\%$&~\cite{tb4_8}\\
\hline
CLEO&20K&1.9K&0.7K&$(-1.1\pm2.5\pm1.4)\%$&~\cite{tb_9}\\
\hline
BaBar&158K&16.5K&8.4K&$(1.4\pm1.0^{+0.6}_{-0.7})\%$&~\cite{tb_10}\\
\hline
Belle&214K&18.3K&$-$&$(-0.5\pm1.0\pm0.8)\%$&~\cite{tb4_11}\\
\hline BaBar (New)&220K&26K&12.8K&$(0.8\pm
0.4\pm^{+0.5}_{-0.4})\%$&~\cite{meadows}\\
\hline
\end{tabular}
\end{center}
\end{table}

\begin{figure}
\centerline{
\psfig{figure=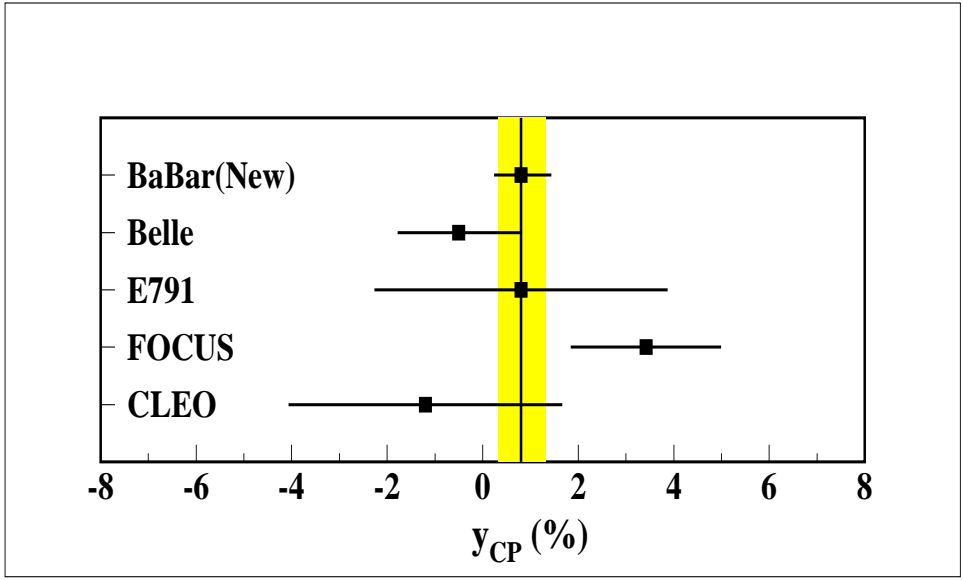,height=3.0in,width=5.0in,angle=0}}
\caption{
Comparison of experimental determinations of $y_{CP}$. 
The average value is indicated by the shaded band. We calculate the
average of the measurements to be $\langle y_{CP}\rangle
= (0.8\pm 0.5)\%.$}
\label{allycp}
\end{figure}

\begin{figure}
\centerline{\psfig{figure=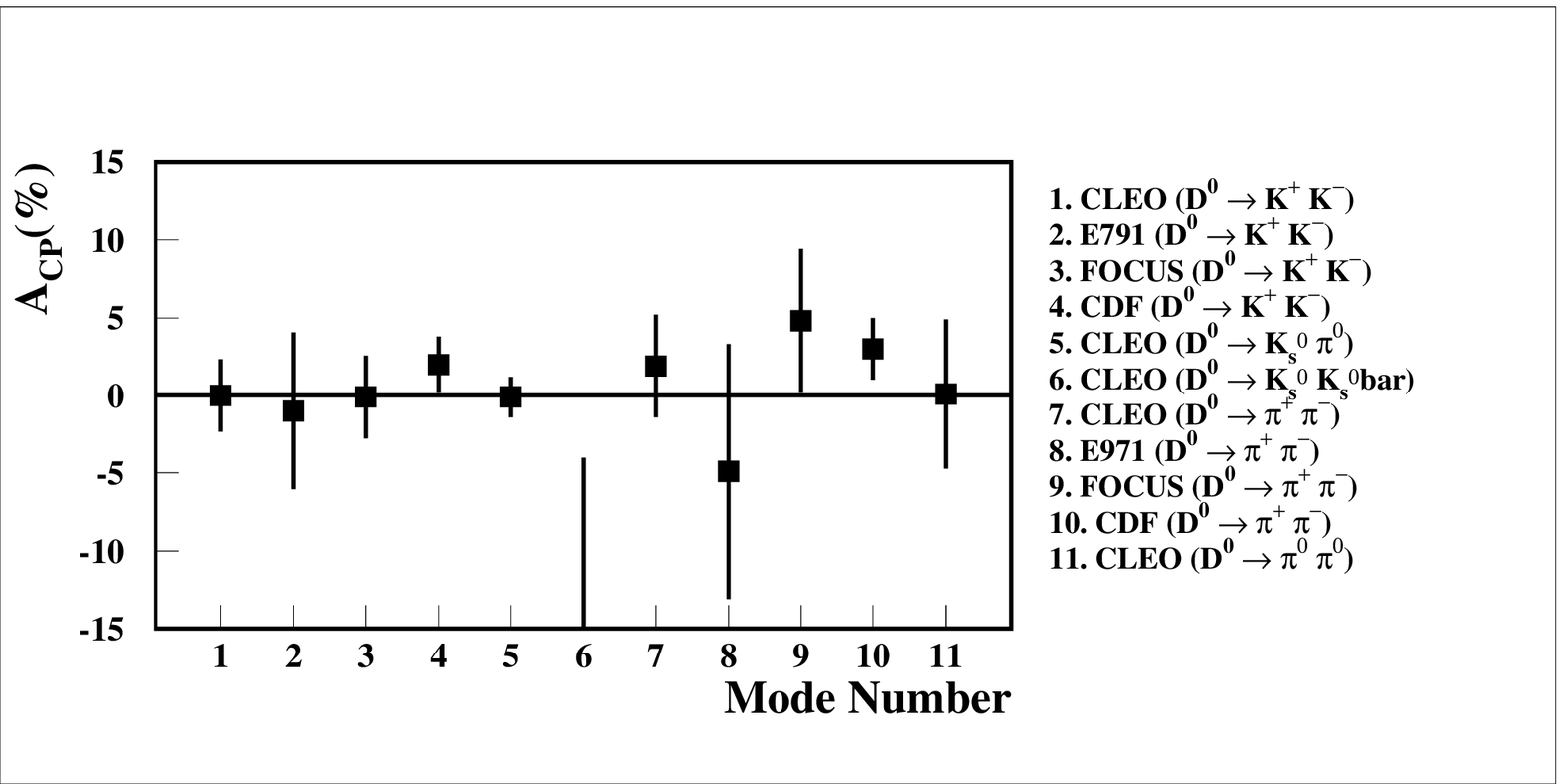,width=13.0cm}}
\caption{Measurements of $A_{CP}$ for decay modes of the $D^0$. }
\label{fig:ACP_graph1}
\end{figure}%41

\begin{table}
\begin{center}
\caption{Comparison of measurements in $A_{CP}$ for $D^0$ modes,
from E791~\cite{tb6_16},
FOCUS~\cite{tb_17},CDF~\cite{CDF_ACP}, and CLEO~\cite{tb_9}}
\label{table:acp_dzero_summary}
\vspace*{0.3cm}
\begin{tabular}{|c|c|c|c|c|}
\hline
&Mode& $A_{CP}$&Mode& $A_{CP}$\\
\hline
CLEO&$D^0 \rightarrow K^+K^-$&$(0.0\pm2.2\pm0.8)\%$&$D^0\rightarrow \pi^+ \pi^-$&$(1.9\pm3.2\pm0.8)\%$\\
\hline
E791&$D^0 \rightarrow K^+ K^-$&$(-1.0\pm4.9\pm1.2)\%$&$D^0 \rightarrow \pi^+ \pi^-$&$(-4.9\pm7.8\pm2.5)\%$\\
\hline
FOCUS&$D^0 \rightarrow K^+K^-$&$(-0.1\pm2.2\pm1.5)\%$&$D^0 \rightarrow \pi^+ \pi^-$&$(4.8\pm3.9\pm2.5)\%$\\
\hline
CDF&$D^0 \rightarrow K^+K^-$&$(2.0\pm1.7\pm0.6)\%$&$D^0 \rightarrow \pi^+\pi^-$&$(3.0\pm1.9\pm0.6)\%$\\
\hline
CLEO&$D^0 \rightarrow K^0_S \pi^0$&$(0.1\pm1.3)\%$&$D^0 \rightarrow \pi^0 \pi^0$&$(0.1\pm4.8)\%$\\
\hline
CLEO&$D^0 \rightarrow K_S \bar K^0_S$&$(-23\pm19)$\%&&\\
\hline
\end{tabular}
\end{center}
\end{table}

\begin{figure}
\centerline{\psfig{figure=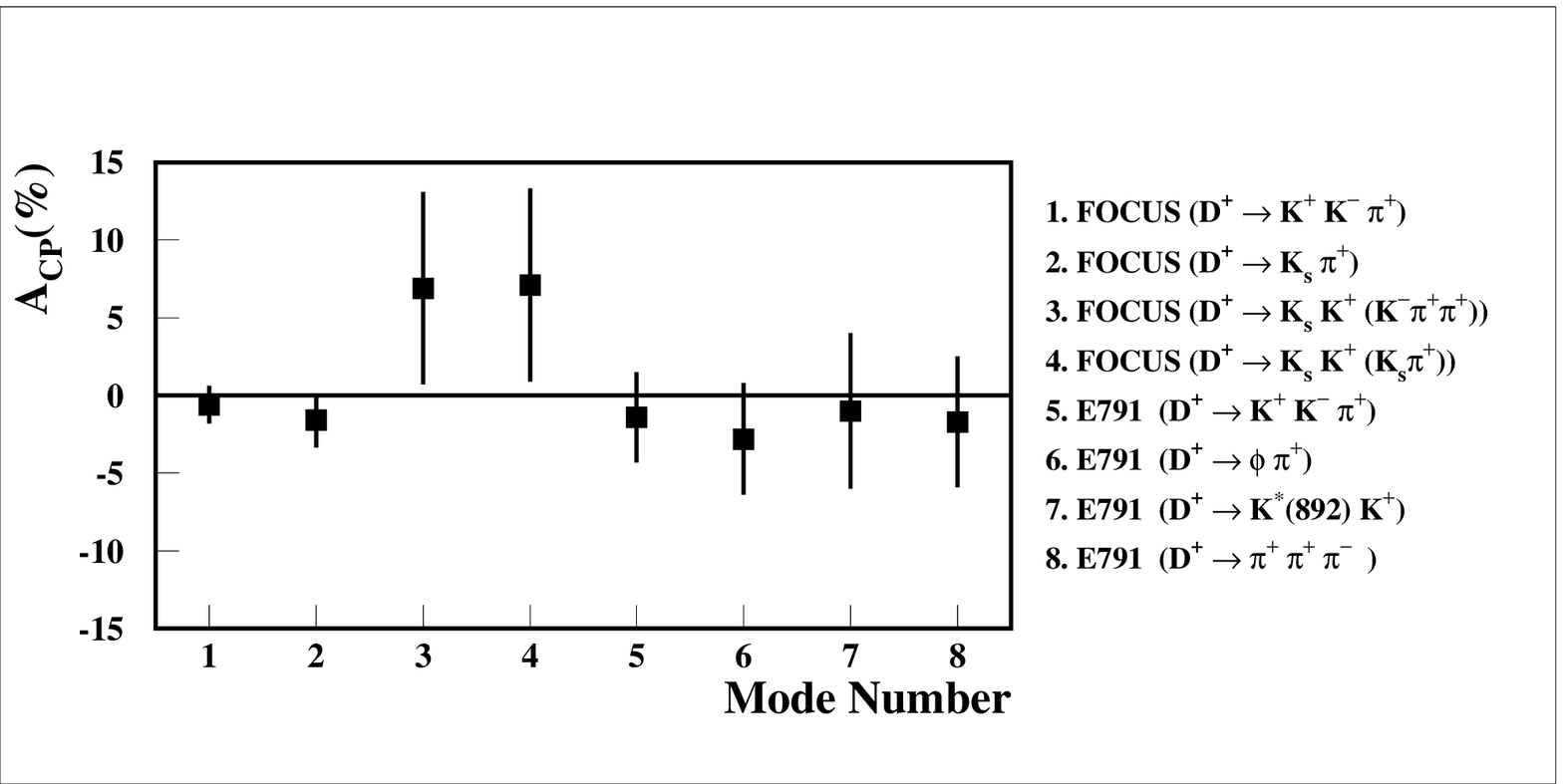,width=13.0cm}}
\caption{Measurements of $A_{CP}$ for decay modes of the $D^+$.}
\label{fig:ACP_graph2}
\end{figure}%41

\begin{table}
\begin{center}
\caption{Comparison of measurements in $A_{CP}$ for $D^+$ modes,
from E791~\cite{tb6_16} and FOCUS~\cite{tb_17}}
\label{table:acp_dcharged_summary}
\vspace*{0.3cm}
\begin{tabular}{|c|c|c|}
\hline
&Mode& $A_{CP}$\\
\hline
FOCUS&$D^+ \rightarrow K^-K^+ \pi^+$&$(0.6\pm1.1\pm0.5)\%$\\
\hline
&$D^+\rightarrow K_S \pi^+$&$(-1.6\pm1.5\pm0.9)\%$\\
\hline
&$D^+ \rightarrow K_S K^+$ w.r.t. $(K^-\pi^+\pi^+)$&$(6.9\pm6.0\pm1.5)\%$\\
\hline
&$D^+\rightarrow K_S K^+$ w.r.t.$(K_S\pi^+)$&$(7.1\pm6.1\pm1.2)\%$\\
\hline
E791&$D^+ \rightarrow K^-K^+ \pi^+$&$(-1.4\pm2.9)\%$\\
\hline
&$D^+\rightarrow \phi\pi^+$&$(-2.8\pm3.6)\%$\\
\hline
&$D^+ \rightarrow K^{*0}(892)K^+$&$(-1.0\pm5.0)\%$\\
\hline
&$D^+\rightarrow \pi^+\pi^-\pi^+$&$(-1.7\pm4.2)\%$\\
\hline
\end{tabular}
\end{center}
\end{table}

\begin{table}
\begin{center}
\caption{Yields for direct $CP$ violation in $\psi(3770)\rightarrow$
($CP$ Eigenstate 1)($CP$ Eigenstate 2) in one year at
CLEO-c \cite{YB}}\label{tab:tabYB18pg110}
\vspace*{0.3cm}
\begin{tabular}{|c|c|c|}
\hline
$CP$ Eigenstate 1&$CP$ Eigenstate 2&Events for 100\% $CP$\\
\hline\hline
$K^+K^-$    &$K^+K^-$           &$174$\\
$K^+K^-$    &$\rho^0\pi^0$      &$171$\\
$K_S \pi^0$ &$K_S \pi^0$        &$183$\\
$K_SK_S$    &$K^+K^-$           &$136$\\
\hline\hline
\end{tabular}
\end{center}
\end{table}

\begin{table}
\begin{center}
\caption{Direct $CP$-violation asymmetry $\delta A$ reach from $\psi
(3770) \rightarrow$~(flavor tag)~($CP$ eigenstate), in one year 
at CLEO-c \cite{YB}}\label{tab:tabYB19pg110}
\vspace*{0.3cm}
\begin{tabular}{|c|c|c|}
\hline\hline
$CP$ Eigenstate&Flavor-Tagged Sample&$\delta A$\\
\hline
$K^+K^-$&$10200$&$0.01$\\
$K_{S}\pi^0$&$10400$&$0.01$\\
$K_S \omega$&$3500$&$0.02$\\
\hline\hline
\end{tabular}
\end{center}
\end{table}

\begin{table}
\begin{center}
\caption{Projected $CP$ asymmetry and sensitivity to rare decays for
the full data set of  existing experiments.
%%
%% COMP: Please turn the rest of the caption into a footnote to the table.
%%
For next-generation
experiments, the sensitivities and rates are per year of $10^7$~s. 
The yields at charm factories are computed without
requiring a charm tag. Super Charm is not a proposed facility but
is included for comparison. All entries are estimates by the authors.
}\label{REACH}
\vspace*{0.3cm}
\begin{tabular}{|c|c|c|c|}
\hline
Experiment &   $\delta A_{CP}$ &  $D \rightarrow \pi \ell^+ \ell^-$  limit &   
$D \rightarrow \pi \ell^+ \ell^-$ 
yield\\
\hline
BaBar &  $3 \times 10^{-3}$  &   $1 \times 10^{-6}$& 6  \\
\hline
Belle &  $3 \times 10^{-3}$ &  $1 \times 10^{-6}$    & 6\\
\hline CDF (Run II)    & $1 \times 10^{-3}$ &  $2 \times 10^{-6}$ & 4 \\
\hline CLEO-c & 0.01 &   $4 \times 10^{-6}$ & 4 \\
\hline BES~III & $3 \times 10^{-3}$ &   $4 \times 10^{-7}$ & 40\\
\hline Super Charm &  $2 \times 10^{-4}$  &   $2 \times 10^{-8}$&  700 \\
\hline Super-KEK-B & $1 \times 10^{-3}$ &  $5 \times 10^{-7}$ & 60 \\
\hline SuperBaBar & $6 \times 10^{-4}$ &  $3 \times 10^{-7}$ & 300 \\
\hline BTeV & $3 \times 10^{-4}$&   $4 \times 10^{-7}$ & 200  \\
\hline
\end{tabular}
\end{center}
\end{table}

\clearpage

\begin{figure}
\centerline{\psfig{figure=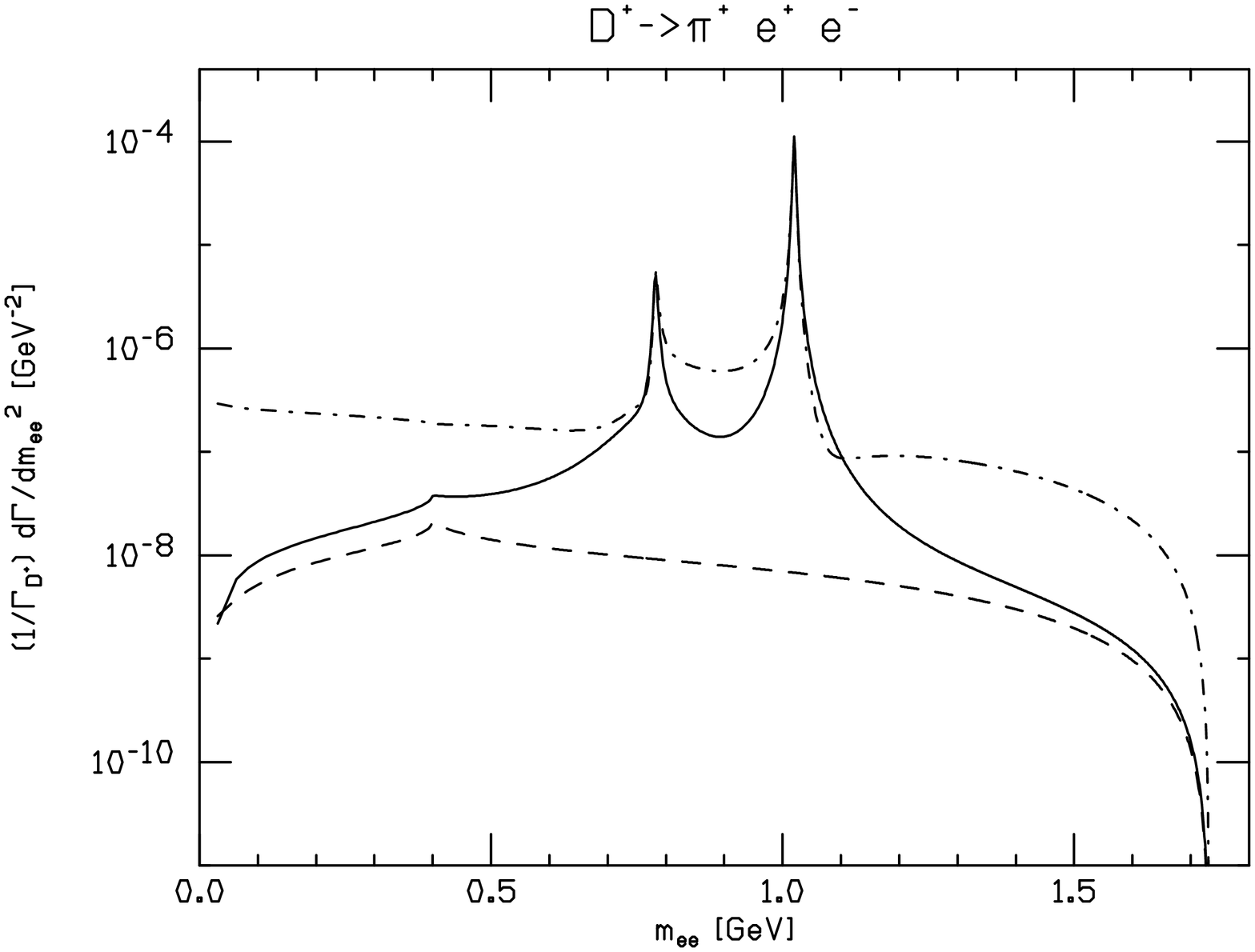,height=3.2in,angle=90}}
\caption{The dilepton mass distribution for the 
$D^+\to\pi^+ e^+e^-$, normalized to $\Gamma_{D^+}$. 
The solid line shows the sum of the short- and long-distance 
standard-model contributions.
The dashed line represents the short-distance contribution only.
The  dotted-dashed line includes the contribution of  $R$-parity-violating terms in 
SUSY (see Section~\ref{sec_rpv}).}
\label{pill}
\end{figure}

\begin{figure}
\centerline{\psfig{figure=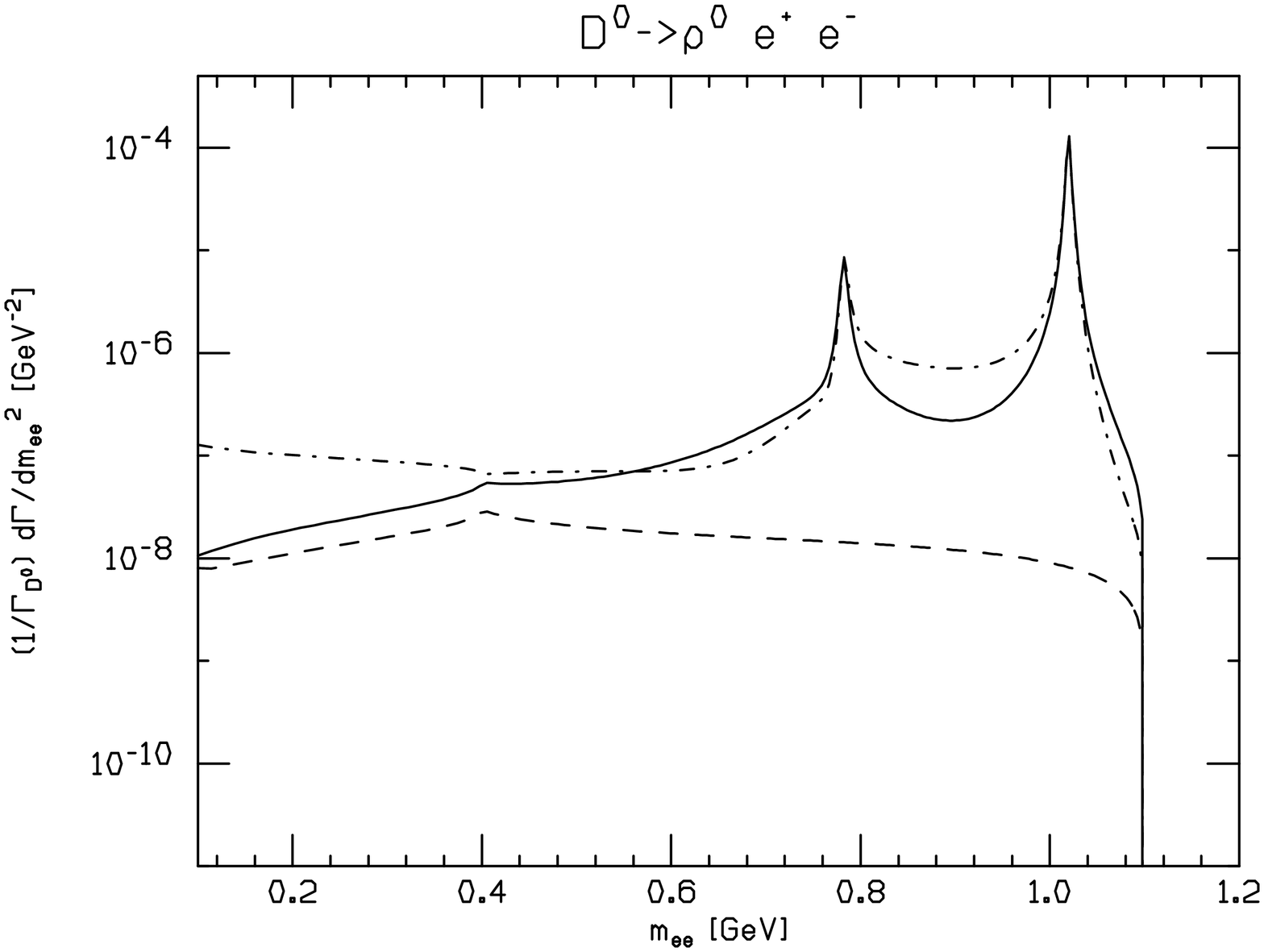,height=3.2in,angle=90}}
\caption{The dilepton mass distribution for the 
$D^0\to\rho^0 e^+e^-$, normalized to $\Gamma_{D^0}$. 
The solid line shows the sum of the  short- and long-distance standard-model contributions.
The dashed line represents the short-distance contribution only.
The dotted-dashed line includes the contribution of  $R$-parity-violating terms in 
SUSY (see Section~\ref{sec_rpv}).} 
\label{rholl}
\end{figure}

\begin{table}
\caption{Theoretical predictions for charm radiative  decays ($Br\times10^6$)}
\label{drad}
\vspace*{0.3cm}
\begin{center}
\begin{tabular}{@{}|c|c|c|c|@{}}
\hline\hline
$D\to V\gamma$ & Reference~\cite{bghp} & Reference~\cite{fajld2} & Reference~\cite{ksw}\\
\toprule
$D_s^+\to\rho^+\gamma$ & $6-38$ & $20-80$ & $4.4$ \\
$D^0\to\bar{K}^{*0}\gamma$ & $7-12$ & $6-36$ & $0.18$ \\
$D^0\to\rho^0\gamma$ & $0.1-0.5$ & $0.1-1$ & $0.38$ \\
$D^0\to\omega^0\gamma$ & $\simeq 0.2$ & $0.1-0.9$ & $-$ \\
$D^0\to\phi^0\gamma$ & $0.1-3.4$ & $0.4-0.9$ & $-$ \\
$D^+\to\rho^+\gamma$ & $2-6$ & $0.4-6.3$ & $0.43$ \\
$D_s^+\to K^{*+}\gamma$ & $0.8-3$ & $1.2-5.1$ & $-$ \\
\botrule
\end{tabular}
\end{center}
\end{table}

\begin{figure}
\centerline{
\psfig{figure=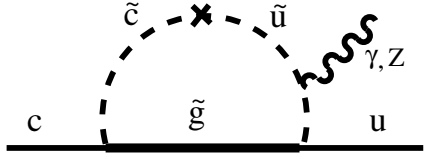,height=3.8in,angle=-90}}
\vskip-2.5cm
\caption{A typical contribution to $c\to u$ FCNC transitions in 
the MSSM. The cross denotes one  mass insertion 
$(\delta^{12})_{\lambda\lambda'}$, with $\lambda,\lambda'=L,R$.}
\label{mssm}
\end{figure} 

\begin{figure}
%\hskip 3.0cm
\centerline{\psfig{figure=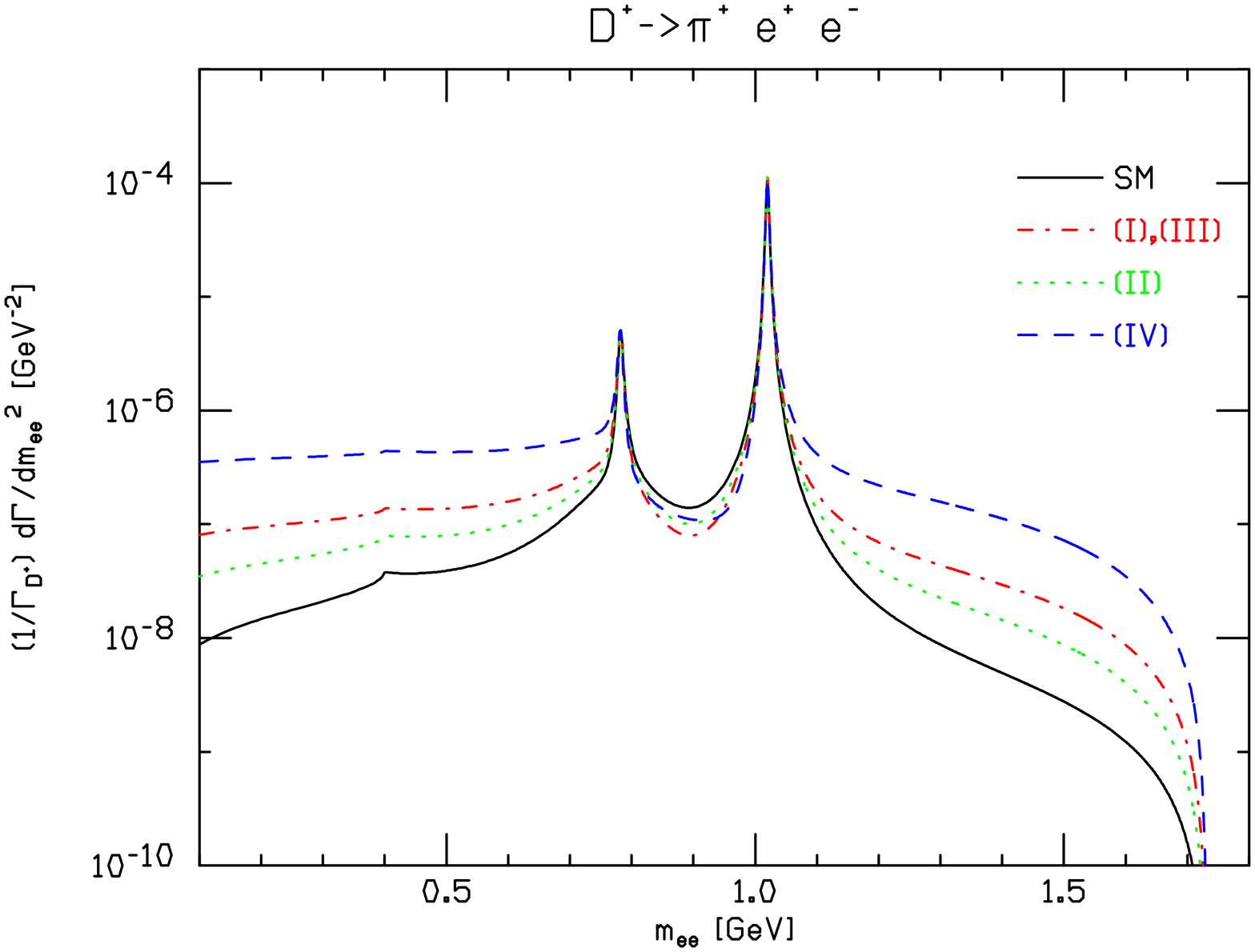,height=3.2in,angle=90}}
\caption{The dilepton mass distribution for  
$D^+\to\pi^+ e^+e^-$ (normalized to $\Gamma_{D^+}$) 
in the MSSM with nonuniversal 
soft breaking effects.
The solid line is the standard model. (I) $M_{\tilde g}=M_{\tilde q}=250$~GeV; 
(II) $M_{\tilde g}=2\,M_{\tilde q}=500$~GeV; 
(III) $M_{\tilde g}=M_{\tilde q}=1000$~GeV; 
(IV) $M_{\tilde g}=(1/2)\,M_{\tilde q}=250$~GeV.
}
\label{pill_mssm}
\end{figure}

\begin{figure}
\centerline{
\psfig{figure=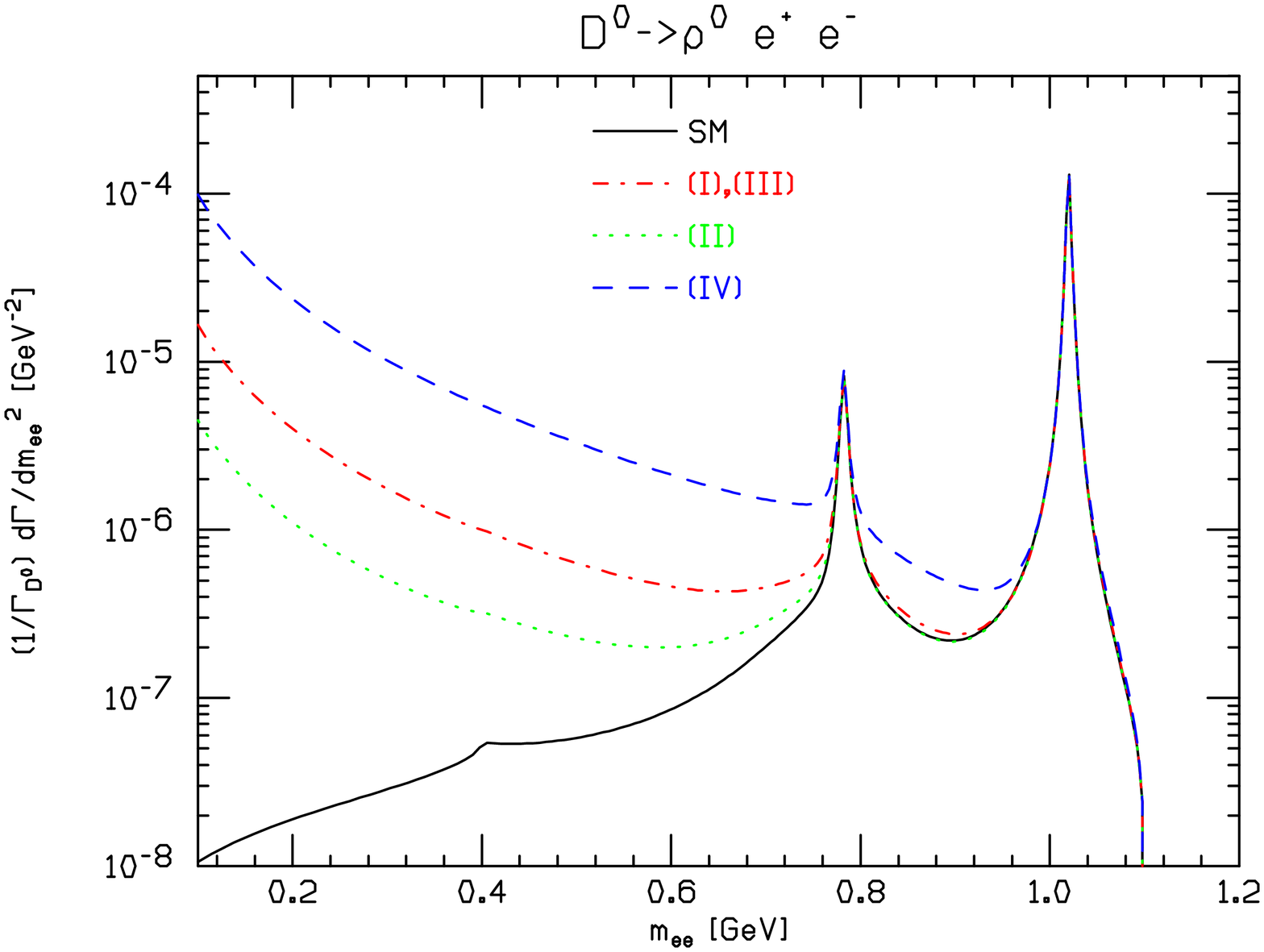,height=3.2in,angle=90}}
\caption{The dilepton mass distribution for  
$D^0\to\rho^0 e^+e^-$ (normalized to $\Gamma_{D^0}$)
in the MSSM with nonuniversal 
soft breaking effects.
The solid line is the standard model. (I) $M_{\tilde g}=M_{\tilde q}=250$~GeV; 
(II) $M_{\tilde g}=2\,M_{\tilde q}=500$~GeV; 
(III) $M_{\tilde g}=M_{\tilde q}=1000$~GeV; 
(IV) $M_{\tilde g}=(1/2)\,M_{\tilde q}=250$~GeV.
}
\label{rholl_mssm}
\end{figure}

\begin{table}
\caption
{Most stringent ($2\sigma$) bounds for the $R$-parity violation couplings 
entering 
in rare $D$ decays, from ($a$) charged-current universality,
($b$) $R_\pi$, and ($c$) $D\to K\ell\nu^*$.
%%
%% COMP: Please turn the rest of the caption into a footnote to the table.
%%
See Ref.~\cite{rpv1}
for details. All numbers should be multiplied by 
$(m_{\tilde{d}^k_R}/100{\rm ~GeV)}$.
}
\label{rpvbounds}
\begin{center}
\begin{tabular}{@{}|c|c|c|c|@{}} \hline\hline
 & & & \\
$\tilde{\lambda}'_{11k}$ & $\tilde{\lambda}'_{12k}$ & 
$\tilde{\lambda}'_{21k}$ & $\tilde{\lambda}'_{22k}$\\ \hline
$0.02^{\rm (a)}$ & $0.04^{\rm (a)}$ & $0.06^{\rm (b)}$ & $0.21^{\rm (c)}$ \\
\hline
\end{tabular}
\end{center}
\end{table}

\begin{table}
\caption{Comparison of various decay modes between the standard model and 
$R$-parity violation.
%%
%% COMP: Please turn the rest of the caption into a footnote to the 3rd-column heading
%%
The third column shows how large the $R$-parity--violating effect can be. 
}
\centering
\vspace*{0.3cm}
\begin{tabular}{|l|c|c|} \hline\hline
Decay Mode & Standard Model & ${\not R_p}$ \\ \hline
$D^+\to\pi^+e^+e^-$ & $2.0\times10^{-6}$ & $2.3\times10^{-6}$
\\ \hline
$D^0\to\rho^0 e^+ e^-$ & $1.8\times10^{-6}$ & $5.1\times10^{-6}$ 
\\ \hline
$D^+\to\pi^+\mu^+\mu^-$ &  $1.9\times10^{-6}$ & $1.5\times10^{-5}$ 
\\ \hline
$D^0\to\rho^0 \mu^+ \mu^-$ & $1.8\times10^{-6}$ & $8.7\times10^{-6}$ \\ \hline
$D^0\to\mu^+\mu^-$ & $3.0\times10^{-13}$ & $3.5\times10^{-7}$ \\ \hline
$D^0\to e^+e^-$ & $10^{-23}$ & $1.0\times10^{-10}$\\ \hline
$D^0\to\mu^+ e^-$ & $0$ & $1.0\times10^{-6}$ \\ \hline
$D^+\to\pi^+\mu^+e^-$ & $0$ & $3.0\times10^{-5}$  \\ \hline
$D^0\to\rho^0 \mu^+ e^-$ & $0$ & $1.4\times10^{-5}$ \\
\hline
\end{tabular}
\label{rpv_table}
\end{table}

\begin{table}
\caption{FOCUS results with incorporated systematic errors for the
modes shown (from Reference~\cite{ybA})
%%
%% COMP: Please turn the rest of the caption into a footnote to the table.
%%
Each number represents a $90\%$ upper limit for the
branching ratio of the decay mode listed.
} 
\label{tab:Will}.
\begin{center}
\begin{tabular}{@{}|c|c|@{}}
\hline\hline
Decay Mode & Result\\
\hline
$D^+ \rightarrow K^+ \mu^+ \mu^-$ & $9.2 \times 10^{-6}$\\
$D^+ \rightarrow K^- \mu^+ \mu^+$ & $13.1 \times 10^{-6}$\\
$D^+ \rightarrow \pi^+ \mu^+ \mu^-$ & $8.8 \times 10^{-6}$\\
$D^+ \rightarrow \pi^- \mu^+ \mu^+$ & $4.8 \times 10^{-6}$\\
$D^{+}_{S} \rightarrow K^+ \mu^+ \mu^-$ & $3.6 \times 10^{-5}$\\
$D^{+}_{S} \rightarrow K^- \mu^+ \mu^+$ & $1.3 \times 10^{-5}$\\
$D^{+}_{S} \rightarrow \pi^+ \mu^+ \mu^-$ & $2.6 \times 10^{-5}$\\
$D^{+}_{S} \rightarrow \pi^- \mu^+ \mu^+$ & $2.9 \times 10^{-5}$ \\
\botrule
\end{tabular}
\end{center}
\end{table}

\begin{figure}
\centerline{\psfig{figure=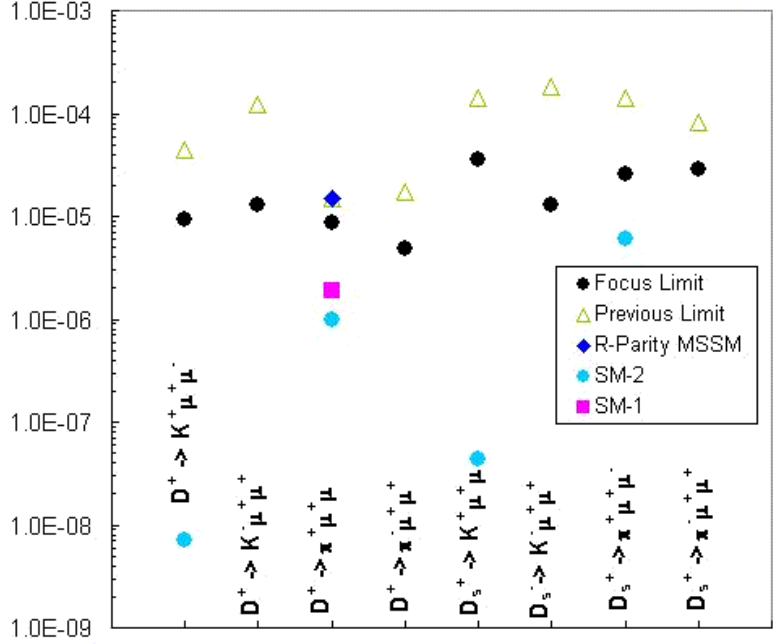,height=3.5in}}
\caption{FOCUS limits and sensitivity
%~\cite{sense} 
from Reference~\cite{ybA}, 
compared to other experiments and theory.
(For a treatment of sensitivity, see \cite{sense}.)
The previous limits, except for the E687 $D^+ \to K^- \mu^+ \mu^+
$ \cite{yb40}, are from the Fermilab experiment E791~\cite{yb35}. The
theory estimates come from Reference~\cite{bghp2} ($R$-parity MSSM
and standard model--1) and Reference~\cite{fajfer2} (standard model--2). 
The results for 
four-body decays measured by
E791 \cite{Aitala:2000kk} are not shown.} 
\label{fig:Will}
\end{figure}

\begin{table}%table 8
\begin{center}
\caption{Current and projected 90\%-CL upper limits in parts per million on rare $D^0$
decay modes.}\label{tab:YB20pg112}
\vspace*{0.3cm}
\begin{tabular}{|c|cc|cc|cc|}\hline\hline
Mode &\multicolumn{2}{|c|}{$\Upsilon(4S)$CLEO} &\multicolumn{2}{|c|}
{Best Upper Limit}&$B$ Factory&CLEO-c\\
    &Reference& &Reference&\\

\hline
$e^+e^-$       &~\cite{yb34}&$13$&~\cite{yb35}&$6.2$& $0.6$&$1.2$\\
$\mu^+\mu^-$   &~\cite{yb34}&$34$&~\cite{CDF_ACP}&$2.4$& $3.4$&$1.2$\\
$\mu^\pm e^\mp$&~\cite{yb34}&$19$&~\cite{yb35}&$8.1$&$1.9$&$1.2$\\
$\pi^0 e^+e^-$&~\cite{yb34}&$45$&               &     &$1.0$&$4$\\
$\pi^0\mu^+\mu^-$&~\cite{yb34}&$540$&~\cite{yb37}&$180$&$53$&$4$\\
$\pi^0 \mu^\pm e^\mp$&~\cite{yb34}&$86$&&&$8.5$&$4$\\
$\bar K^0e^+e^-$     &~\cite{yb34}&$110$&&&$2.6$&$7$\\
$\bar K^0 \mu^+\mu^-$&~\cite{yb34}&$670$&~\cite{yb37}&$260$&$46$&$7$\\
$\bar K^0 \mu^\pm e^\mp$&~\cite{yb34}&$100$&&&$2.3$&$7$\\
$\eta e^+e^-$&~\cite{yb34}&$110$&&&$1.1$&$10$\\
$\eta \mu^+\mu^-$&~\cite{yb34}&$530$&&&$5.1$&$10$\\
$\eta \mu^\pm e^\mp$&~\cite{yb34}&$100$&&&$1$&$10$\\
\botrule
\end{tabular}
\end{center}
\end{table}

\begin{table}%table 9
\begin{center}
\caption{Current and projected 90\%-CL upper limits in parts per million on rare $D^0$
decay modes, (with the exception of $\phi \gamma$, 
where observation has been recently reported)}\label{tab:YB20pg112b}
\vspace*{0.3cm}
\begin{tabular}{|c|cc|cc|cc|}\hline\hline
Mode &\multicolumn{2}{|c|}{$\Upsilon(4S)$} &\multicolumn{2}{|c|}{Best Upper Limit}&$B$ Factory&CLEO-c\\
    &Reference& &Reference&    \\

\hline

$\rho e^+e^-$&~\cite{yb34}&$100$&&&$9.9$&$2.5$\\
$\rho \mu^+\mu^-$&~\cite{yb34}&$490$&~\cite{yb37}&$230$&$48$&$2.5$\\
$\rho \mu^\pm e^\mp$&~\cite{yb34}&$49$&&&$4.8$&$2.5$\\
$\omega e^+e^-$&~\cite{yb34}&$180$&&&$17$&$9$\\
$\omega \mu^+\mu^-$&~\cite{yb34}&$830$&&&$57$&$9$\\
$\omega \mu^\pm e^\mp$&~\cite{yb34}&$120$&&&$11$&$9$\\
$K^*(892)^0 e^+e^-$&~\cite{yb34}&$140$&~\cite{yb38}&$24$&$14$&$4$\\
$K^*(892)^0 \mu^+\mu^-$&~\cite{yb34}&$1180$&~\cite{yb38}&$47$&$54$&$4$\\
$K^*(892)^0 \mu^\pm e^\mp$&~\cite{yb34}&$100$&~\cite{yb38}&$83$&$6.9$&$4$\\
$\phi e^+e^-$&~\cite{yb34}&$52$&&&$5.1$&$5$\\
$\phi \mu^+\mu^-$&~\cite{yb34}&$410$&~\cite{yb38}&$31$&$28$&$5$\\
$\phi \mu^\pm e^\mp$&~\cite{yb34}&$34$&&&$3.4$&$5$\\
$\pi^+\pi^-\pi^0\mu^+\mu^-$&$\rm
none$&&~\cite{yb37}&$810$&&$6$\\\hline
$\rho \gamma$      &~\cite{yb39}  &$240$  &&&  $26$&$1.5$\\
$\omega \gamma$    &~\cite{yb39}  &$240$  &&&  $26$&$8$\\
$\phi \gamma$      &~\cite{Yabsley}  &$26^{+7 +1.5}_{-6-1.7}$  &&&  $21$&$4$\\
$K^*(892)^0 \gamma$&~\cite{yb39}  &$760$  &&&  $380$&$12$\\
$\gamma \gamma$    & ~\cite{Yongsheng_gamgam} & 29&&&
&$1.2$\\\hline\hline
\end{tabular}
\end{center}
\end{table}

%\clearpage
\begin{table}
\begin{center}
\caption{Current and projected 90\%-CL upper limits on rare $D^+$
decay modes.}\label{tab:YB21pg113}
\vspace*{0.3cm}
\begin{tabular}{|c|cc|c|}\hline\hline
Mode        &   \multicolumn{2}{|c|}{Best Upper Limit}       &   CLEO-c\\
        &Reference   &   & \\
\hline\hline
$\pi^+e^+e^-$&~\cite{yb35} &$52$&1.5\\
$\pi^+\mu^+\mu^-$&~\cite{ybA}&$8.8$&$1.5$\\
$\pi^+\mu^\pm e^\mp$&~\cite{yb35}&$34$&$1.5$\\
$\pi^- e^+ e^-$&~\cite{yb35}&$96$&$1.5$\\
$\pi^- \mu^+\mu^+$&~\cite{ybA}&$4.8$&$1.5$\\
$\pi^- \mu^+ e^+$&~\cite{yb35}&$50$&$1.5$\\
$K^+e^+e^-$&~\cite{yb35}&$200$&$1.5$\\
$K^+\mu^+\mu^-$&~\cite{ybA}&$9.2$&$1.5$\\
$K^+\mu^\pm e^\mp$&~\cite{yb35}&$68$&$1.5$\\
$K^-e^+e^+$&~\cite{yb40}&$120$&$1.5$\\
$K^-\mu^+\mu^+$&~\cite{ybA}&$13.1$&$1.5$\\
$K^-\mu^+e^+$&~\cite{yb40}&$130$&$1.5$\\
\botrule
\end{tabular}
\end{center}
\end{table}

%\clearpage
\begin{table}
\caption{Current and projected 90\%-CL upper limits in parts per million on rare $D^+$
decay modes.}\label{tab:YB21pg113b}
\begin{center}
\begin{tabular}{|c|cc|c|}\hline\hline
Mode        &   \multicolumn{2}{|c|}{Best Upper Limit}       &   CLEO-c\\
        &Reference   &     &   \\
\hline\hline
$\rho^+e^+e^-$&&&$7$\\
$\rho^+\mu^+\mu^-$&~\cite{yb37}&$560$&$7$\\
$\rho^+\mu^\pm e^\mp$&&&$7$\\
$\rho^-e^+e^+$&&&$7$\\
$\rho^-\mu^+\mu^+$&~\cite{yb37}&$560$&$7$\\
$\rho^-\mu^+e^+$&&&$7$\\
$K^*(892)^+e^+e^-$&&&$10$\\
$K^*(892)^+\mu^+\mu^-$&&&$10$\\
$K^*(892)^+\mu^\pm e^\mp$&&&$10$\\
$K^*(892)^- e^+ e^+$&&&$10$\\
$K^*(892)^-\mu^+\mu^+$&~\cite{yb37}&$850$&$10$\\
$K^*(892)^-\mu^+e^+$&&&$10$\\\hline\hline
\end{tabular}
\end{center}
\end{table}

\begin{figure}
\centerline{
\epsfig{figure=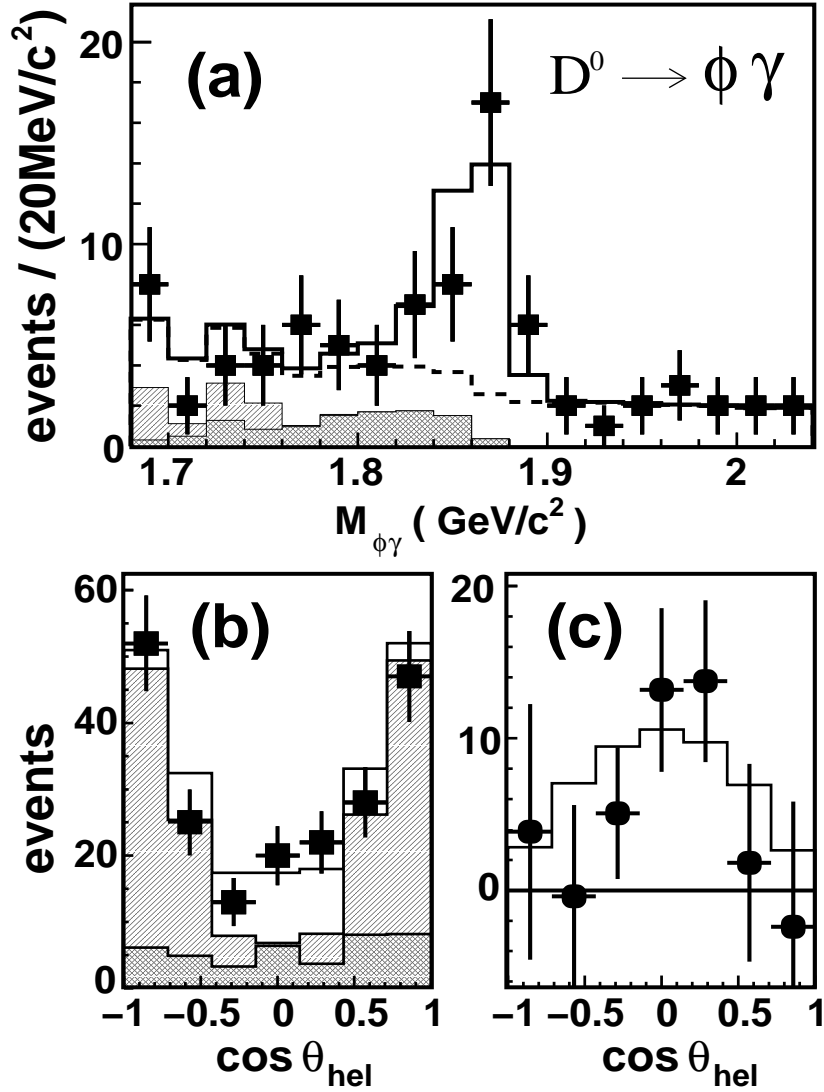,height=5.7in,width=4.3in}}
\caption{The observation of $D^0 \rightarrow \phi\gamma$, $\phi 
\rightarrow K^+ K^-$ by the
BELLE collaboration (\cite{Yabsley}). 
(a) The invariant mass distribution of $\phi\gamma$ pairs for data (points 
with error bars), the fit to the data (line), the background component 
of the fit (dashed line), $\phi \pi^0$
background (dark shaded histogram), and the sum of
$\phi \eta$ and $D^+ \rightarrow \pi^+ \pi^0$ backgrounds (light shaded 
histogram). (b) The helicity angle $\theta_{hel}$ distribution in the 
signal region, where $\theta_{hel}$ is defined as the angle between
the direction of the $K^+$ momentum vector and the direction of the 
$D^0$ momentum vector in the rest frame of the $\phi$ meson.
Data (points with error bars), MC prediction (line), total background
(light shaded histogram), and non $\phi \pi^0$ background (dark shaded 
histogram. (c) Background subtracted  $\theta_{hel}$ distribution in the 
signal region (points with error bars). The MC prediction is the line. 
The data are consistent with a $\sin ^2 \theta_{hel}$ distribution as 
expected for $D^0 \rightarrow \phi \gamma$.}
\label{d2figamma}
\end{figure}

\begin{figure}
\centerline{\psfig{figure=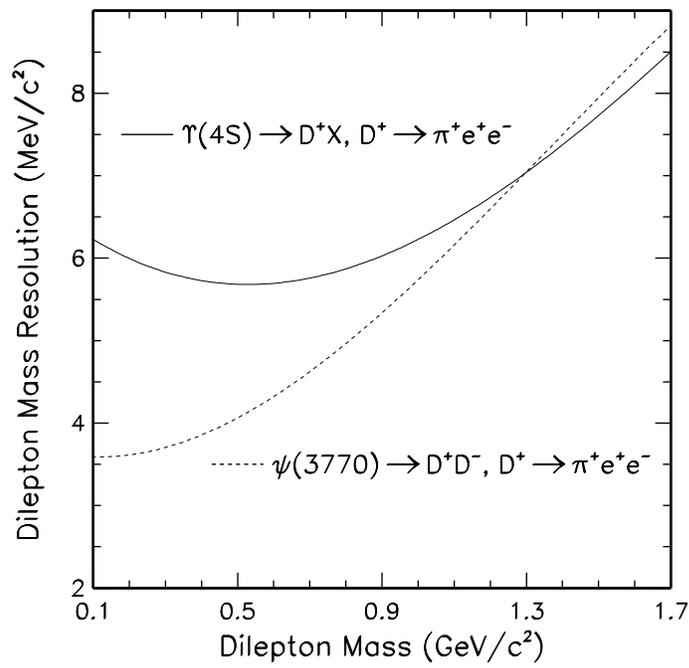,width=9.0cm}}
\caption{Dilepton mass resolution in $D^+\to\pi^+e^+e^-$ for 
$D$s produced at the $\Upsilon(4S)$ (solid) and at the 
$\psi(3770)$ (dashed).} 
\label{dlmass_sens}
\end{figure}

\end{document}